%% file: mopra_co_iii.tex
\documentclass{pasa}%

\title[Mopra CO Data Release 3]{The Mopra Southern Galactic Plane CO Survey --- \\Data Release 3}
\author[Braiding et al.]{Catherine~Braiding$^{1,8}$, G.~F.~Wong$^{1,2}$, N.~I.~Maxted$^{1,2}$, D.~Romano$^1$, M.~G.~Burton$^{1,3}$, R. Blackwell$^4$, M.~D. Filipovi\'c$^2$, 
M.~S.~R.~Freeman$^{1}$, B.~Indermuehle$^5$, J.~Lau$^4$, 
D.~Rebolledo$^{6,7}$, G.~Rowell$^4$, C.~Snoswell$^4$,  N.~F.~H.~Tothill$^2$, F.~Voisin$^4$, and P.~de~Wilt$^4$\\
    \affil{$^1$School of Physics, University of New South Wales, Sydney, NSW 2052, Australia}
    \affil{$^2$School of Computing Engineering and Mathematics, Western Sydney University, Locked Bay 1797, Penrith, NSW 2751, Australia}
    \affil{$^3$Armagh Observatory and Planetarium, College Hill, Armagh BT61 9DG, UK}
    \affil{$^4$School of Physical Sciences, University of Adelaide, Adelaide, SA 5005, Australia}
    \affil{$^{5}$CSIRO Astronomy \& Space Science, Australia Telescope National Facility, PO Box 76, Epping, NSW 2121, Australia}
    \affil{$^6$Joint ALMA Observatory, Alonso de C\'{o}rdova 3107, Vitacura, Santiago, Chile}
    \affil{$^7$National Radio Astronomy Observatory, 520 Edgemont Road, Charlottesville, VA 22903, USA}
    \affil{$^{8}$Email: catherine.braiding@gmail.com}}%
\jid{PASA}
\doi{10.1017/pasa.\the\year.xxx}
\jyear{\the\year}

\usepackage{aas_macros}
\usepackage{graphicx}
\usepackage{hyperref} 
\hypersetup{colorlinks,citecolor=blue,linkcolor=blue,urlcolor=blue}
\usepackage{natbib}
\usepackage{amsmath}
\usepackage{amssymb}
\usepackage{xspace}
\usepackage{gensymb}
\usepackage{supertabular,booktabs}
\usepackage{multicol}
\usepackage{longtable}
\usepackage{lipsum}
\usepackage{pdflscape} 
\newcommand{\kms}{km\,s$^{-1}$}
\newcommand{\jcap}{JCAP} 

\newcommand{\cootto}{C$^{18}$O }

\hypersetup{draft}

\begin{document}

\begin{frontmatter}
\maketitle

\begin{abstract}
We present observations of fifty square degrees of the Mopra carbon 
monoxide (CO) survey of the Southern Galactic Plane, covering 
Galactic longitudes $l = 300$--$350^\circ$ and latitudes $|b| \le 0.5^\circ$. These data have been taken at 0.6\,arcminute spatial resolution and 0.1\,\kms spectral resolution, providing 
an unprecedented view of the molecular clouds and gas of the Southern Galactic 
Plane in the 109--115\,GHz $J = 1$--0 transitions of $^{12}$CO, $^{13}$CO, C$^{18}$O and C$^{17}$O.

We present a series of velocity-integrated maps, spectra and position-velocity plots that illustrate  Galactic arm structures and trace masses on the order of $\sim$10$^{6}$\,M$_{\odot}$ per square degree; and include a preliminary catalogue of \cootto clumps located between $l=330$--$340$\degree. Together with information about the noise statistics of the survey these data can be retrieved from the Mopra CO website, the PASA data store and the Harvard Dataverse.

\end{abstract}
\end{frontmatter}

\input{Section_Intro}

\input{Section_previousRelease}

\input{Section_DR3}
\input{Section_Results}
\input{Section_SampleScience}

\input{Section_Summary}

\bibliographystyle{pasa-mnras}
\bibliography{mopra_co_iii}
\begin{appendix}

\input{appendix_a}

\input{table_c18o_clumps_longtable}

\end{appendix}

\end{document}

%% file: Section_Intro.tex
\section{INTRODUCTION}\label{sec:intro}
Since its astronomical discovery in 1970 \citep{Wilson:1970}, the carbon monoxide (CO) molecule has been utilised as a key tracer of molecular gas in the Galaxy. Second in numerical abundance only to the H$_2$ molecule, the large abundance ($\sim$10$^{-4}$) and low electric dipole moment (0.122\,D)  of CO allows cold molecular gas to be well-traced by the J=1--0 rotational transition. CO surveys of the molecular gas in the Galactic Plane can help shape an understanding of star formation, dynamics, and chemistry within the Milky Way.

\begin{figure*}[ht]
\begin{center}
\includegraphics[width=\textwidth]{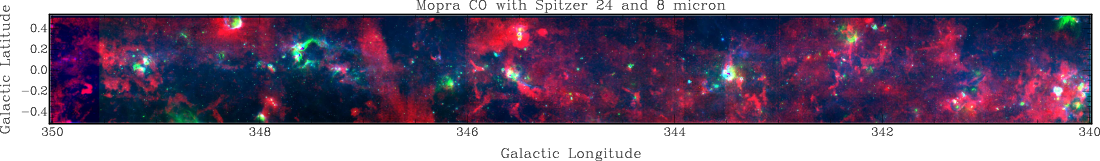}
\includegraphics[width=\textwidth]{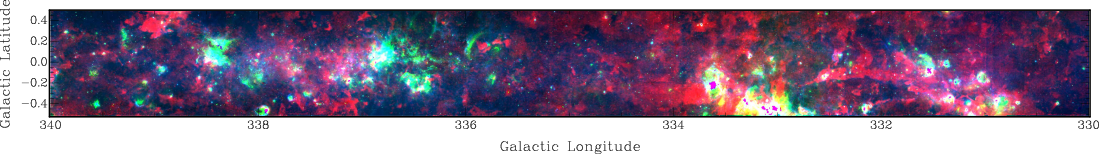}
\includegraphics[width=\textwidth]{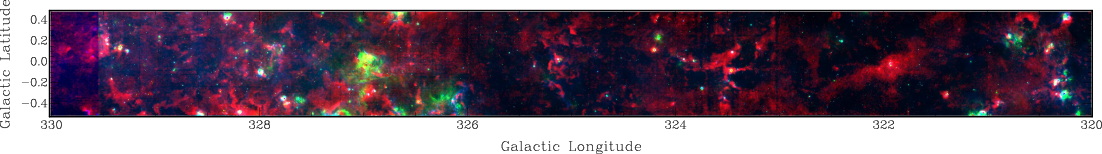}
\includegraphics[width=\textwidth]{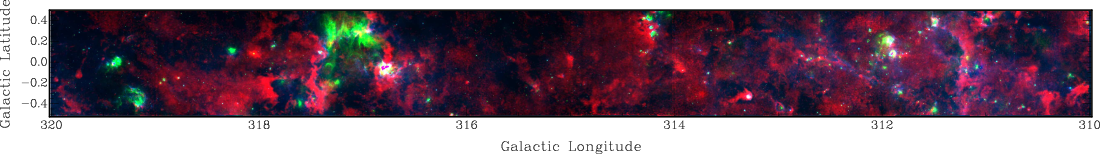}
\includegraphics[width=\textwidth]{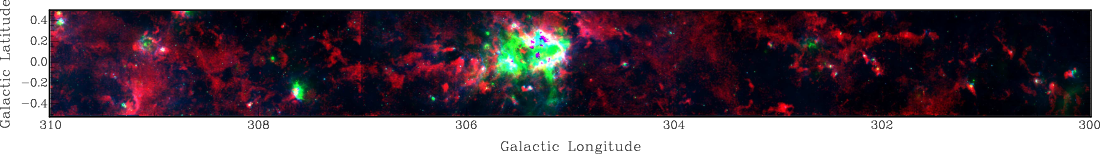}
	\caption{Three-colour image of the Galactic Plane from the Mopra CO survey (red; peak intensity image from $v = -150$ to $+50$\,km\,s$^{-1}$) 
	together with the Spitzer MIPSGAL 24\,$\mu$m \citep[green;][]{Carey:2009} and GLIMPSE 8\,$\mu$m \citep[blue;][]{Churchwell:2009,Benjamin:2003}
	surveys.}\label{Pretty-12CO}
\end{center}
\end{figure*}

The Mopra Galactic Plane CO survey\footnote{http://newt.phys.unsw.edu.au/mopraco/} is the highest resolution large-scale 3-dimensional molecular gas survey of the inner Galactic Plane so far. With this paper, we publicly release CO\footnote{To avoid confusion we henceforth refer to this isotopologue as $^{12}$CO, using `CO' to refer to the molecule more generally.}, $^{13}$CO and C$^{18}$O(1--0) data\footnote{We have also surveyed the C$^{17}$O(1--0) transition, but it is too weak to be detected above a 1-$\sigma$ level over most of the surveyed region and will be excluded from our discussion; nevertheless the fully-reduced data is being made publicly available in DR3.} for the region from $l=$ 300 to 350$^{\circ}$, and $b=\pm$0.5$^{\circ}$ (see Figure \ref{Pretty-12CO}). This release encompasses updated reductions of data contained within the previous Mopra CO data releases \citep{Burton:2013,Braiding:2015} and, along with the Mopra CO maps of the Carina/Gum\,31 region \citep{Rebolledo:2016} and the Central Molecular Zone \citep[CMZ,][]{Blackwell:inprep}, will eventually be incorporated into a larger mosaic with coverage spanning longitude $l\sim$260 to +11$^{\circ}$ and a latitudinal range of $\pm$1$^{\circ}$. Additional targeted extensions beyond 1$^{\circ}$ will also be included.

The Columbia CO(1--0) survey \citep{Dame:2001} has been a revolutionary tool for the study of the Milky Way;  its 8$^{\prime}$-resolution has helped enable a large number of important astronomical results. Some examples include the constraining of the CO X-factor and Galactic cloud masses, the identification of Galactic structures \citep[e.g.][]{Dame:2011,Vallee:2014} and implications for dark matter research \citep[e.g.][]{Drlica:2014}.

The Mopra Galactic Plane CO survey will enable new Galactic science. ALMA offers arcsecond-level spatial resolution in the CO(1--0) transition \citep[e.g.][]{Mathews:2013}, and the very fine field of view of ALMA is well-complemented by a large sub-arcminute-scale map which can allow the identification of molecular gas components for observations. Furthermore, comprehensive investigations of the `missing' gas of the Galaxy \citep[as inferred from dust and gamma-ray observations, e.g.][]{Ade:2015} require molecular clouds to be imaged at a resolution able to separate these `dark' components, complementing observatories such as the Cherenkov Telescope Array \citep[CTA,][]{CTA:2017} and the High Elevation Antarctic Telescope \citep[HEAT,][]{Walker:2004}. Our survey also goes one step further than historical $^{12}$CO surveys of the Galactic Plane by mapping multiple CO isotopologues that probe the regions where CO(1--0) becomes optically thick, tracing additional mass components. 

In this paper, we present 50 square degrees of Mopra $^{12}$CO, $^{13}$CO, C$^{18}$O and C$^{17}$O data\footnote{available at doi:10.7910/DVN/LH3BDN}. Section~\ref{sec:DR3} is an update on the data-processing techniques used by the survey, descriptions of the data quality, and a list of regions affected by contaminated reference positions. A series of images and characterisations of Mopra data are presented in Section~\ref{sec:results} alongside a new catalogue of \cootto clumps located within the region $l = 330$--$340$\degree. Finally, we outline recent region and source-specific investigations that utilise Mopra CO isotopologue data and propose additional work to be performed, in conjunction with other surveys, to determine the evolution of molecular clouds, and the origins of cosmic rays (Section~\ref{sec:samsci} \& \ref{ss:gamma}).

%% file: Section_previousRelease.tex
\section{PREVIOUS DATA RELEASES}\label{s:otherdata}

The Mopra CO Galactic Plane survey was first described in \citet{Burton:2013}, which presented the results of the pilot observations undertaken in March 2011 of the single square degree, $l=323$--324$\degree$, $b=\pm$0.5$\degree$. The aims and the approach of the Mopra CO survey were also discussed in detail, with science goals of comparing the emission of Gamma and Cosmic rays to emission from nearby molecular clouds, and characterising the formation of molecular clouds. The initial observations were used to describe a number of giant molecular clouds along the G323 line of sight, demonstrating how the multiple isotopologue CO survey data could be used in conjunction with \textsc{Hi} data from the Southern Galactic Plane Survey \citep[SGPS;][]{McClure:2005} to determine the mass and column densities within these clouds. This data remains some of the noisiest in the Mopra CO survey (see Figures \ref{RMS-12CO}--\ref{RMS-C18O}), as the initial observations were conducted during summer months.


The first data release by \citet[][hereafter DR1]{Braiding:2015}, presented the first ten degrees of the Mopra CO Survey, covering Galactic longitudes $l=320$--330$\degree$ and latitude $b=\pm$0.5$\degree$, and $l=327$--330$\degree$, $b=-0.5$--+1.0$\degree$. The maps possessed a typical rms noise of $\sim$1.2\,K in the $^{12}$CO data and $\sim$0.5\,K in the other isotopologue frequency bands. It was noted that the line intensities from the Mopra CO survey are $\sim$1.3 times higher than those of \citet[][see also \S\ref{ss:spectra}]{Dame:2001}, although the average line profiles for each square degree show similar qualitative features. The calculated mass of the molecular clouds in the ten square degrees of DR1 was found to be $\sim 4 \times 10^7$\,M$_{\odot}$, and included many complex filamentary structures from the tangent point to the Norma spiral arm of the Galaxy. Data release 3 contains an improved reduction of these ten square degrees, and although the changes are very minor ($\sim1\%$ of voxels differ from their previous values, most by $\ll0.1$\,K), it is recommended that current Mopra CO users download the newer files for future work. 


Data release 2 \citep[DR2;][]{Rebolledo:2016} comprised 8 square degrees of observations covering the Carina Nebula and Gum 31 molecular complex. In addition to the molecular gas distribution derived from  $^{12}$CO and $^{13}$CO line emission maps, this paper presented total gas column density and dust temperature maps. The Mopra CO data were complemented by dust maps from Herschel \citep{Molinari:2010}, allowing the calculation of dust column densities within the nebula of between 6.3$\times$10$^{20}$ and 1.4$\times$10$^{23}$ cm$^{-2}$, and dust temperatures within a range of 17 to 43 K. The distribution of the total gas is well represented by a lognormal function, but a tail at high densities is clearly observed, which is a common feature of nebulae with active star formation \citep{Pineda:2010,Rebolledo:2016}. Regional variations in the masses recovered from the Mopra CO emission and Herschel dust maps revealed the different evolutionary stages of the molecular clouds that comprise the Carina Nebula--Gum 31 complex. 

The Central Molecular Zone (CMZ) of the Galaxy contains dense \citep[e.g.][]{Jones:2012,Jones:2013}, high-temperature \citep[e.g.][]{Ao:2013} molecular clouds with a total cumulative mass of $\sim$5$\times$10$^{7}$\,M$_{\odot}$ \citep[e.g.][]{Pierce:2000}. The fourth data release of the Mopra CO Survey will cover the Galactic Centre region\citep[$358.5 \le l \le 3.5$\degree, $-0.5 \le b \le +1.0$\degree;][]{Blackwell:inprep}, which has required the development of new observing and data processing techniques to produce high quality spectral maps. As the star formation behaviour of molecular clouds in the CMZ is linked to their orbital dynamics \citep[e.g.][]{Kruijssen:2015}, the high resolution Mopra CO maps will enable new studies of the inner Galaxy gas dynamics and triggered star formation.

%% file: Section_DR3.tex
\section{DATA RELEASE 3}\label{sec:DR3}

Observations for this data release were carried out over the Southern Hemisphere winters of 2011--2016 using Mopra, a 22\,m single-dish radio telescope located $\sim$450\,km north-west of Sydney, Australia (at 31$^{\circ}$16$^{\prime}$04$^{\prime\prime}$ S, 149$^{\circ}$05$^{\prime}$59$^{\prime\prime}$\,E, 866\,m a.s.l.). Observations utilised the UNSW Digital Filter Bank (the UNSW-MOPS) in its `zoom' mode, with 8$\times$137.5 MHz dual-polarization bands. These bands were tuned to target the J=1--0 transition of $^{12}$CO, $^{13}$CO, C$^{18}$O and C$^{17}$O (with central frequencies of 115271.202, 110201.354, 109782.176 and 112358.988\,MHz, respectively). Three bands were tuned to sample a CO(1--0) velocity coverage of $\sim$1,100\,kms$^{-1}$, while the $^{13}$CO(1--0) and C$^{18}$O(1--0) transitions each had two dedicated bands to sample a velocity-coverage of $\sim$770\,km\,s$^{-1}$. One band was allocated to the least-abundant molecular transition, C$^{17}$O(1--0), resulting in $\sim$380\,km\,s$^{-1}$ of velocity coverage. The targeted central frequencies were offset according to Galactic Longitude to maximise overlap with expected Galactic rotation velocities. As in DR1, the observations from $l=323$--$331\degree$, $b=\pm0.5\degree$ cover a smaller velocity range, as only 4 `zoom' bands were utilized during the 2011 observing period, one for each isotopologue. Our observations across the full 50 square degrees of DR3 suggest that it is unlikely that there are any high-velocity clouds in this region, as all of the $^{12}$CO clouds detected to date have been within the velocity range of the primary zoom band.

As described in \citet{Burton:2013}, Fast-On-The-Fly (FOTF) mapping has been carried out in 1 square degree segments, with each segment being comprised of orthogonal scans (in longitudinal and latitudinal directions) to reduce scanning artefacts. The exposure of each square degree is at least $\sim$20\,hrs, more in places where poor weather or technical issues interrupt observations and triggered a repeat of the affected scans.

The angular resolution of the Mopra beam at 115\,GHz is 33-arcsec FWHM \citep{Ladd:2005}; after a median filter convolution is applied during the data reduction process, the dataset possesses a beam size of around 0.6~arcmin. Since the survey region is spatially fully-sampled, the extended beam efficiency, $\eta _\textrm{XB} = 0.55$ at 115~GHz, should be used to convert brightness temperatures into line fluxes \citep[rather than the main beam efficiency $\eta_\textrm{MB} = 0.42$;][]{Ladd:2005}.

DR3 has been made available at the survey website and in the PASA data archive as a series of fits files covering a square degree for each isotopologue, separated by half a degree along the Galactic Plane. Although the full range of velocity space available for $^{12}$CO is [$-$1000,$+$1000]\,km\,s$^{-1}$ (with file sizes of $\sim800$\,MB per square degree), it is expected that most users would prefer to download the smaller ($\sim$300\,MB per square degree) files covering the velocity range [$-150$, $+50$]\,km\,s$^{-1}$ illustrated in this paper. No emission has as yet been detected outside this range, although closer to the Galactic centre this range will be extended. Also available are the T$_\textrm{sys}$, beam coverage and RMS noise images for each line.

\subsection{Data Processing}\label{ss:dp}

From September 2015 observations were affected by a recurring synchronisation loss between the telescope and spectrometer, creating a spike in the central channel of $\sim$0.03\% of spectra. This was not automatically flagged by software during observations or using the data reduction processes from DR1 and DR2, and if left unchecked these artefacts would propagate throughout the reduction, particularly when spikes appeared in the sky reference spectra. A new IDL routine was created to mask the affected channels in the raw data from the telescope; by coincidence this also flags other poor spectra caused by clouds or system fluctuations. 

The rest of the data reduction remains unchanged from DR1 and DR2, involving six main stages of processing \citep{Braiding:2015,Rebolledo:2016}. Firstly, \texttt{LIVEDATA}\footnote{Both \texttt{LIVEDATA} and \texttt{GRIDZILLA} can be downloaded from: www.atnf.csiro.au/computing/software/livedata/index.html} \citep{Gooch:1996} was used to perform the band-pass correction and background subtraction using data from the nearest reference position to create the calibrated spectra. These are then analysed using an IDL routine that flags any spectra that are poorly-calibrated or particularly noisy (mostly in the $^{12}$CO bands where the system temperature is higher). The data are then interpolated onto a uniform grid using \texttt{GRIDZILLA} \citep{Gooch:1996}, employing a median filter for the interpolation of the oversampled data onto each grid location. \texttt{GRIDZILLA} excludes from the grid those spectra for which the T$_\textrm{sys}$ values are above 1000\,K for the $^{12}$CO cube, and 700\,K for the other lines (see \S\ref{ss:dq}). The output of these processes is a \texttt{fits} data cube in ($l$, $b$, $V_\textrm{LSR}$) coordinates.

Noisy outlier pixels, rows and columns are then cleaned using an IDL routine that first detects voxels that differ from their neighbours by more than 5 standard deviations, then rows and columns where the mean spectra differs by more than 3 standard deviations from the median of the entire data set (typically caused by the passage of clouds during the sky reference measurement). These `bad' data are replaced by interpolating the values of their neighbouring pixels, as described in \citet{Burton:2013}, and the process is repeated until no further `bad' data are detected.

The data are then continuum-subtracted using first a fourth-order and then a seventh-order polynomical fit, to yield data cubes of the brightness temperature for each spectral line as a function of Galactic coordinate and $V_\textrm{LSR}$ velocity. Finally, data cubes that have been affected by contamination from faint molecular clouds present within the reference position (see \S\ref{ss:badsky}), are corrected by modelling the contaminant component and adding that back into the final clean spectrum \citep{Braiding:2015}.

\subsection{Data Quality}\label{ss:dq}

\begin{figure*}[ht]
\begin{center}
\includegraphics[width=\textwidth]{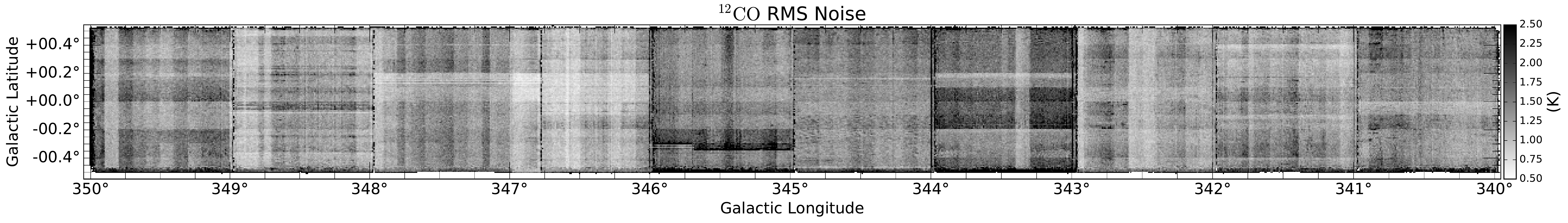}
\includegraphics[width=\textwidth]{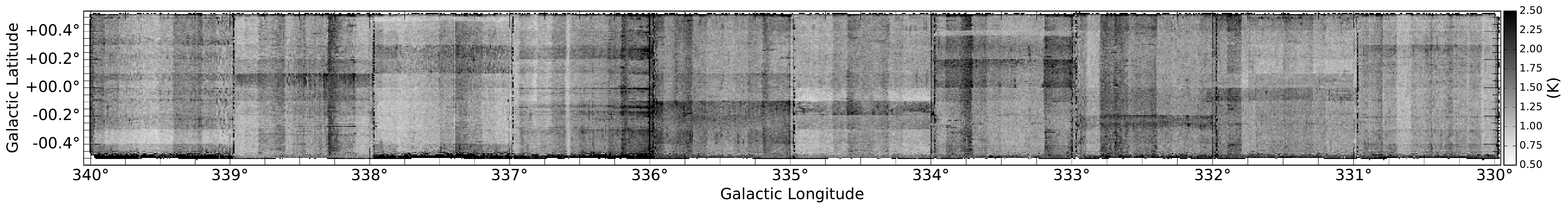}
\includegraphics[width=\textwidth]{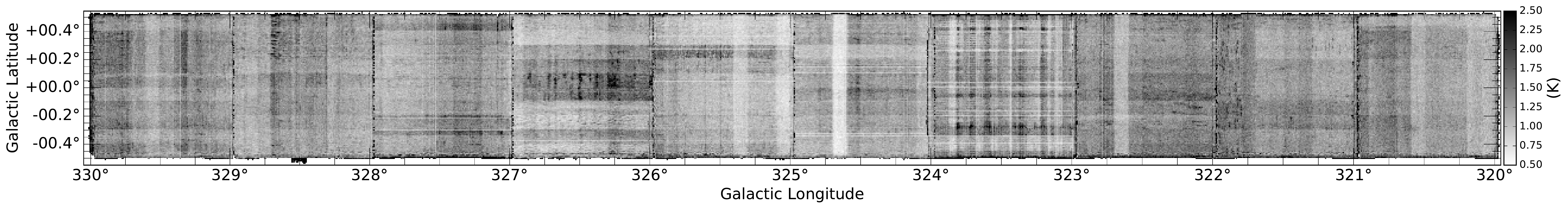}
\includegraphics[width=\textwidth]{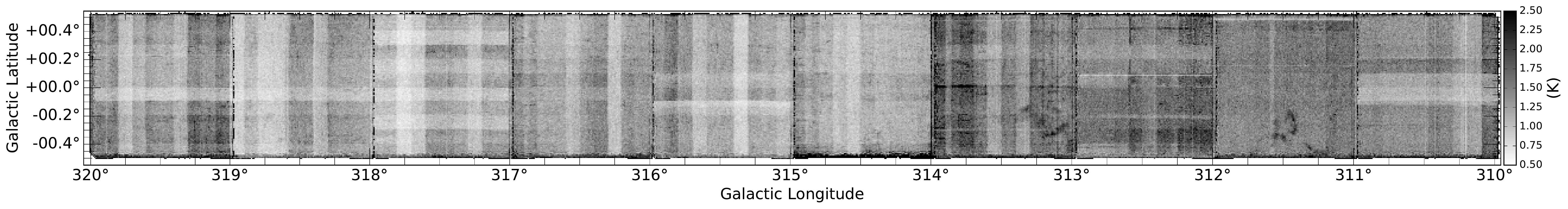}
\includegraphics[width=\textwidth]{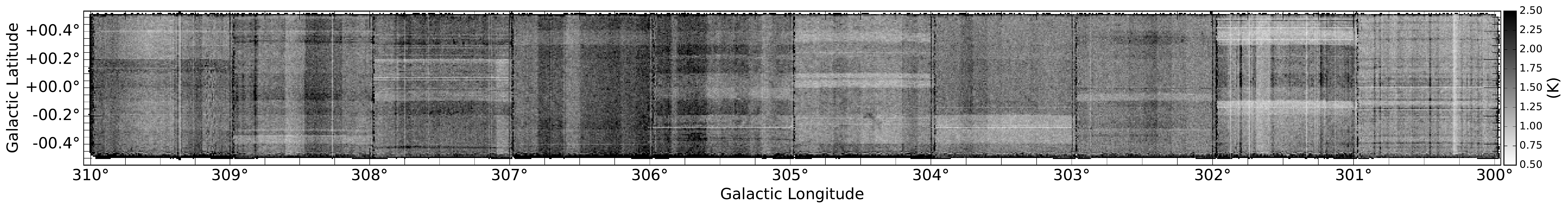}
	\caption{
	1\,$\sigma$ noise images for the $^{12}$CO data covering the full 50 square degrees from 
	$l = 300$--$350^\circ$ in units of T$_\textrm{A}^*$ (K) (as indicated by the scale bars). The 
	striping pattern is inherent to the data set, resulting from scanning in the $l$ and $b$ 
	directions in variable observing conditions.}\label{RMS-12CO}
\end{center}
\end{figure*}
\begin{figure*}[ht]
\begin{center}
\includegraphics[width=\textwidth]{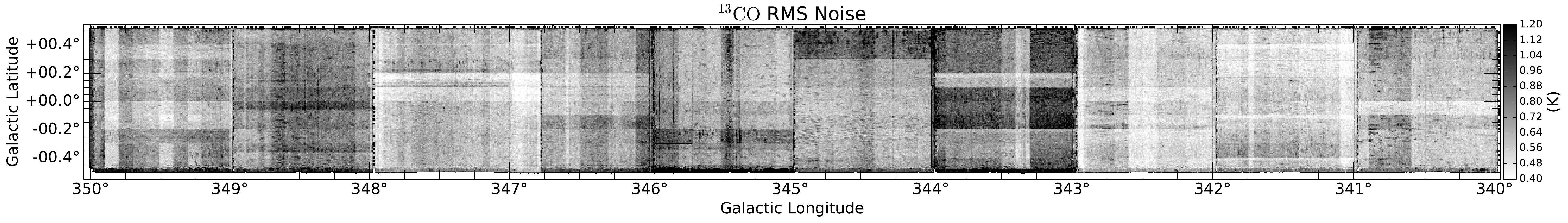}
\includegraphics[width=\textwidth]{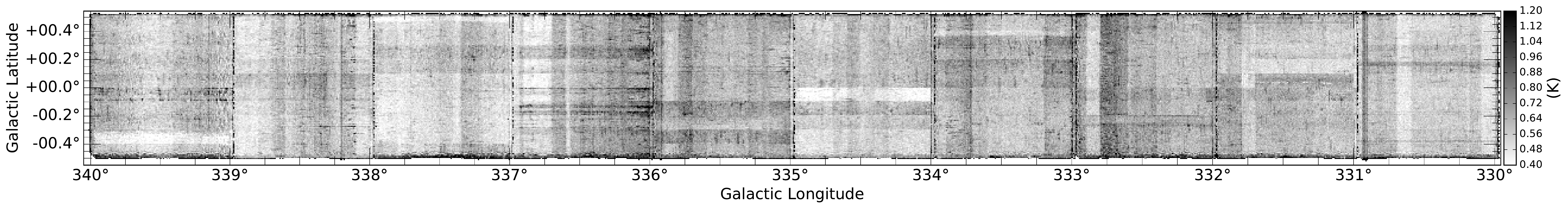}
\includegraphics[width=\textwidth]{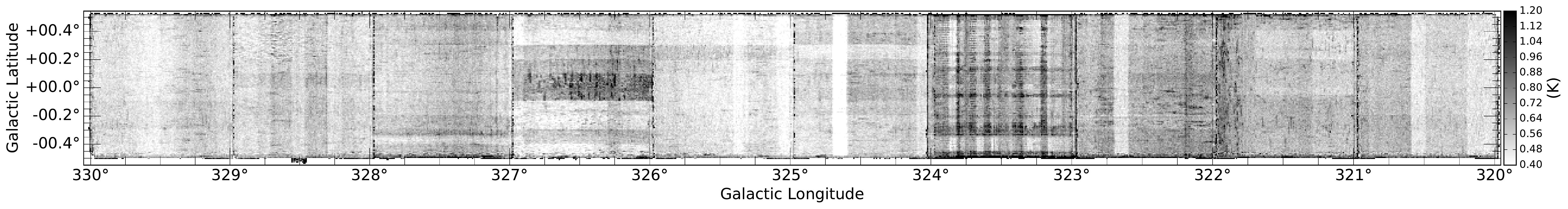}
\includegraphics[width=\textwidth]{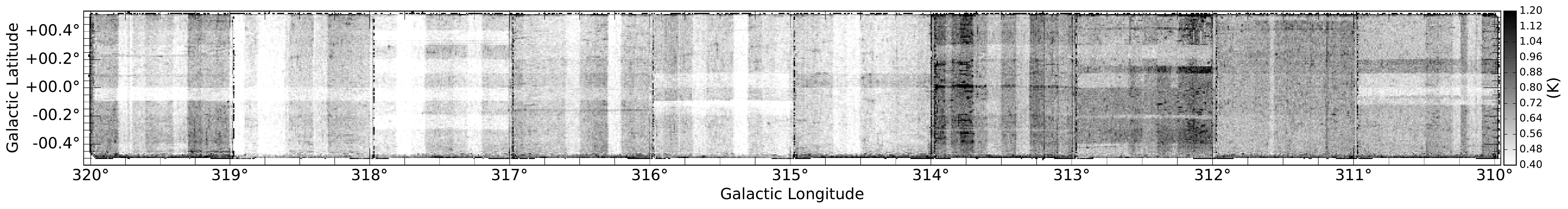}
\includegraphics[width=\textwidth]{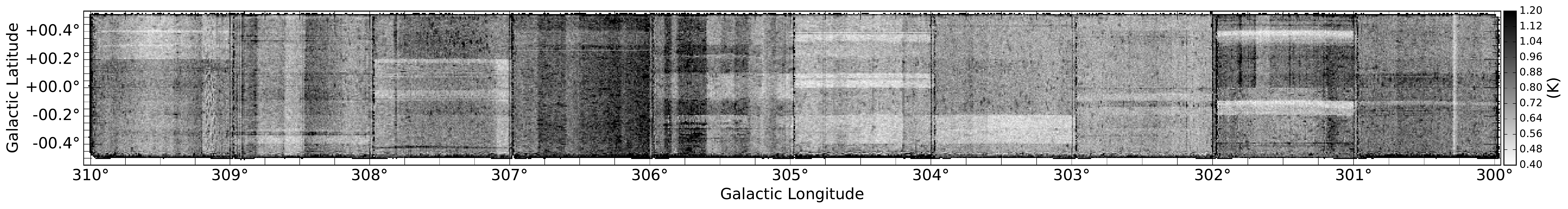}
	\caption{
	As in Figure \ref{RMS-12CO}, these are the 1\,$\sigma$ noise images for the $^{13}$CO 
	data covering the full 50 square degrees from $l = 300$--$350^\circ$ in units of 
	T$_\textrm{A}^*$ (K) (as indicated by the scale bars).}\label{RMS-13CO}
\end{center}
\end{figure*}
\begin{figure*}[ht]
\begin{center}
\includegraphics[width=\textwidth]{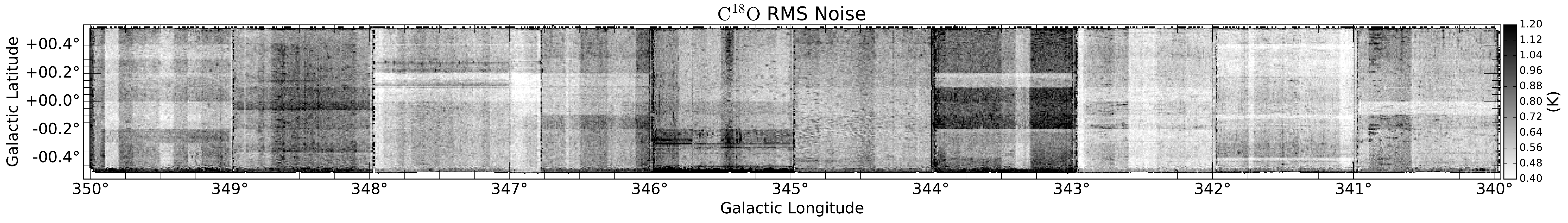}
\includegraphics[width=\textwidth]{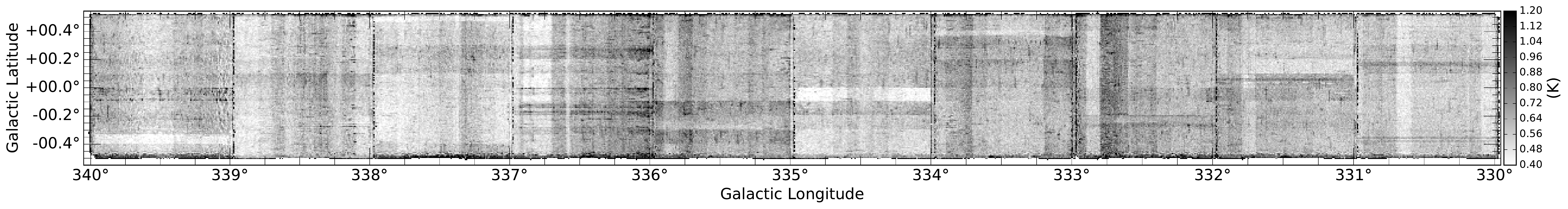}
\includegraphics[width=\textwidth]{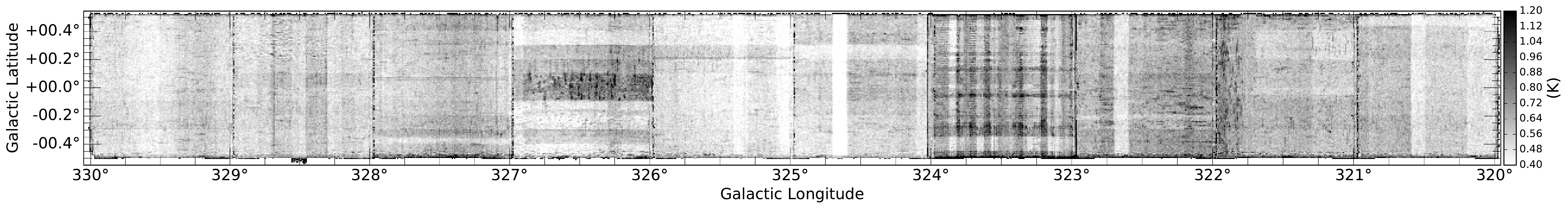}
\includegraphics[width=\textwidth]{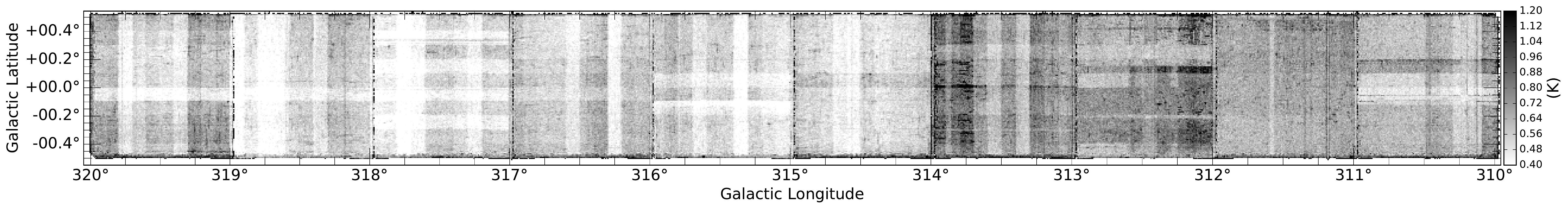}
\includegraphics[width=\textwidth]{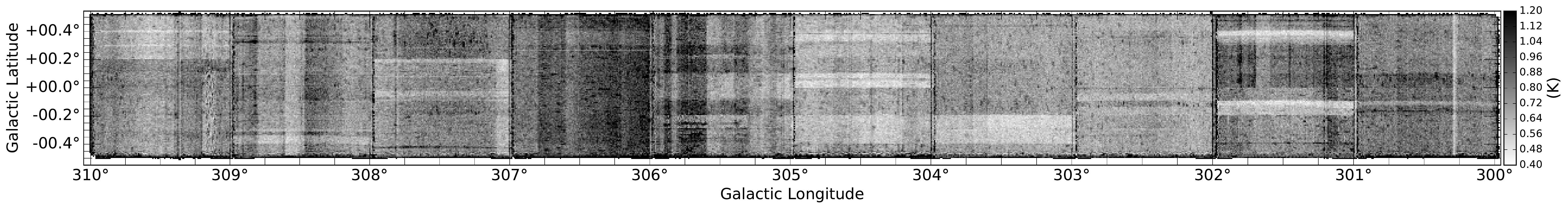}
	\caption{
	As in Figure \ref{RMS-12CO}, these are the 1\,$\sigma$ noise images for the C$^{18}$O 
	data covering the full 50 square degrees from $l = 300$--$350^\circ$ in units of 
	T$_\textrm{A}^*$ (K) (as indicated by the scale bars).}\label{RMS-C18O}
\end{center}
\end{figure*}

The observations for DR3 were carried out over a wide variety of weather conditions, ranging 
from those of late Southern Hemisphere summer (March, 2011, 2014, and 2015) when the system temperature was high 
($\sim$900\,K for $^{12}$CO) to deep winter (2011--2016) where the conditions are quite stable and 
T$_\textrm{sys}$ in the $^{12}$CO band is quite stable at around 400\,K. This is reflected in the noise 
maps of the $^{12}$CO, $^{13}$CO, and C$^{18}$O data, which are are presented in Figures \ref{RMS-12CO}, 
\ref{RMS-13CO}, and \ref{RMS-C18O} respectively. The noise value for each pixel was determined 
from the standard deviation of the continuum channels for line-of-sight velocities containing no 
apparent signal. The thick striping seen within these maps is an effect of scanning in $l$ and $b$ 
directions under different system temperatures and observing conditions. Thinner stripes and 
spots of low noise indicate the positions of `bad' pixels, rows, or columns that were identified and 
interpolated over \citep[see \S\ref{ss:dp};][]{Burton:2013}.

The system temperature of Mopra (T$_\mathrm{sys}$, discussed further in Appendix \ref{app:Tsys}) reflects the total level of signal received from the instrument, telescope, and sky, in addition to that from the source (which contributes a near-insignificant portion to T$_\mathrm{sys}$). It is calibrated with reference to an ambient temperature paddle every 30 minutes, allowing us to control for variations in the weather. Observations with an average T$_\mathrm{sys}$ greater than 1000\,K were repeated under better weather conditions, while any individual measurement which possessed a T$_\mathrm{sys}$ greater than 1000\,K was filtered out during the imaging stage of the data reduction (see \S\ref{ss:dp}).

To examine the Mopra beam resolution, observations of SiO(2-1,v=1) maser emission from the red supergiant variable star, AH Scorpii, were taken in November 2017 (Figure\,\ref{fig:Beam} in \S\ref{app:beam}). The data were integrated over all masers in velocity-space, and gaussian line profiles were fitted to the diametral distribution in both right-ascension and declination. The measured FWHM was then deconvolved with the map smoothing by subtracting the smoothing kernal FWHM in quadrature. The resultant Mopra beam FWHM was calculated to be $35^{\prime\prime}\pm1^{\prime\prime}$ and $34^{\prime\prime}\pm1^{\prime\prime}$ for the Mopra beam axis at 86\,GHz. This is consistent with previous calibration measurements that characterised the Mopra beam FWHM as 36$^{\prime\prime}\pm$3$^{\prime\prime}$ \citep{Ladd:2005}.

\subsection{Sky Reference Positions}\label{ss:badsky}

Table \ref{tab-sky} lists coordinates of the sky reference positions used for each square degree of observations. Several of these reference positions contained $^{12}$CO, $^{13}$CO and in at least one instance C$^{18}$O line emission, which propagate into the data cubes as negative ``emission''. The line of sight velocity and peak intensity of the reference position $^{12}$CO line emission are included in Table \ref{tab-sky}; the $^{13}$CO line emission is typically fainter than the $^{12}$CO line by a factor of 5, and similarly C$^{18}$O emission is only detected when the $^{12}$CO contamination is exceptionally bright.
\begin{table*}
\caption{Sky reference beam positions for each square degree of DR3. Several cubes were contaminated by 
	negative $^{12}$CO, $^{13}$CO and C$^{18}$O line emission from the sky reference position; the central velocity and intensity (in 
	$T_A^\star$\,[K] units) of the $^{12}$CO contamination are noted here.}
\begin{center}
\vspace{-20.2pt}
\begin{tabular}{@{}ccccc@{}}
\hline\hline
Cube & Reference & Reference & Contamination          & Contamination Peak\\
     & Longitude & Latitude  & Velocity (km\,s$^{-1}$) & Intensity ($T_A^\star$\,[K]) \\
\hline%
 G300  & 300.500 & $+2.000$    \\
 G301  & 301.375 & $+2.375$    \\
 G302  & 302.875 & $-1.875$    & $-6.2$  & $-0.7$ \\
 G303  & 303.000 & $-2.000$    & $-5.4$  & $-0.5$ \\
 G304  & 304.000 & $-2.000$    \\
 G305  & 305.000 & $-2.000$    \\ 
 G306  & 306.000 & $-2.000$    & $+38.3$ & $-0.3$ \\ 
 G307  & 307.500 & $-2.500$    & $-46.2$ & $-0.6$ \\
 G308  & 309.000 & $-2.120$    \\
 G309  & 309.375 & $-2.500$    \\
 G310  & 310.500 & $-2.500$    \\
 G311  & 311.000 & $-2.500$    \\ 
 G312  & 312.500 & $-2.500$    \\ 
 G313  & 313.500 & $-2.500$    \\
 G314  & 314.875 & $-2.000$    & $-4.9$  & $-0.4$ \\
 G315  & 315.500 & $-2.000$    \\
 G316  & 316.500 & $+2.000$    \\
 G317  & 318.000 & $+2.000$    \\
 G318  & 318.750 & $-2.000$    \\
 G319  & 319.000 & $-2.125$    & $-22.9$ & $-0.3$ \\
 G320  & 320.375 & $-3.000$    \\
 G321  & 321.500 & $-2.500$    & $-1.7$  & $-3.2$ \\
 G322  & 322.500 & $-2.000$    & $-56.1$, $-13.0$ & $-1.3$, $-0.4$ \\ 
 G323  & 323.500 & $-2.000$    \\ 
 G324  & 324.500 & $-2.000$    & $+0.9$  & $-0.8$ \\ 
 G325  & 325.500 & $-2.200$    & $-46.6$ & $-0.5$ \\ 
 G326  & 326.500 & $-2.000$    & $-19.5$ & $-0.3$ \\ 
 G327  & 327.500 & $-2.200$    \\ 
 G328  & 328.500 & $-2.200$    & $+2.7$  & $-0.4$ \\
 G329  & 329.500 & $-2.200$    \\ 
 G330  & 329.500 & $-2.200$    \\
 G331  & 331.500 & $-2.000$    \\
 G332  & 332.500 & $-2.000$    \\
 G333  & 333.500 & $-2.000$    \\
 G334  & 333.500 & $-2.000$    \\
 G335  & 333.500 & $-2.000$    \\
 G336  & 336.500 & $-4.500$    & $-61.8$ & $-0.46$ \\
 G337  & 337.500 & $-2.600$    \\
 G338  & 337.500 & $-2.600$    \\
 G339  & 339.500 & $+2.000$    \\
 G340  & 339.500 & $+2.000$    \\
 G341  & 342.500 & $-3.000$    & $-0.89$ & $-0.26$ \\
 G342  & 342.500 & $-3.000$    & $-10.1$ & $-0.22$ \\
 G343  & 343.500 & $+2.500$    \\
 G344  & 344.500 & $+3.000$    \\
 G345  & 345.635 & $-3.125$    & $-18.7$, $-28.4$ & $-0.81, -0.63$\\
 G346  & 346.250 & $-3.000$    \\
 G347  & 347.500 & $+2.625$    \\
 G348  & 349.250 & $-2.000$    \\
 G349  & 349.250 & $-2.000$    \\
\hline\hline
\end{tabular}
\end{center}
\label{tab-sky}
\end{table*}

As mentioned in \S\ref{ss:dp}, data affected by contaminated reference positions are corrected in the last stage of data processing. An area of the cube which does not contain emission within $\pm 2$\,kms$^{-1}$ of the contamination is averaged and a gaussian profile is fitted to the spectrum, creating a clean model of the line emission at the reference position. The gaussian profile is then subtracted from each pixel, taking into account frequency shifts relating to Doppler tracking corrections across the cube. In some instances when the observations have been taken many months apart and the Doppler corrections are large, the data are processed into separate cubes for each time period, and these are combined into a single cube after the reference contamination has been corrected. This technique was developed in response to reference-contaminated Mopra CO spectra from the Central Molecular Zone \citep[CMZ,][]{Blackwell:inprep}.

%% file: Section_Results.tex
\section{RESULTS}\label{sec:results}

We present here images from DR3, comparing the increased resolution of Mopra CO data to the lower spatial (8\,arcmin) and spectral (1.3\,km\,s$^{-1}$) resolution Columbia survey \citep{Dame:2001}. We note consistent features between the two CO datasets, and illustrate the structure of the Milky Way through a series of position-velocity and mass-distribution plots.

\subsection{Spectra \& Moment Maps}\label{ss:spectra}

The average line profiles of $^{12}$CO and $^{13}$CO for each central square degree of the Mopra Southern Galactic Plane CO survey between longitudes 300$\degree$ and 350$\degree$ are presented in Appendix\,\ref{app:AvSpec}. The $^{12}$CO average spectra from the \citet{Dame:2001} CO survey are included for comparison and are consistent with the features of Mopra CO survey data. We note a systematic shift in the brightnesses between the Mopra and Columbia surveys of a factor of $\sim1.35$ as discussed in \citet{Braiding:2015}, which is calculated per square degree in Table \ref{tab-ratio} of Appendix \ref{app:AvSpec}. The same shift is also apparent in the Nanten2 CO Survey data, which has very similar fluxes to the Mopra CO Survey (H. Sano, March 2017,  private communication).

Integrated intensity maps from DR3 are presented in Appendix\,\ref{app:IntMaps}, calculated over an interval of 10\,\kms\ in the velocity range of $-150$ to $+50$\,\kms. We note similar features to those present in the Columbia CO survey \citep{Dame:2001}, but also scanning artefacts of $1\degree$-long line features in longitude or latitude near the edges of individual square degrees.

\subsection{Velocity \& Mass Distribution}\label{ss:mass}

\begin{figure*}[htp]
\includegraphics[width=0.99\textwidth]{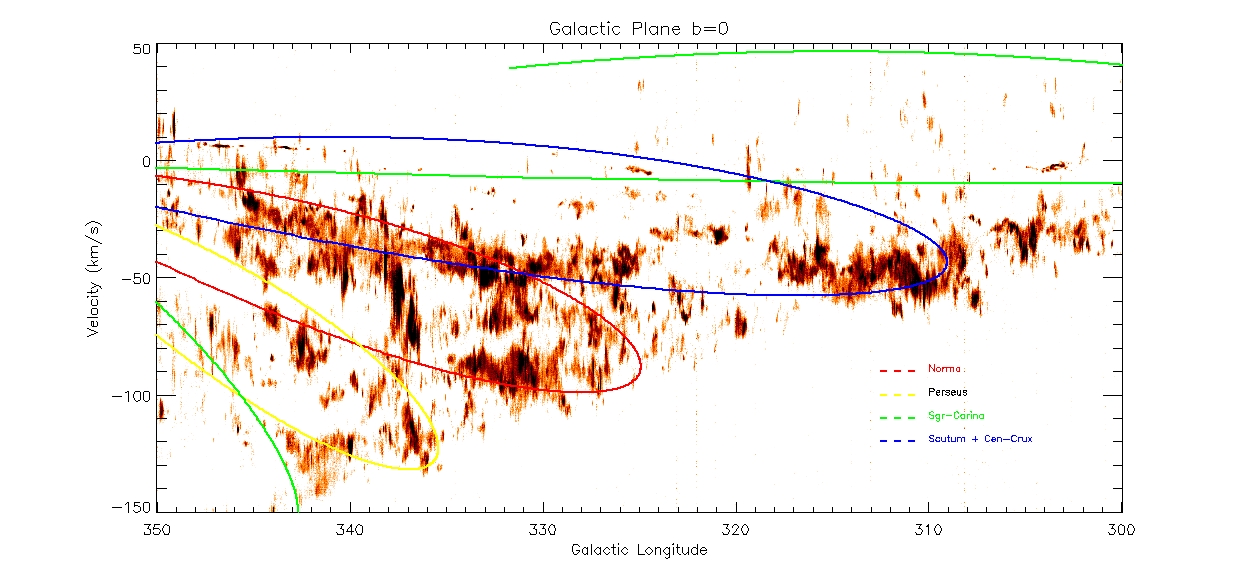}
\includegraphics[width=0.99\textwidth]{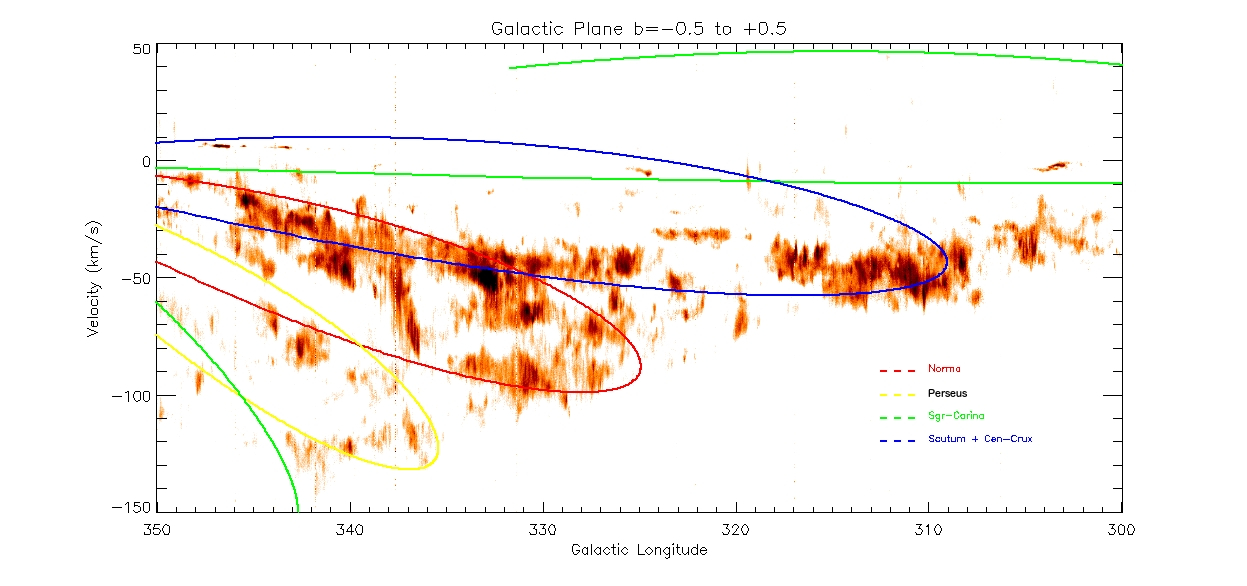}
	\caption{$^{12}$CO position-velocity plots as a function of Galactic longitude. $l$ is on the $x$-axis, $V_\textrm{LSR}$ (the radial velocity) in km\,s$^{-1}$ is on the $y$-axis. The data has been averaged between latitudes $b = -0.05\degree$ and $+0.05\degree$ (top) and $b=-0.5\degree$ to $+0.5$ (bottom), respectively representing the midpoint of the plane and the entire surveyed plane. The solid lines are the model positions for the centres of the four spiral arms in the model of \citet{Vallee:2016}. Note also that some artefacts are visible as features extending across the entire velocity direction, resulting from residual systematic effects.}\label{fig:PV_MGB1}
\end{figure*}

\begin{figure*}[htp]
\includegraphics[width=0.99\textwidth]{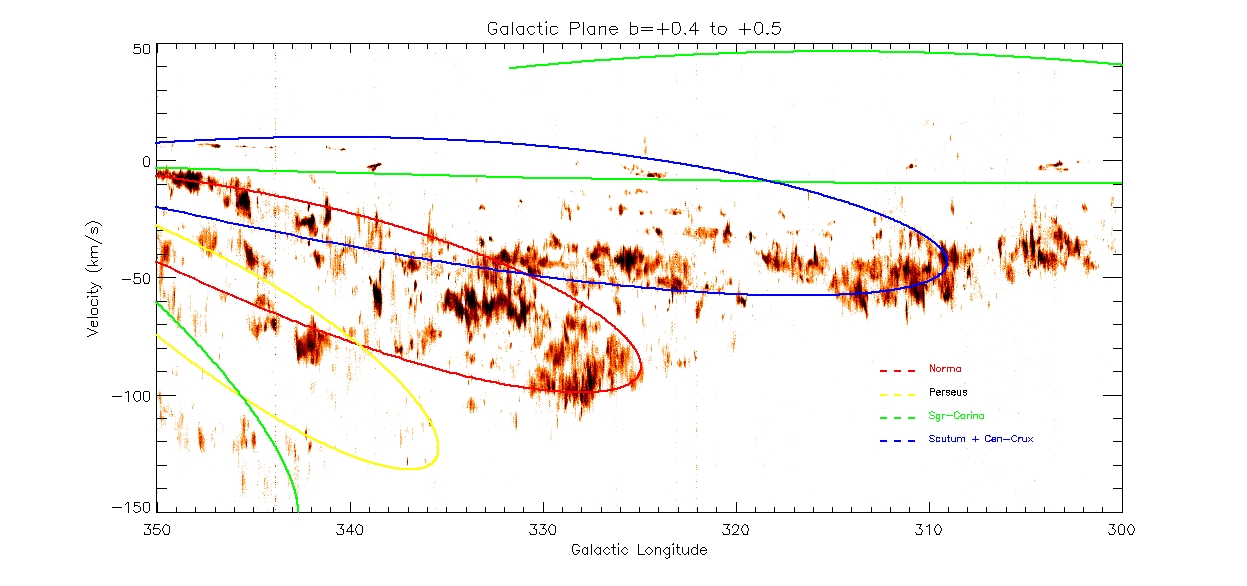}
\includegraphics[width=0.99\textwidth]{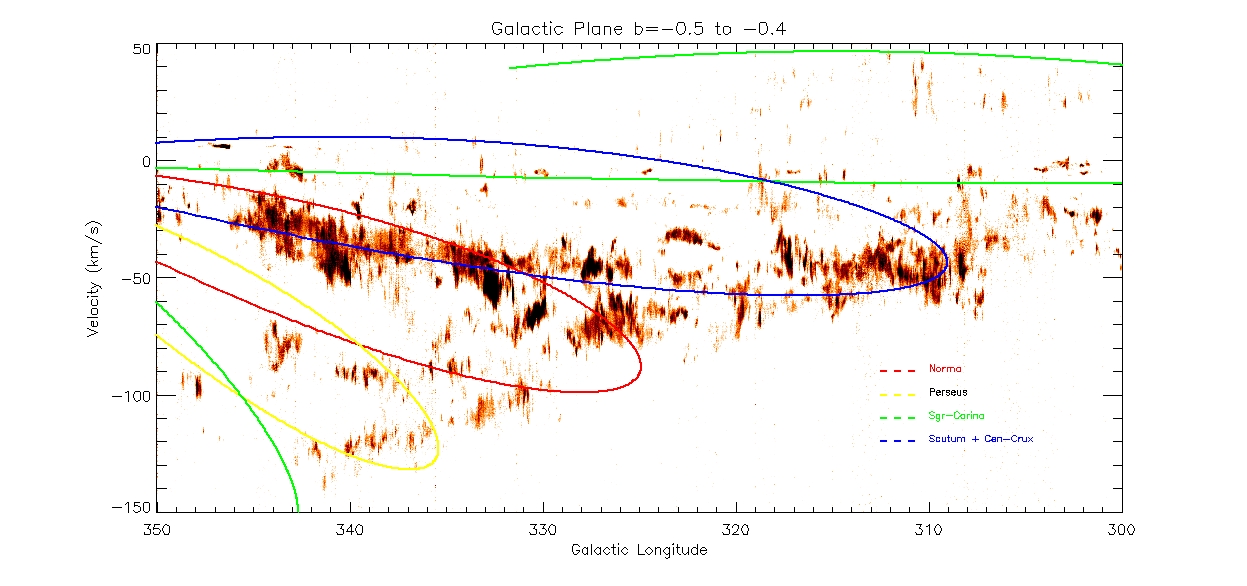}
	\caption{$^{12}$CO position-velocity plots as a function of Galactic longitude, $l$, on the $x$-axis and the radial velocity, $V_\textrm{LSR}$, in km\,s$^{-1}$ on the $y$-axis. The data has been averaged between latitudes $b=+0.5\degree$ and $+0.4\degree$ (top) and $b=-0.5\degree$ to $-0.4\degree$ (bottom), respectively representing the top and the bottom of the surveyed plane. The solid lines are the model positions for the centres of the four spiral arms in the model of \citet{Vallee:2016}. Note also that some artefacts of residual systematic effects are visible as features extending across the entire velocity direction.}\label{fig:PV_MGB2}
\end{figure*}

\begin{figure}[ht]
\includegraphics[width=0.48\textwidth]{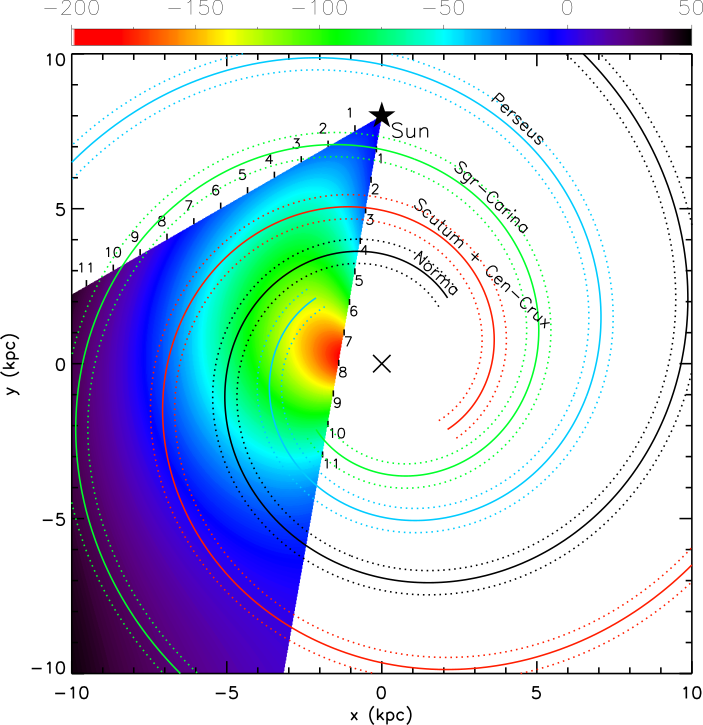}
	\caption{A schematic of the 4 spiral arms of the Milky Way Galaxy. Arm names, the Galactic Centre, and the location of the Sun are indicated. A colour-scale displays the expected kinematic line-of-sight velocity from the reference point of the Sun. The Galaxy is modelled as having a pitch angle of $13.1\degree$, a central bar length of 2.2\,kpc inclined at $-30\degree$ to our sightline, a Galactic Centre distance of 8.0\,kpc and a flat rotation curve with velocity 220\,km\,s$^{-1}$ \citep{Vallee:2016}. }\label{fig:GalRot}
\end{figure}

Figures\,\ref{fig:PV_MGB1} and \ref{fig:PV_MGB2} show position-velocity (PV) slices of the CO emission averaged over 4 different ranges in latitude overlaid with the positions of four spiral arms from the Galaxy model of \citep[][i.e. a pitch angle of $13.1\degree$, central bar length of 2.2\,kpc inclined at $-30\degree$ to our sightline, Galactic Centre distance of 8.0\,kpc and a flat rotation curve with velocity 220\,km\,s$^{-1}$]{Vallee:2016}. Figure\,\ref{fig:PV_MGB1} shows the central plane $|b|< \pm 0.05\degree$ and the entire data range of $|b|< \pm 0.5\degree$, while Figure\,\ref{fig:PV_MGB2} shows the very top ($b=+0.4$ to $+0.5\degree$) and bottom ($b=-0.5$ to $-0.4\degree$) of the surveyed field. The spiral arm model and its associated distance-velocity relationship are shown in Figure \ref{fig:GalRot}.

The observed features generally correspond to giant molecular cloud complexes within the spiral arms, with their velocity widths corresponding to internal motions within the complexes rather than the differential distances that could be inferred by applying a Galactic rotation model. In a few instances prominent outflows are evident as molecular emission that is quite extended in velocity but narrow in the longitudinal direction; these could also be giant molecular clouds at far Galactic distances, but such clouds are generally expected to be somewhat extended in longitude. 

However, while the molecular gas does tend to follow the trajectory of the spiral arm model in Figure \ref{fig:GalRot}, the detailed agreement is only modest. The differences between the PV-plots for the different latitude ranges are also striking, showing how discrete molecular gas is even within the Galactic Plane. Little $^{12}$CO(1--0) emission is evident at `forbidden' velocities, i.e. those that are more negative than the tangent point velocity at each latitude, and in particular the most negative velocities for each of the 4 model spiral arm positions.

\begin{figure}[ht]
\includegraphics[width=0.48\textwidth]{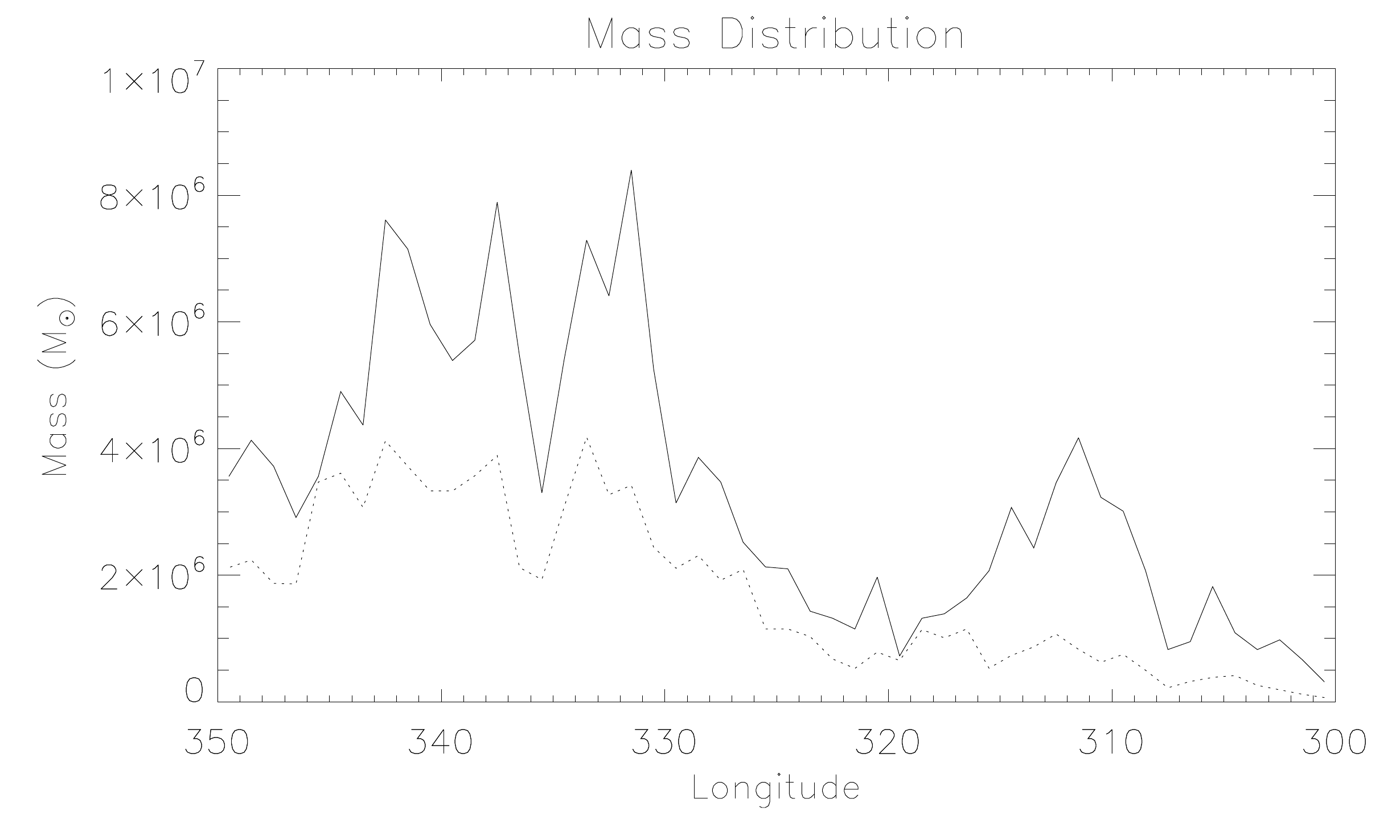}
	\caption{Molecular mass distribution for the inner degree of latitude across the Galaxy, extending from $l=300$--$350\degree$.  Masses are calculated using a $^{12}$CO X-factor of 2.7$\times$10$^{20}$cm$^{-2}$ (K\,km\,s$^{-1}$)$^{-1}$ to convert line fluxes to column densities for all gas detected within the solar circle (i.e. that with negative velocities). The solid line illustrates the mass calculated using the assumption of near-side distances; the dotted line assumes far-side distances (and is divided by a factor of 10 for presentation).}\label{fig:massDistro}
\end{figure}
Figure \ref{fig:massDistro} shows the distribution of molecular mass with Galactic longitude, averaged for each square degree of the survey range. This calculation is as in \citet{Braiding:2015} for $l=320$--$330\degree$, now extended to $l=300$--$350\degree$ and covering the entirety of DR3: applying the CO X-factor to convert $^{12}$CO flux to column density using a conversion factor of $2.7\times$10$^{20}$cm$^{-2}$ (K\,km\,s$^{-1}$)$^{-1}$, the value derived for $|b|<1\degree$ in \citet{Dame:2001}. Though larger than the $1.8\times$10$^{20}$cm$^{-2}$ commonly used for $|b|<5\degree$, it is within the uncertainties discussed by \citet{Bolatto:2013} from their meta-analysis of X-factor values derived from a variety of observational and theoretical studies. The mass is calculated from the column density by applying a Galactic rotation curve to convert velocity to distance. This included making assumptions regarding whether gas velocities relate to near- or far-side distances. Comparison with the spiral arm positions suggest that the bulk of the gas observed does arise at near distances, similar to the assumption in \citet{Braiding:2015}. However, we now present mass estimates based on the two extremes, of all the gas being at the near-distance and all at the far-distance. We have not included, however, molecular gas at positive velocities, which lie beyond the Solar Circle in this calculation. While such gas is clearly present at some PV-plot positions, the low fluxes when averaged over an entire degree, combined with their large distances, result in large absolute errors in the mass determination. Hence, we have not included them in Figure \ref{fig:massDistro} (but see also the note at the end of the next paragraph).  
We find that typically there are a few$\times10^6$\,M$_{\odot}$ of molecular gas per square degree from $l=300\degree$ to $l=330\degree$. The mass then rises to a peak of around $10^7$\,M$_{\odot}$ per square degree between $l=330\degree$ and $340\degree$, before falling by a factor of 2 at $l=350\degree$.  The total molecular gas within $\pm 0.5\degree$ of the Galactic Plane, in the 50 degrees from $l=300$-$350\degree$, is found to be $\sim2\times 10^8$\,M$_{\odot}$ (if all were at the near distance) and $\sim9 \times 10^8$\,M$_{\odot}$ (all at the far distance). While not included in the plot, the total amount of gas lying beyond the solar circle is estimated to be around $4 \times 10^7$\,M$_{\odot}$. This would contribute an additional 5 to 20\% to the total molecular gas content calculated here. From the moment maps in Appendix\,\ref{app:IntMaps}, it is clear that we have observed not only $^{12}$CO, but also $^{13}$CO at far distances with velocities $>+30$\,\kms.

\subsection{C${^{18}}$O clump catalogue}\label{ss:c18o}

When the faint C$^{18}$O line is detected it is almost certainly optically thin due to the low abundance of the molecule with respect to $^{12}$CO, and so provides a sightline that penetrates entirely through clouds. Thus the highest column density molecular clumps are detected in C$^{18}$O. In order to investigate the distribution of \cootto clumps across the Milky Way, we isolate individual clumps by applying the \texttt{Clumpfind}\footnote{\url{http://www.ifa.hawaii.edu/users/jpw/clumpfind.shtml}} algorithm \citep{Williams:1994} to the region between $330< l <340\degree$ where $-0.5<b<+0.5$, and $-140 < V_\textrm{LSR}< +50$\kms. In order to efficiently run this calculation on the Mopra CO data cubes, a number of simplifications needed to be performed: firstly the data was binned by 10 spectral channels (reducing the velocity resolution to $\sim$0.92 \kms) and secondly it was box smoothed using the IDL routine \texttt{smooth}\footnote{We employed a box smoothing kernel of dimensions [2,2,1]; utilising 2 voxels in the angular directions and 1 along the velocity axis.}. The relative 1$\sigma$ emission level of the \cootto data cube is then ${\sim}0.10$\,K (derived by computing the standard deviation of the spectra in an emission-free region). 

\texttt{Clumpfind} operates by finding local peaks of emission and then contouring around them to decompose the data cube into a set of clumps of \cootto emission. It requires two input parameters to work: the increment between contour levels, \texttt{inc}, and the desired lower limit of clump emission, \texttt{low}; after a series of tests both of these were assigned the value of $0.45$\,K, equivalent to the $4.5\sigma$ emission level. 

The resulting bright \cootto clumps are presented in Table \ref{tab-c18o} of Appendix \ref{app:c18o}; for each of these we report the coordinates of the clump centroid, the average and peak main beam brightness temperature, the relative FWHM extensions in arcsec and \kms, and its radius in arcseconds. Prior to computing the physical properties of each clump, we compare the C$^{18}$O and $^{13}$CO spectral profiles for each line of sight, rejecting those false detections that are noisy spikes in the C$^{18}$O band without a clear peak in both CO isotopologues. We also check the \texttt{clumpfind} ID assignation to ensure each clump is physically distinct (as it lists nearby local maxima within the same physical structure as independent clumps); within the catalogue the minimum separation between two centroids is 9 voxels. 

Using the model described above in  \S\ref{ss:mass} we calculate near and far distances to each clump, which are also listed in Table \ref{tab-c18o}. Following \citet{Wilson:2009} we derive the C$^{18}$O column density for each clump, assuming that the gas is in local thermodynamic equilibrium (LTE) at a fixed T$_{ex}=10$\,K, to be:
\begin{equation}
\centering
N(\mathrm{C^{18}O})= 3.0\times10^{14}\frac{T_\mathrm{ex}}{1-e^{-(T_{0}^{18}/T_\mathrm{ex})}}\int{\tau_{18}d\mathrm{v} },
	\label{eq:N13}
\end{equation}
where T$_0^{18}=5.27$\,K is the energy level of the {$J=1$--$0$} C$^{18}$O transition, and $d\mathrm{v}$ is the velocity interval in \kms. The optical depth, $\tau_{18}$, is:
\begin{equation}
	\tau_{18} = -\ln\biggr(1-\frac{T^{18}_\textrm{MB}}{5.27}\biggr[\big[e^{5.27/T_\textrm{ex}}
	-1\big]^{-1}-0.16\biggr]^{-1}\biggr).
	\label{eq:T18}
\end{equation}
The column density of H$_2$, N(H$_2$) is then calculated for each clump using the abundance ratio of [H$_2$/C$^{18}$O]$ = 5.88 \times 10^6$ from \citet{Frerking:1982}. Finally, we are able to derive near and far masses for each clump.
Assuming that the clumps are all located at the near distance, the total mass traced by C$^{18}$O in this region is $\sim 2.5 \times 10^6$\,M$_\odot$, equivalent to 10\% of the total molecular gas mass traced by $^{12}$CO in the same portion of Galactic Plane, as only the highest density regions are traced by C$^{18}$O. Nevertheless, the C$^{18}$O molecule provides an important optically thin tracer of the molecular gas.


%% file: Section_SampleScience.tex
\section{ISM Survey Synergies}\label{sec:samsci}

The Mopra CO survey is a timely addition to several Galactic Plane surveys currently under way. In particular, we have observed to a Galactic longitude of $l=11\degree$, giving 2 square degrees of overlap with the Northern Hemisphere FOREST Unbiased Galactic plane Imaging survey with the Nobeyama 45-m telescope \citep[FUGIN;][]{Minamidani:2015,Umemoto:2017}, a similarly high resolution $^{12}$CO(1--0) survey covering from $l=10$--$50\degree$ and $l=198$--$236\degree$ where $|b|<1\degree$. Preliminary work comparing these surveys is now underway, but as mentioned in \S\ref{sec:results}, similar comparisons with the Nanten2 CO survey suggest that the observed fluxes and structures in the overlapping region will be equivalent (H. Sano, March 2017, pers.comm.).

Combining Mopra CO with data from other surveys presents the opportunity to fully characterise regions of interest in order to probe the origins and history of molecular clouds \citep{Burton:2013}. Recently, \citet{Sano:2017arXiv} presented work using data from this survey, ASTE and Nanten2 to investigate high-mass star cluster formation triggered by cloud-cloud collisions, in particular focusing on the \mbox{H\,{\sc ii}} region RCW\,36, located in the Vela Molecular Ridge. They suggested that a cloud-cloud collision formed RCW\,36 and its stellar cluster, with a relative velocity separation of $\sim$5\,km\,s$^{-1}$. The estimated cloud separation of $\sim$0.85\,pc was estimated to occur over a time-scale of $\sim$0.2\,Myr, assuming a distance of 1.9\,kpc, consistent with the age of the cluster. The resulting O-type star formation lead to a high measured CO rotational temperature.

CO isotopologue data from DR1 of the Mopra Galactic Plane Survey was exploited in an investigation of the so-called `dark' component of molecular gas towards a cold, filamentary molecular cloud in the G328 sight-line \citep{Burton:2014,Burton:2015}. The authors compared the distribution of atomic \mbox{H\,{\sc i}}-traced gas to atomic C emission (which traces some CO-dark molecular gas) and the CO-traced molecules observed in DR1. They found that $^{13}$CO and [\mbox{C\,{\sc i}}] show similar distributions and velocity profiles, the abundance of atomic C increases towards the molecular cloud edges, consistent with CO photodissociation by an average interstellar radiation field and cosmic-ray ionisation at low optical depths. Further comparison to dust images from Herschel showed that the cloud was IR-dark and that no star formation is yet occurring within the young filament. Extended studies using C$^+$ observations from the Stratospheric Observatory for Infrared Astronomy (SOFIA) will target gas that may be in the transition phase between atomic and molecular forms of Carbon \citep{Risacher:2016}, allowing a better tracing of the dark molecular gas that has been inferred from gamma-ray observations (see \S\ref{ss:gamma}) and to study molecular cloud formation in detail.

The APEX Telescope Large Area Survey of the Galaxy (ATLASGAL) has been used to perform an unbiased census of massive dense clumps in the Galactic Plane at 870\,$\mu$m, using Mopra CO data to provide radial velocities to $\sim500$ of the clumps \citep{Urquhart:2018}. The properties of these clumps were exploited to develop an evolutionary classification scheme showing trends for increasing dust temperatures and luminosities as a function of evolution, showing that most of the clumps are already associated with star formation. We expect to determine a similar evolutionary scheme for molecular cloud cores (including the \cootto cores presented in \S\ref{ss:c18o}), using the Mopra CO and other survey data in future.

The OH $\lambda=18$\,cm lines are complementary to CO and \mbox{H\,{\sc i}} when estimating the mass of H$_2$ in molecular clouds, as at densities well below 10$^3$\,cm$^{-3}$ CO is affected by ultraviolet shielding, interstellar chemistry, and sub-thermal processes \citep{Dickey:2013}. The GASKAP (The Galactic Australian Square Kilometre Array Pathfinder) Survey will trace atomic hydrogen directly through the \mbox{H\,{\sc i}} transition and OH \citep{Dickey:2013}. The GASKAP survey will achieve spatial resolutions of 20--60 arcsec for \mbox{H\,{\sc i}} and 90--180 arcsec for OH, the former comparable with the Mopra CO survey, allowing astronomers to carry out pixel to pixel comparisons between the different tracers. Also using ASKAP, the EMU (Evolutionary Map of the Universe) Survey will provide a similar-resolution high sensitivity map of radio continuum emission in the Galactic Plane, allowing the detection of \mbox{H\,{\sc ii}} regions, supernova remnants and planetary nebulae at similar resolutions to Mopra CO and GASKAP \citep{Norris:2011}. Future improvements to the velocity resolution of ASKAP may bring it closer to that of Mopra.

Towards a better understanding of the motions of the Galactic Centre, \citet{Yan:2017} employed Mopra $^{12}$CO and $^{13}$CO data in conjunction with OH absorption data from the Southern Parkes Large-Area Survey in Hydroxyl \citep[SPLASH;][]{Dawson:2014} to create a 3-dimensional image of the Galactic Centre region at a spatial resolution of 38\,pc. The Galactic Centre was well-described with a bar-like structure at an angle of $67 \pm $2$^{\circ}$ to the line-of-sight. More generally the Mopra CO data from the Central Molecular Zone \citep{Blackwell:inprep} will be key to studies of star formation and its effect on the evolution of the CMZ \citep[e.g.][]{Kruijssen:2015}. 

Interstellar medium surveys are critical to X-ray and other high-energy studies, where they are used to determine distances to supernova remnants, particularly using morphology anti-correlations and photoelectric absorption studies \citep[e.g.][]{Maxted:2018}. \citet{Heinz:2015} observed X-ray light echoes from the X-ray binary Circinus X-1, which demonstrated four well-defined rings with radii from 5 to 13 arcmin. By modelling the dust distribution towards Circinus X-1 and comparing that to $^{12}$CO emission from Data Release 3 they were able to identify the source clouds for the innermost rings and determine the distance to Circinus X-1 to be $\sim$9.4\,kpc for the binary system, ruling out previous near distance estimates of $\sim4$\,kpc.

The Mopra CO survey will also be well-complemented by $^{13}$CO(2-1) data from SEDIGISM \citep[Structure, Excitation, and Dynamics of the Inner Galactic Interstellar Medium;][]{Schuller:2017}, allowing the calculation of rotational temperature estimates for regions that are optically-thick in $^{12}$CO and $^{13}$CO(1--0). Similarly, James Clerk Maxwell Telescope (JCMT) $^{12}$CO(3-2) observations \citep{Dempsey:2013} will be used to refine temperature and column density estimations for molecular clouds of interest. 

Perhaps most importantly the Mopra CO survey data will be used as a finder chart for objects of interest to be observed with ALMA, which offers arcsecond-level spatial resolution in the CO(1--0) transition \citep{Mathews:2013}. We note that Mopra CO data has guided successful ALMA and SOFIA applications for observations towards molecular cores interacting with the young supernova remnant, RX\,J1713.7$-$3946 \citep{Rowell:2017sofia,Sano:2017alma}. The resultant data will reveal the extremely fine details of these structures and provide an important step towards understanding cosmic ray origins.

Future ISM catalogues of sources of interest traced by the Mopra CO survey will be essential for determining the best objects to follow-up with ALMA. A sample catalogue of bright C$^{18}$O clumps is in Appendix\,\ref{app:c18o}, along with characterisations of the location, size, temperature and mass. The complete catalogue will provide supplementary information useful for targeting dense clumps in star formation research.



\section{Gamma-ray Observations}\label{ss:gamma}
Gamma-ray maps provide an important window into the gas distribution of the Galaxy that rely on studies of CO, [\mbox{C\,{\sc i}}], [\mbox{C\,{\sc ii}}] and \mbox{H\,{\sc i}}. 
A large component of gamma-ray emission from the Milky Way comes from three distinct so-called `diffuse' source types \citep[see e.g.][]{Acero:2013,Acero:2017,Dubus:2013}:
\begin{enumerate}
\item Truly diffuse gamma-rays from Cosmic Rays (CRs) circulating in the Galactic Plane for $\gtrsim$10$^6$years, i.e. the diffuse CR `sea';
\item Local gamma-ray enhancements from CRs escaping local accelerators on time-frames $\lesssim$10$^5$\,years; and
\item Unresolved sources of gamma-ray emission across the Galaxy.
\end{enumerate}
Diffuse sources of type (1) and (2) involve cosmic-ray protons interacting with gas in the Galactic Plane via proton-proton collisions that produce charged and neutral pions. The neutral pions quickly decay into gamma-ray photons. Diffuse sources of type (3) may also have a component of gamma-ray emission produced via this mechanism, depending on the source.

Unlike spectral line emission, gamma-rays from pion-decay can be considered a general tracer of gas density independent of temperature or chemistry. As the gamma-ray flux generated by proton-proton interactions is a product of cosmic ray density and gas density, gamma-ray maps can be used to highlight excesses of both gas and CRs. It follows that gas studies and gamma-ray astronomy are intimately tied together.

In order to identify gamma-ray sources in the Galactic Plane maps from Fermi-LAT, which operates in the 0.02-300\,GeV gamma-ray regime \citep{Atwood:2009}, a model of the gamma-ray emission from diffuse source (1) was required. This model was constructed from \mbox{H\,{\sc i}} and CO gas maps of similar resolution \citep[e.g.][]{McClure:2005,Dame:2001} and revealed that previous work underestimated the gamma-ray Galactic contribution at high latitudes and hence overestimated the background CR `sea' \citep[e.g.][]{Strong:2004}. 

Galactic molecular cores resolved at sub-arcminute scales in both gamma-rays and CO(1--0) emission will enable the identification of pion-decay signatures near cosmic-ray sources \citep{Cameron:2012,Acharya:2013}, and may deliver kinematic distance solutions for cosmic-ray sources \citep[][]{CTA:2017}. The signature of this decay will be particularly strong if the gamma-ray spectral index is altered by CR diffusion into molecular cores due to energy-dependent propagation \citep[e.g.][]{Gabici:2007,Maxted:2012}. Indeed, a common theme developing for young gamma-ray shell-type supernova remnants (which are regarded as key PeV CR acceleration candidates) is the existence of sub-arcminute-scale molecular cores embedded in the perimeter of regions blown out by the progenitor star \citep{Fukui:2008,Moriguchi:2005,Sano:2013,Maxted:2013rxj,Fukuda:2014,Kavanagh:2015,Fukui:2017,Sano:2017ApJ}. The next-generation Cherenkov Telescope Array \citep[CTA][]{CTA:2017} is expected to provide gamma-ray maps at a resolution sufficient to identify local peaks due to cloud clumps and filaments. Furthermore, the Mopra CO survey will probe interstellar medium dynamics by revealing broad-line or asymmetric emission indicative of the shock-gas interactions, as is expected of some gamma-ray sources like older supernova shocks and pulsar wind nebulae \citep[e.g.][]{Arikawa:1999}.

Data from the Mopra Galactic Plane CO Survey have already been used to examine the nature of gamma-ray sources observed with the highest resolution gamma-ray telescope, H.E.S.S. \citep[High Energy Stereoscopic System,][]{Aharonian:2006}. Understanding the links between TeV gamma-ray emission and the parent high energy particle populations local to extreme Galactic objects is a step forward in the search for elusive CR hadron acceleration sites; pulsar wind nebulae and SNRs with associated gamma-ray detections are the primary candidates in the pursuit for Galactic CR sources. 

\citet{Lau:2017} exploited Mopra CO and CS isotopologue data to investigate the gamma-ray sources HESS\,J1640$-$465 and HESS\,J1641$-$463, which are positionally coincident with the SNRs G338.3$-$0.0 and G338.5$+$0.1. The authors found coincident diffuse molecular gas, with a bright complex of dense gas and \mbox{H\,{\sc ii}} regions connecting the SNRs. CO-derived masses were then utilised to estimate the density of $>$1TeV CRs assuming that the gamma-ray emission was caused by pion-decay following proton-proton interactions. The values inferred were in the range of 100--1000 times the Solar value \citep{Lau:2017}. Similarly, CO and CS observations that identify filamentary molecular features towards the unidentified objects HESS\,J1614$-$518 and HESS\,J1616$-$508 \citep{Lau:2017PASA} represent the first steps in forming an understanding of the nature gamma-ray emitters.

\citet{Maxted:2018} compared line-of-sight column densities from Mopra CO and the SGPS \mbox{H\,{\sc i}} emission to X-ray absorption column densities for the bright shell-type gamma-ray SNR, HESS\,J1731$-$347. This enabled the calculation of a kinematic distance to the source, and revealed a large associated molecular gas cloud that bridged the region between HESS\,J1731$-$347 and a nearby unidentified gamma-ray source, HESS\,J1729$-$345. Such a discovery adds weight to the suggestion that some unidentified gamma-ray sources are the product of so-called `runaway cosmic rays' that escape a CR acceleration site and diffuse into nearby molecular gas clouds (i.e. diffuse sources of type 2). Similar CR scenarios have also been investigated for pulsar wind nebulae and their progenitor supernova remnants \citep[e.g. HESS\,J1826$-$130,][]{Voisin:2016}.

With the CTA now under construction \citep{CTA:2017}, subarcminute resolution gamma-ray maps will soon become available. As with Fermi-LAT, CTA will require similar-resolution gas maps to characterise the diffuse gamma-ray emission components (1 and 2) and enable studies of gamma-ray sources and dark gas. Furthermore, CTA's improved sensitivity will enable it to detect sources across the Galactic disk, expanding on the rather limited (few kpc for sources of <10\% Crab flux) horizon achieved by current telescopes like H.E.S.S. 
Towards this future, we are preparing molecular gas maps that will fully parametrise Mopra CO, FUGIN and GASKAP data for use in gamma-ray modelling of the Galactic Plane \citep[e.g. using GALPROP, PICARD and DRAGON;][respectively]{Strong:1998,Kissmann:2014,Evoli:2017}.

%% file: Section_Summary.tex
\section{Summary}\label{sec:summary}

We present the third data release of the Mopra Southern Galactic Plane CO Survey, covering 50 square degrees between Galactic longitude $l = 300$--$350^\circ$ and latitude $b = \pm 0.5^\circ$.  
These data are part of a larger campaign to map the molecular gas of the inner Galactic Plane in $^{12}$CO, $^{13}$CO, C$^{18}$O and C$^{17}$O(1--0) emission. 

We find general agreement between the CO spectral line profiles of the Mopra survey and those of the Columbia survey for each square degree of coverage. Furthermore, we find a modest agreement between the Mopra survey position-velocity structure and a recent Galactic Plane model \citep{Vallee:2016}.
A preliminary catalogue of C$^{18}$O clumps presented traces $\sim10\%$ of the observed $^{12}$CO molecular gas, illustrating the importance of this optically thin tracer of molecular gas.

The Mopra CO survey, which is estimated to trace a $\sim$few$\times 10^6$\,M$_\odot$ of molecular gas per square degree, is already making an impact on the search for Galactic cosmic-ray sources and star formation research. We discuss past and future multi-wavelength investigations towards a number of Galactic Plane sources and synergies with other surveys. 



The Mopra CO survey is still ongoing, with the intention of covering the entire fourth quadrant and expanding to Galactic latitudes beyond $b = \pm 1.0^\circ$ by the end of 2018. As the data are published, they are being made available at the survey website, and the PASA data archive.


\begin{acknowledgements}
The Mopra radio telescope is part of the Australia Telescope National Facility. 
Operations support was provided by the University of New South Wales, the University of 
Adelaide and Western Sydney University. Many staff of the ATNF have contributed to the success 
of the remote operations at Mopra.  We particularly wish to acknowledge the contributions of 
David Brodrick, Philip Edwards, Brett Hiscock, and Peter Mirtschin. 

This research made use of Astropy, a community-developed core Python package for Astronomy 
\citep{astropy:2013} and the IDL Astronomy Library \citep{IDLastro:1993}.
	
We are indebted also to the financial support of the \#TeamMopra kickstarter contributors, listed at 
www.mopra.org. The University of New South Wales Digital Filter Bank used for the observations with 
the Mopra Telescope (the UNSW--MOPS) was provided with support from the Australian Research Council 
(ARC). We also acknowledge ARC support through Discovery Project DP120101585 and Linkage, Infrastructure, 
Equipment and Facilities project LE160100094.\end{acknowledgements}

%% file: appendix_a.tex
\section{Supplementary Materials}\label{app:app}

\subsection{The Mopra Beam}\label{app:beam}
\begin{figure}[ht]
\includegraphics[width=0.47\textwidth]{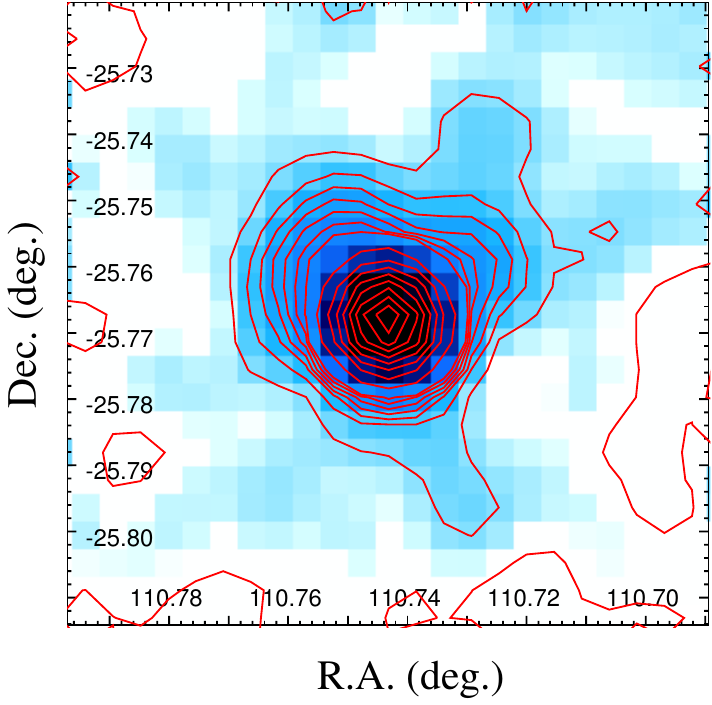}
	\caption{A Mopra SiO(2--1,$v$=1) image of the red supergiant variable star, AH Scorpii, integrated between 2 and 48\,km\,s$^{-1}$. Red contours indicate 1\% steps between 0 to 10\% total intensity, and 10\% steps between 10 and 100\% the total intensity. Observations were carried out on the 14th of November, 2017.}\label{fig:Beam}
\end{figure}

To characterise the Mopra beam resolution, observations of the red supergiant variable star AH Scorpii were performed on the 14th of November, 2017. SiO(2--1,$v$=1) masers from AH Scorpii are point-like, allowing the Mopra beam size and shape to be characterised. The integrated data, described more fully in \S\ref{ss:dq}, are illustrated in Figure\,\ref{fig:Beam}. 

\subsection{System Temperature}\label{app:Tsys}
\begin{figure*}[ht]
\begin{center}
\includegraphics[width=\textwidth]{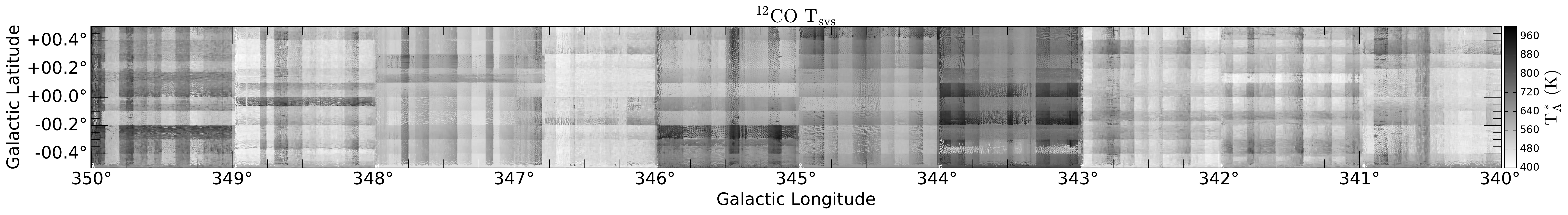}
\includegraphics[width=\textwidth]{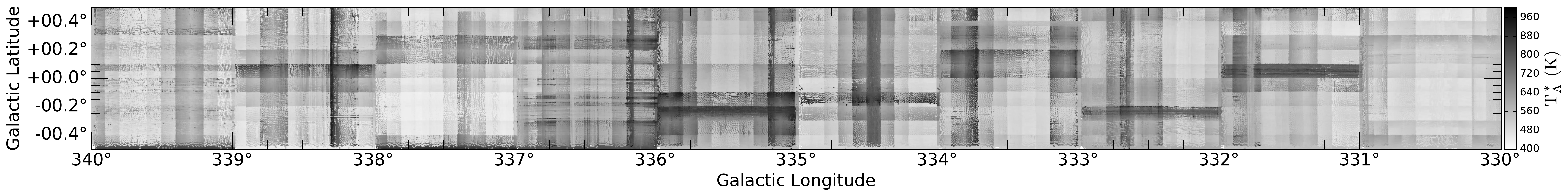}
\includegraphics[width=\textwidth]{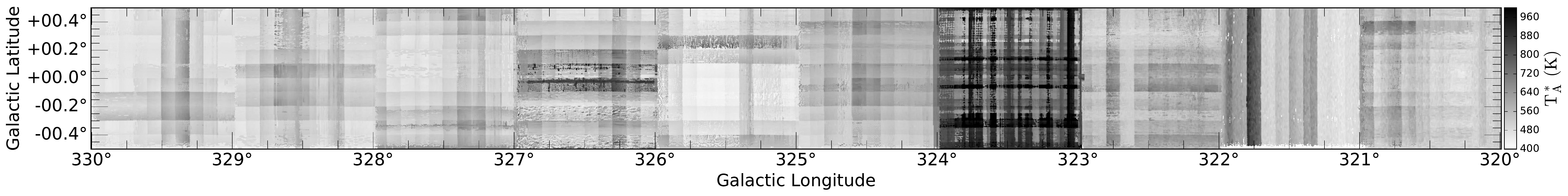}
\includegraphics[width=\textwidth]{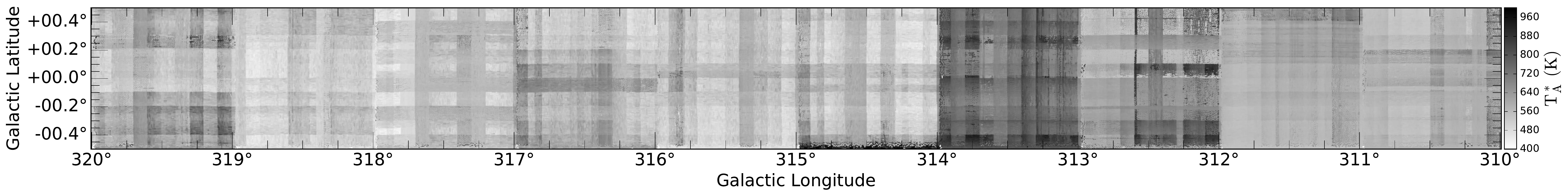}
\includegraphics[width=\textwidth]{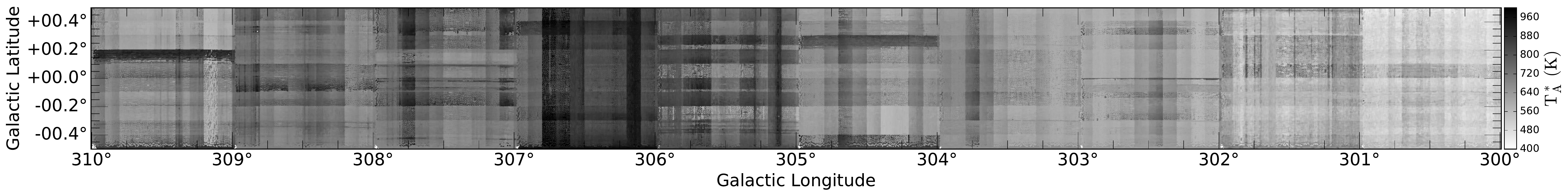}
	\caption{T$_{sys}$ images for the $^{12}$CO data covering the full 50 square degrees from 
	$l = 300$--$350^\circ$ in units of T$_A^*$ (K) (as indicated by the scale bars). The 
	striping pattern is inherent to the data set, resulting from scanning in the $l$ and $b$ 
	directions in variable observing conditions.}\label{Tsys-12CO}
\end{center}
\end{figure*}

\begin{figure*}[ht]
\begin{center}
\includegraphics[width=\textwidth]{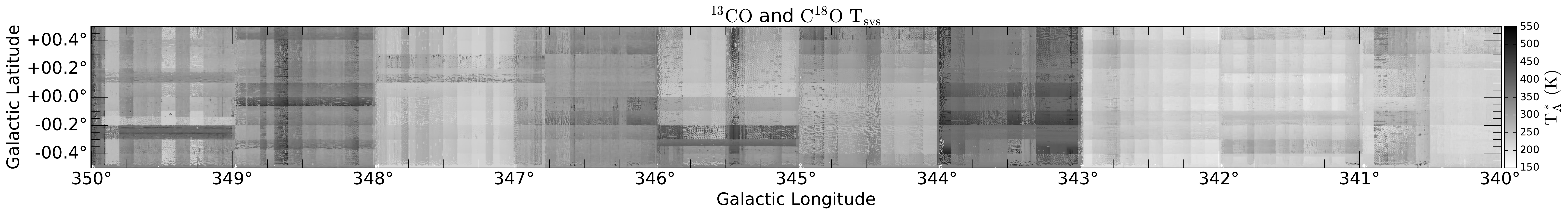}
\includegraphics[width=\textwidth]{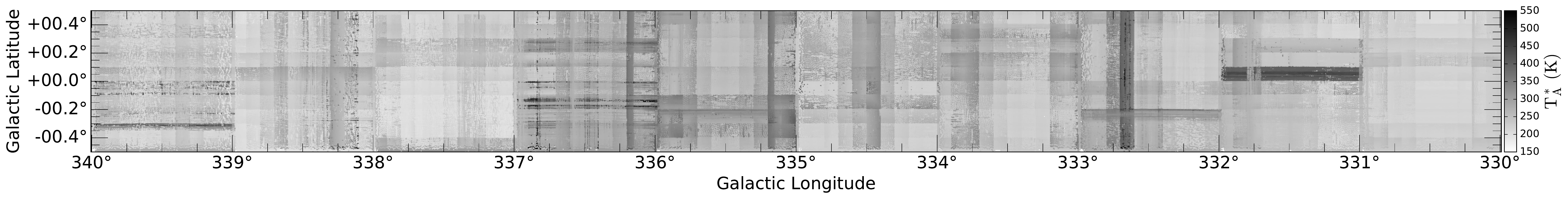}
\includegraphics[width=\textwidth]{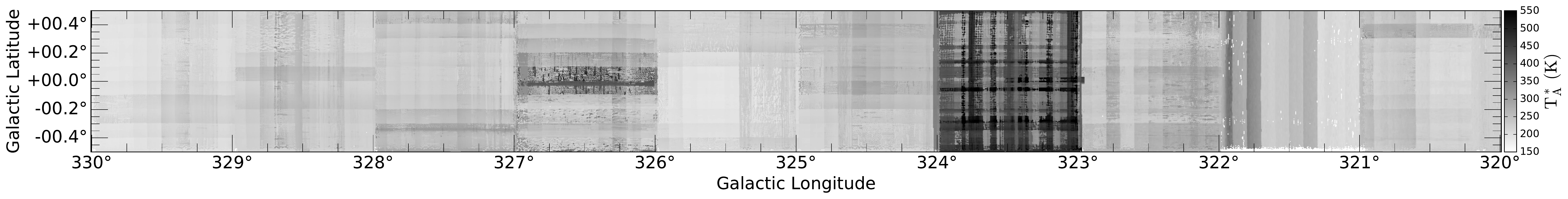}
\includegraphics[width=\textwidth]{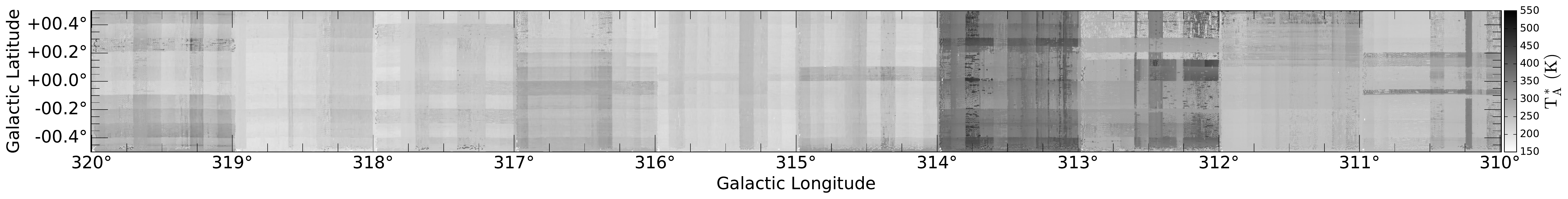}
\includegraphics[width=\textwidth]{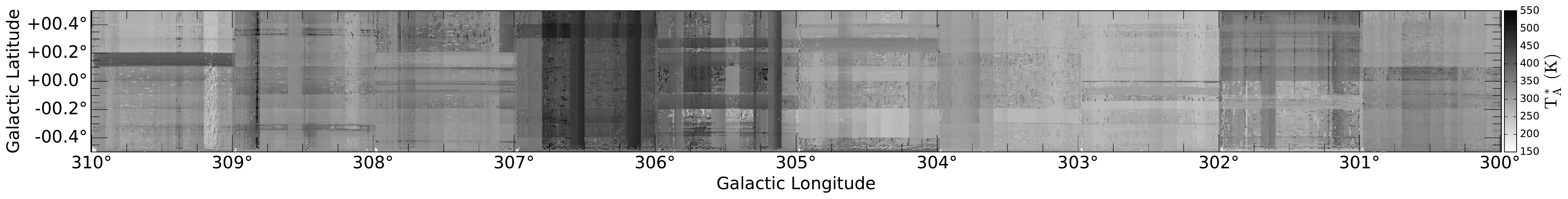}
	\caption{T$_{sys}$ images for $^{13}$CO and C$^{18}$O (which share the same 2 GHz band of 
	the spectrometer) covering the full 50 square degrees from 
	$l = 300$--$350^\circ$ in units of T$_A^*$ (K) (as indicated by the scale bars). The 
	striping pattern is inherent to the data set, resulting from scanning in the $l$ and $b$ 
	directions in variable observing conditions.}\label{Tsys-13CO}
\end{center}
\end{figure*}

	Maps of the system temperature for the $^{12}$CO isotopologue are shown in Figure \ref{Tsys-12CO}, with values approximately twice those of the $^{13}$CO and C$^{18}$O isotopologues shown in Figure \ref{Tsys-13CO} (which occur in the same 2\ GHz band of the spectrometer). Striping in these images occurs because the data is scanned in the $l$ and $b$ directions to minimize artefacts from poor sky conditions, while spots indicate points where high T$_\mathrm{sys}$ values have been removed \citep[see \S\ref{sec:DR3} and][for additional details]{Braiding:2015}. 

\subsection{Averaged Spectra}\label{app:AvSpec}

\begin{figure*}[ht]
\begin{center}
\includegraphics[width=\textwidth]{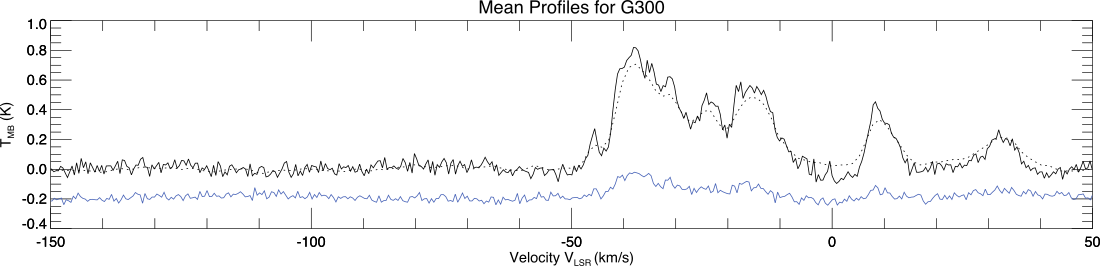}
\includegraphics[width=\textwidth]{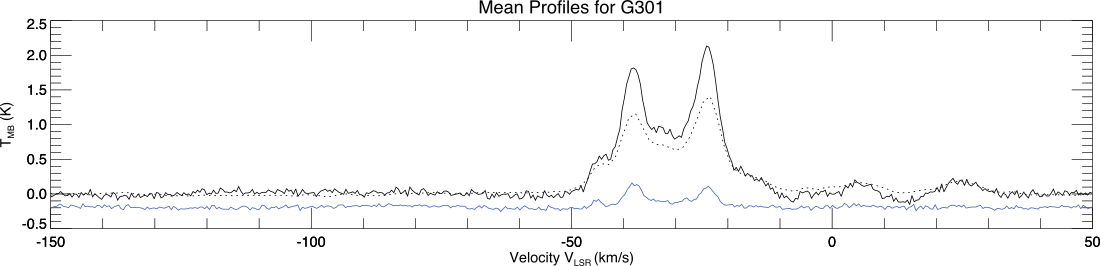}
\includegraphics[width=\textwidth]{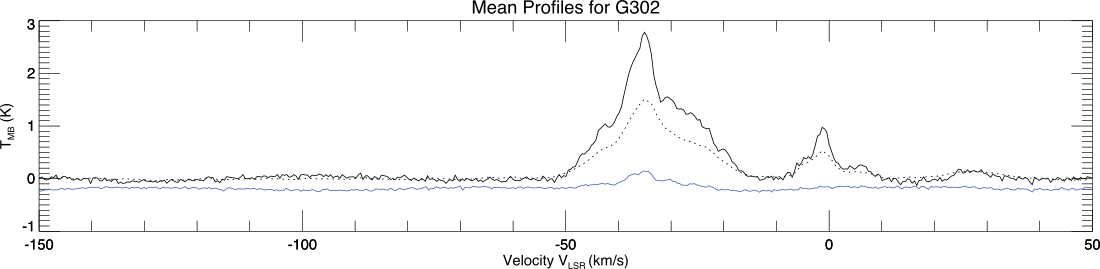}
\includegraphics[width=\textwidth]{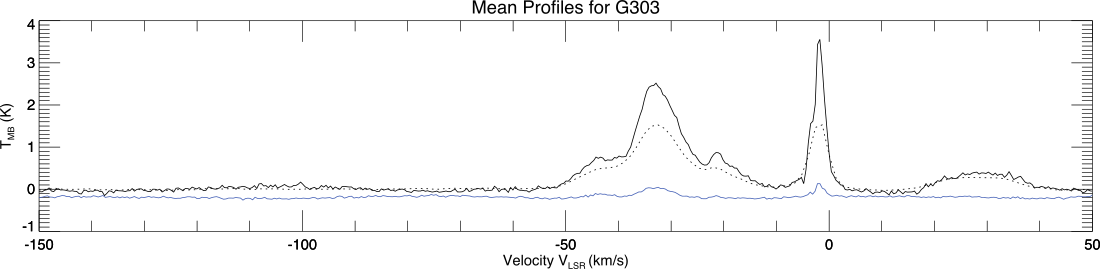}
\includegraphics[width=\textwidth]{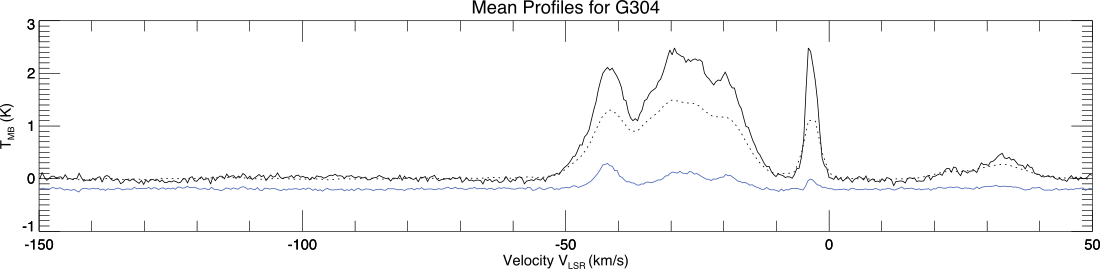}
	\caption{Average spectra for each square degree along the Galactic Plane between $l=300$--$305\degree$, labelled by their lower longitude limit, from $-150 < V_\textrm{LSR} < +50$\kms. The dark solid lines are the Mopra $^{12}$CO data, while the blue is the $^{13}$CO emission offset by $-0.2$\,K. The dashed lines are the equivalent average $^{12}$CO spectra from the Columbia CO Survey \citep{Dame:2001}, which show systematic lower line intensities by a factor of $\sim1.35$ (see Table\,\ref{tab-ratio} for degree-specific values).}\label{Spectra-300}
\end{center}

\end{figure*}

\begin{figure*}[ht]
\begin{center}
\includegraphics[width=\textwidth]{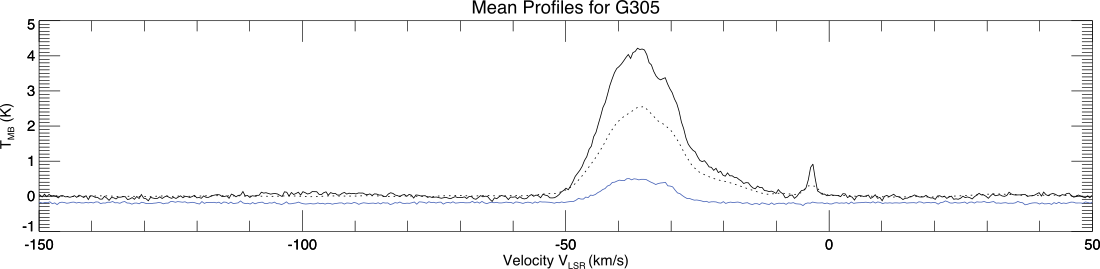}
\includegraphics[width=\textwidth]{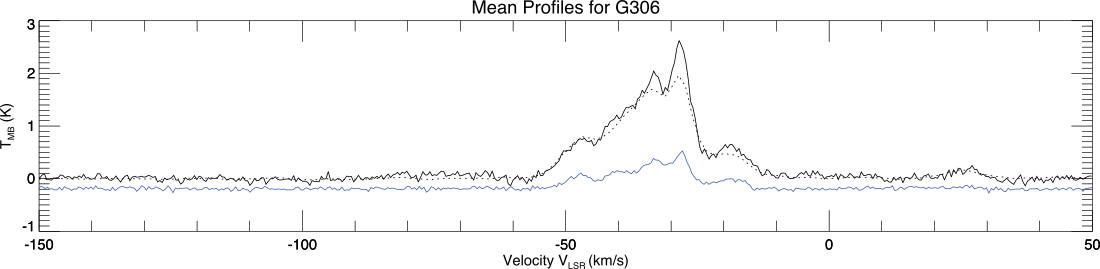}
\includegraphics[width=\textwidth]{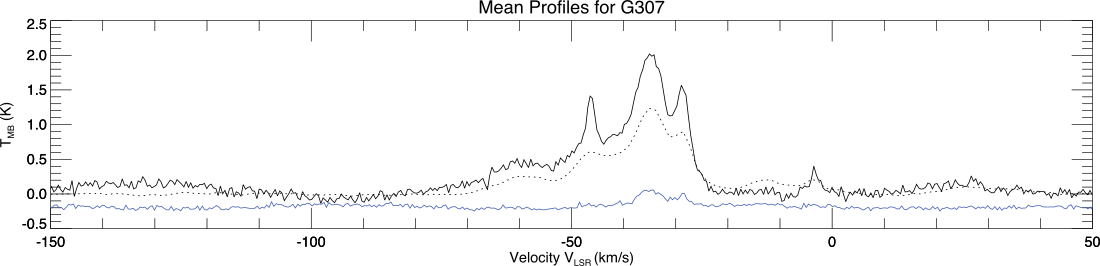}
\includegraphics[width=\textwidth]{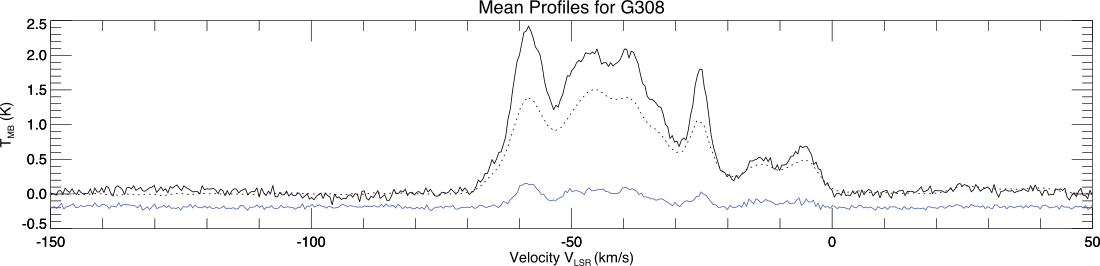}
\includegraphics[width=\textwidth]{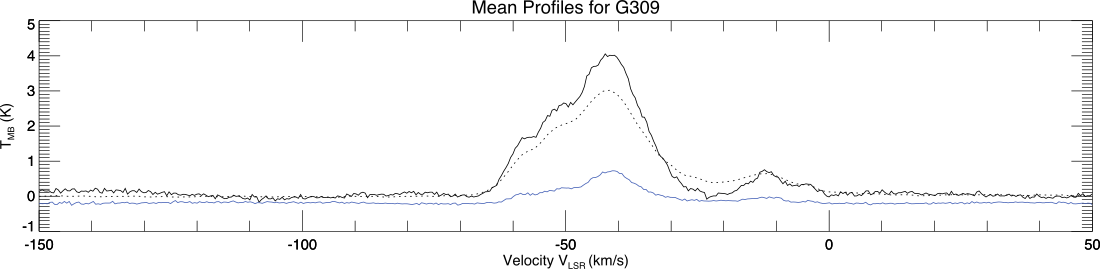}
	\caption{Average spectra for each square degree along the Galactic Plane between $l=305$--$310\degree$, labelled by their lower longitude limit, from $-150 < V_\textrm{LSR} < +50$\kms. The dark solid lines are the Mopra $^{12}$CO data, while the blue is the $^{13}$CO emission offset by $-0.2$\,K. The dashed lines are the equivalent average $^{12}$CO spectra from the Columbia CO Survey \citep{Dame:2001}, which show systematic lower line intensities by a factor of $\sim1.35$ (see Table\,\ref{tab-ratio} for degree-specific values).}\label{Spectra-305}
\end{center}
\end{figure*}

\begin{figure*}[ht]
\begin{center}
\includegraphics[width=\textwidth]{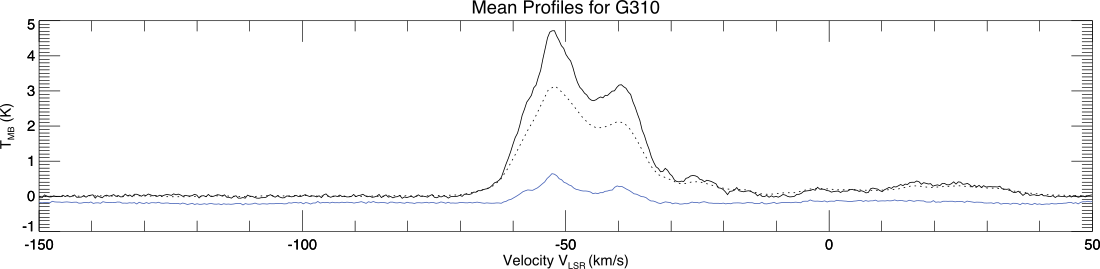}
\includegraphics[width=\textwidth]{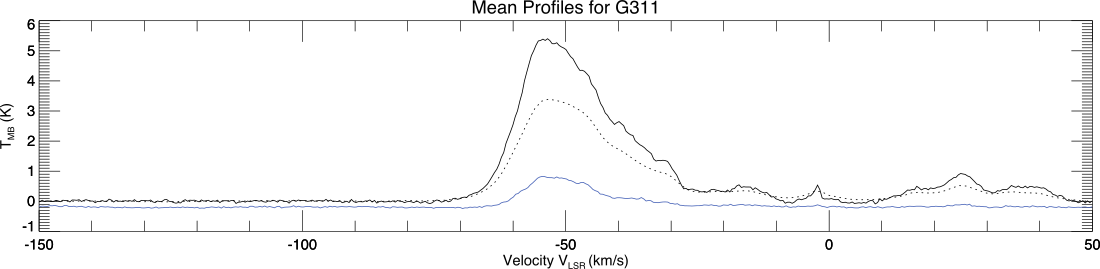}
\includegraphics[width=\textwidth]{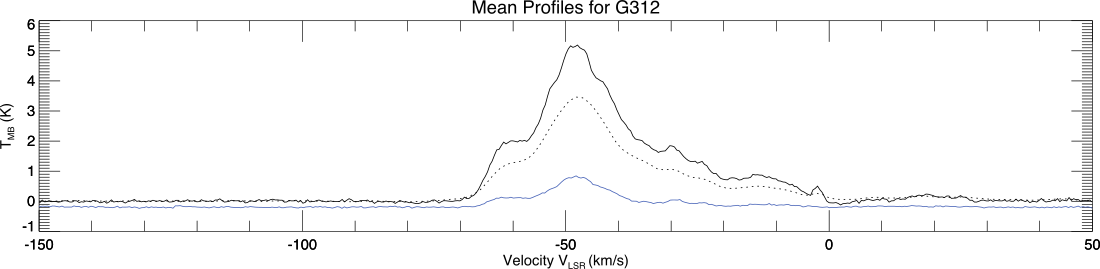}
\includegraphics[width=\textwidth]{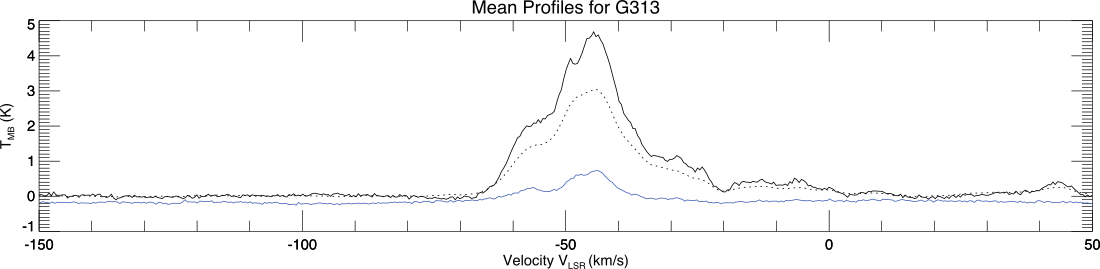}
\includegraphics[width=\textwidth]{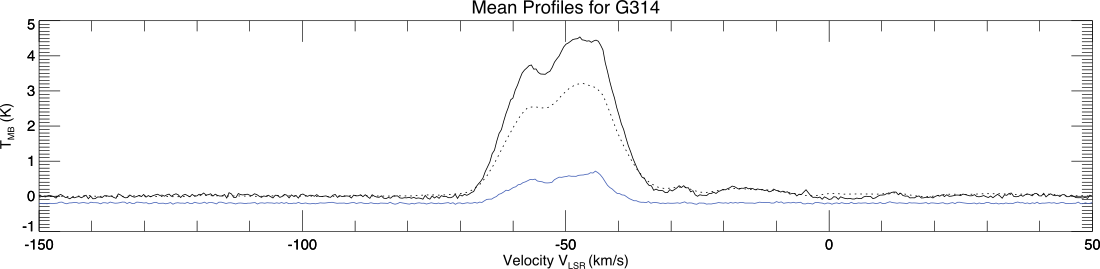}
	\caption{Average spectra for each square degree along the Galactic Plane between $l=310$--$315\degree$, labelled by their lower longitude limit, from $-150 < V_\textrm{LSR} < +50$\kms. The dark solid lines are the Mopra $^{12}$CO data, while the blue is the $^{13}$CO emission offset by $-0.2$\,K. The dashed lines are the equivalent average $^{12}$CO spectra from the Columbia CO Survey \citep{Dame:2001}, which show systematic lower line intensities by a factor of $\sim1.35$ (see Table\,\ref{tab-ratio} for degree-specific values).}\label{Spectra-310}
\end{center}
\end{figure*}

\begin{figure*}[ht]
\begin{center}
\includegraphics[width=\textwidth]{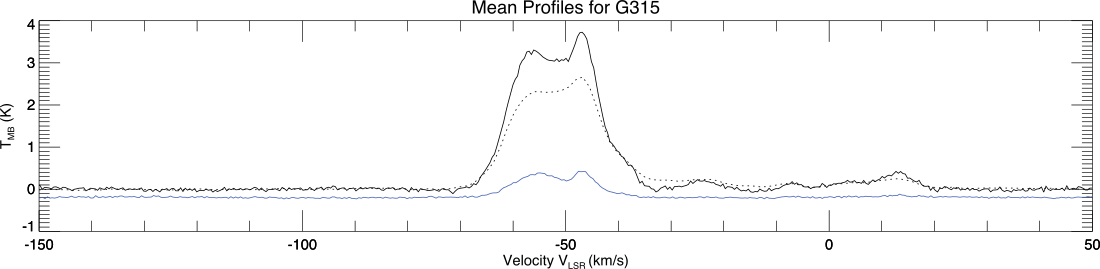}
\includegraphics[width=\textwidth]{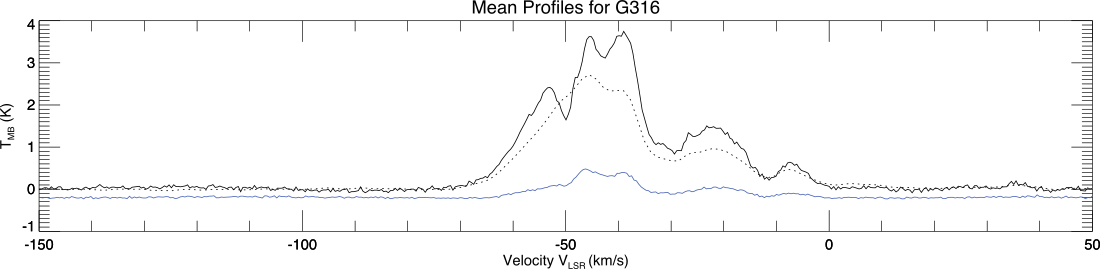}
\includegraphics[width=\textwidth]{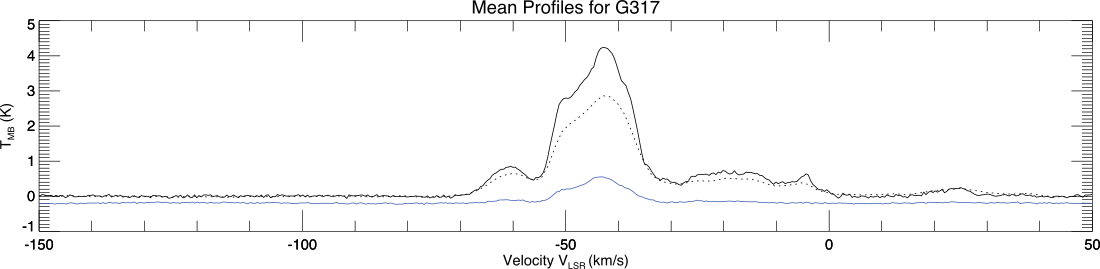}
\includegraphics[width=\textwidth]{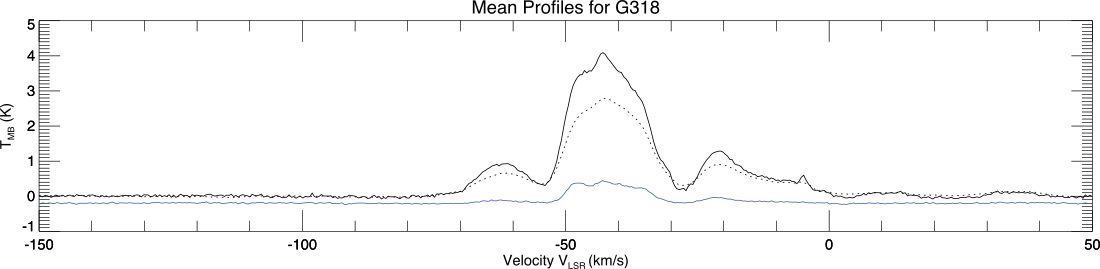}
\includegraphics[width=\textwidth]{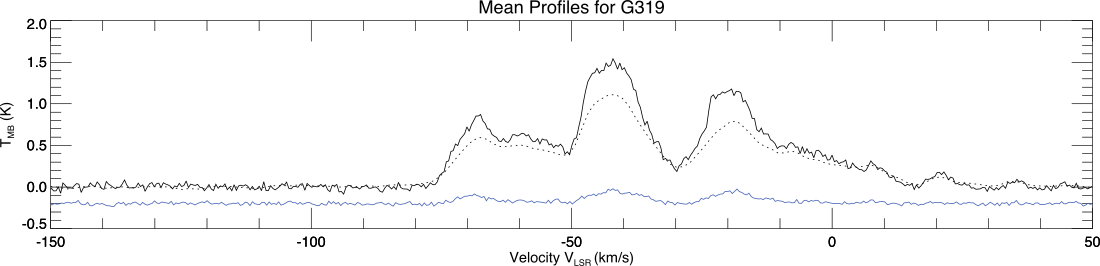}
	\caption{Average spectra for each square degree along the Galactic Plane between $l=315$--$320\degree$, labelled by their lower longitude limit, from $-150 < V_\textrm{LSR} < +50$\kms. The dark solid lines are the Mopra $^{12}$CO data, while the blue is the $^{13}$CO emission offset by $-0.2$\,K. The dashed lines are the equivalent average $^{12}$CO spectra from the Columbia CO Survey \citep{Dame:2001}, which show systematic lower line intensities by a factor of $\sim1.35$ (see Table\,\ref{tab-ratio} for degree-specific values).}\label{Spectra-315}
\end{center}
\end{figure*}

\begin{figure*}[ht]
\begin{center}
\includegraphics[width=\textwidth]{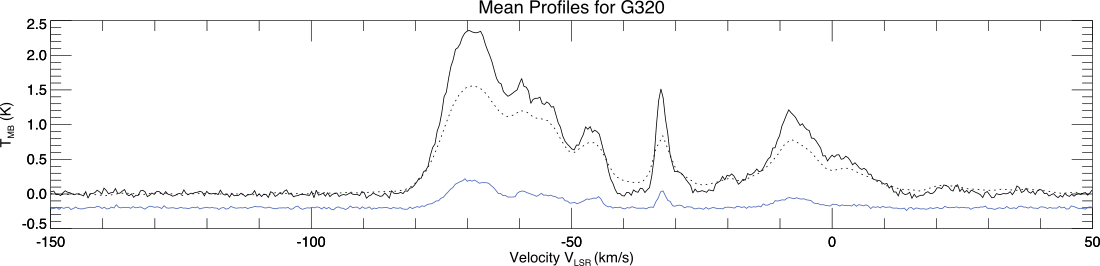}
\includegraphics[width=\textwidth]{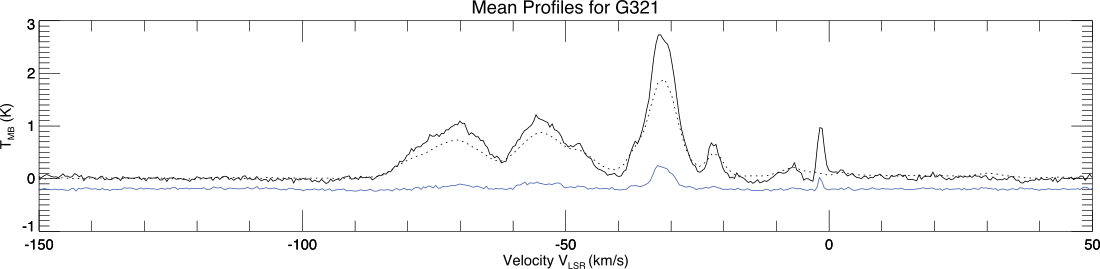}
\includegraphics[width=\textwidth]{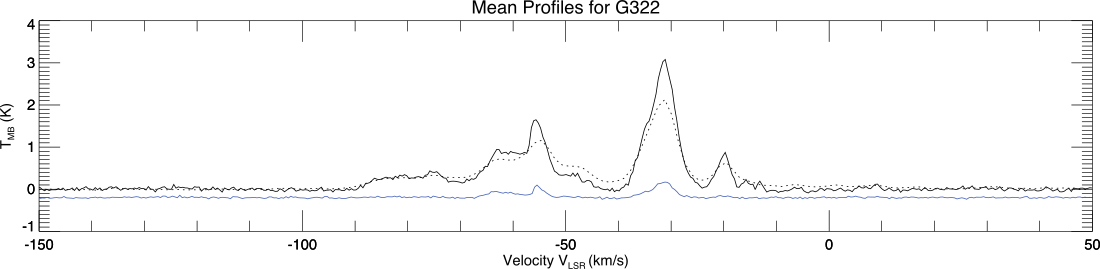}
\includegraphics[width=\textwidth]{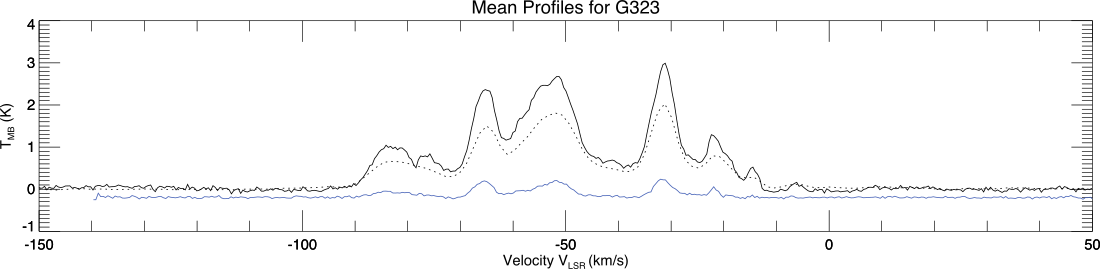}
\includegraphics[width=\textwidth]{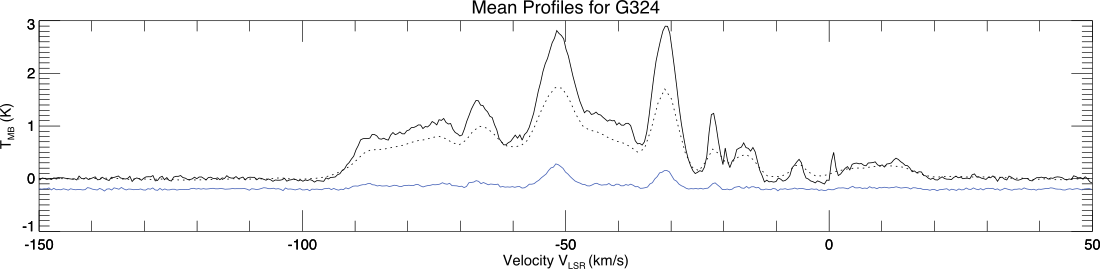}
	\caption{Average spectra for each square degree along the Galactic Plane between $l=320$--$325\degree$, labelled by their lower longitude limit, from $-150 < V_\textrm{LSR} < +50$\kms. The dark solid lines are the Mopra $^{12}$CO data, while the blue is the $^{13}$CO emission offset by $-0.2$\,K. The dashed lines are the equivalent average $^{12}$CO spectra from the Columbia CO Survey \citep{Dame:2001}, which show systematic lower line intensities by a factor of $\sim1.35$ (see Table\,\ref{tab-ratio} for degree-specific values).}\label{Spectra-320}
\end{center}
\end{figure*}

\begin{figure*}[ht]
\begin{center}
\includegraphics[width=\textwidth]{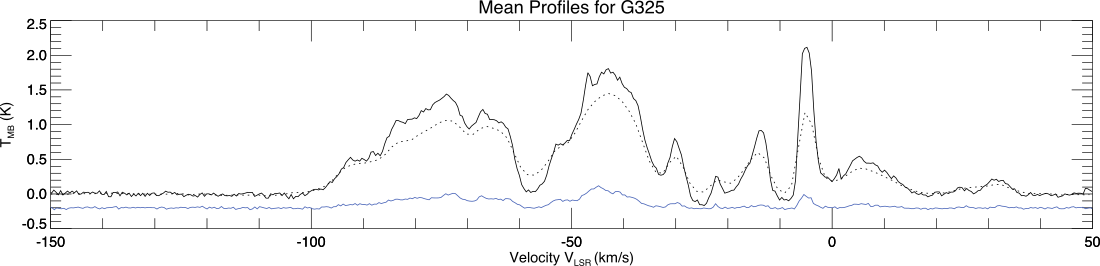}
\includegraphics[width=\textwidth]{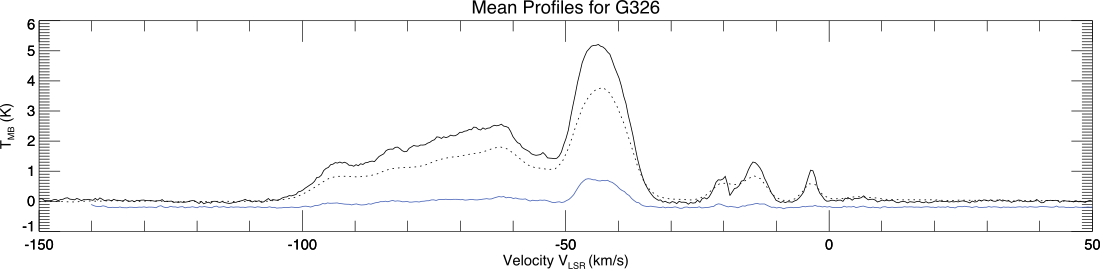}
\includegraphics[width=\textwidth]{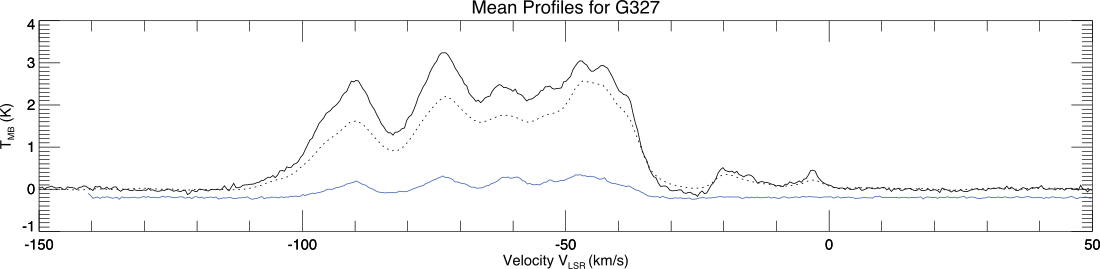}
\includegraphics[width=\textwidth]{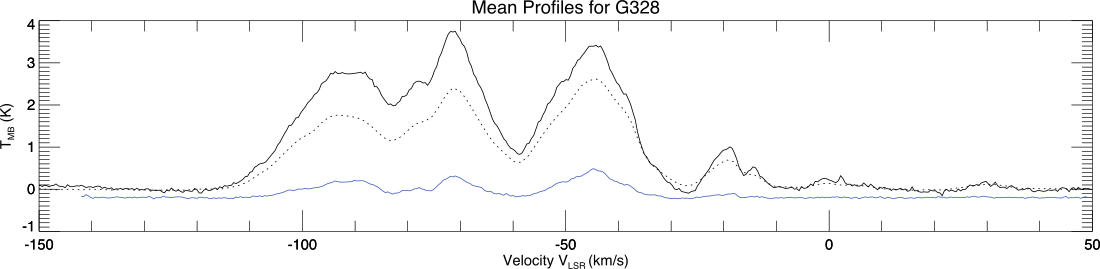}
\includegraphics[width=\textwidth]{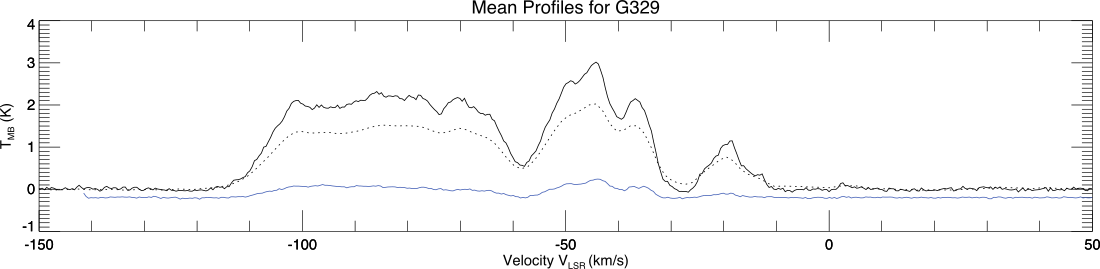}
	\caption{Average spectra for each square degree along the Galactic Plane between $l=325$--$330\degree$, labelled by their lower longitude limit, from $-150 < V_\textrm{LSR} < +50$\kms. The dark solid lines are the Mopra $^{12}$CO data, while the blue is the $^{13}$CO emission offset by $-0.2$\,K. The dashed lines are the equivalent average $^{12}$CO spectra from the Columbia CO Survey \citep{Dame:2001}, which show systematic lower line intensities by a factor of $\sim1.35$ (see Table\,\ref{tab-ratio} for degree-specific values).}\label{Spectra-325}
\end{center}
\end{figure*}

\begin{figure*}[ht]
\begin{center}
\includegraphics[width=\textwidth]{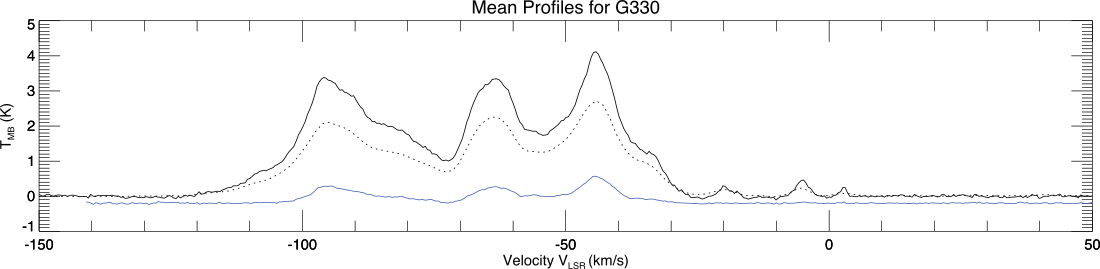}
\includegraphics[width=\textwidth]{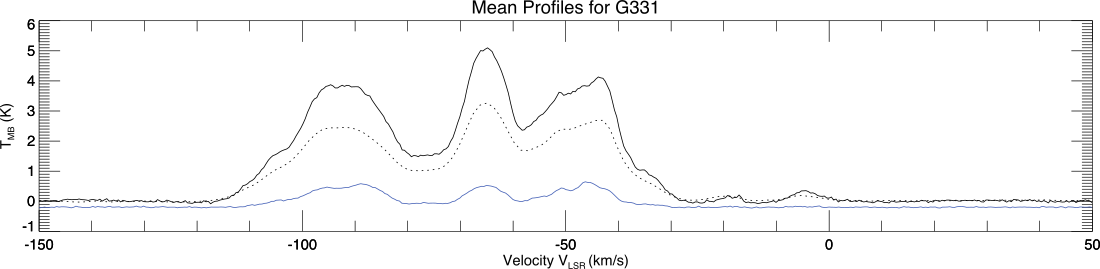}
\includegraphics[width=\textwidth]{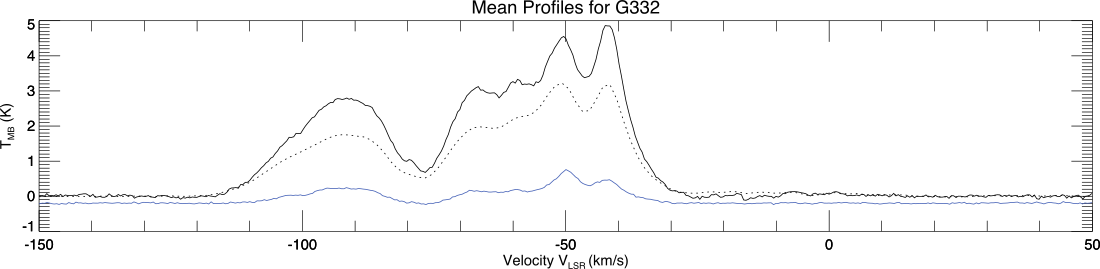}
\includegraphics[width=\textwidth]{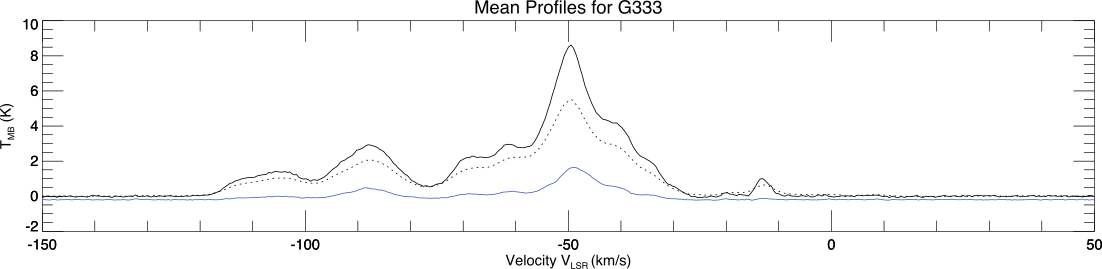}
\includegraphics[width=\textwidth]{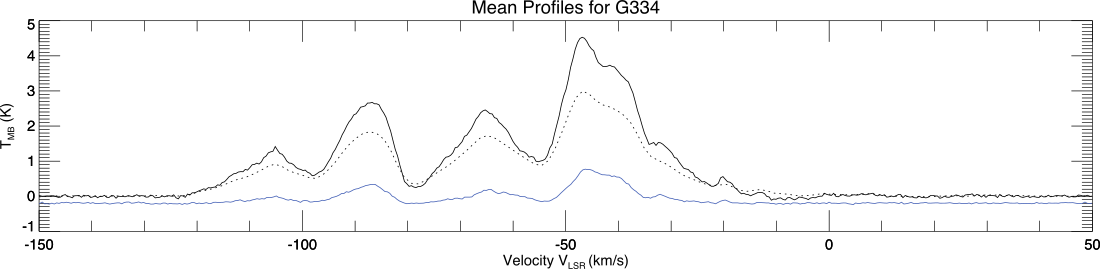}
	\caption{Average spectra for each square degree along the Galactic Plane between $l=330$--$335\degree$, labelled by their lower longitude limit, from $-150 < V_\textrm{LSR} < +50$\kms. The dark solid lines are the Mopra $^{12}$CO data, while the blue is the $^{13}$CO emission offset by $-0.2$\,K. The dashed lines are the equivalent average $^{12}$CO spectra from the Columbia CO Survey \citep{Dame:2001}, which show systematic lower line intensities by a factor of $\sim1.35$ (see Table\,\ref{tab-ratio} for degree-specific values).}\label{Spectra-330}
\end{center}
\end{figure*}

\begin{figure*}[ht]
\begin{center}
\includegraphics[width=\textwidth]{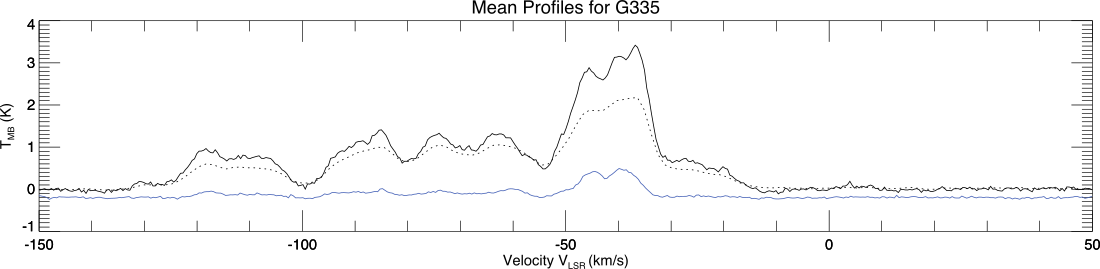}
\includegraphics[width=\textwidth]{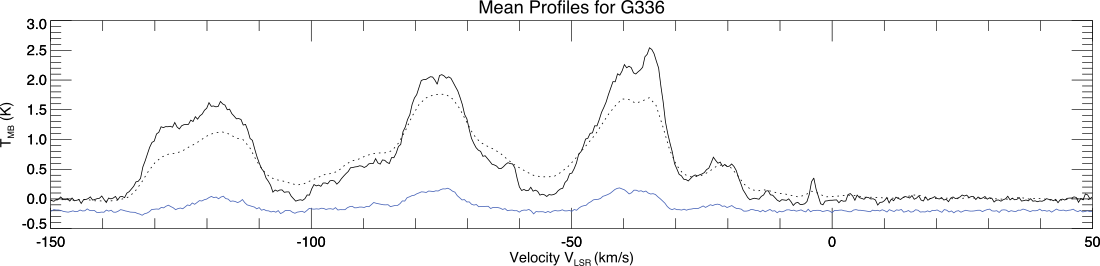}
\includegraphics[width=\textwidth]{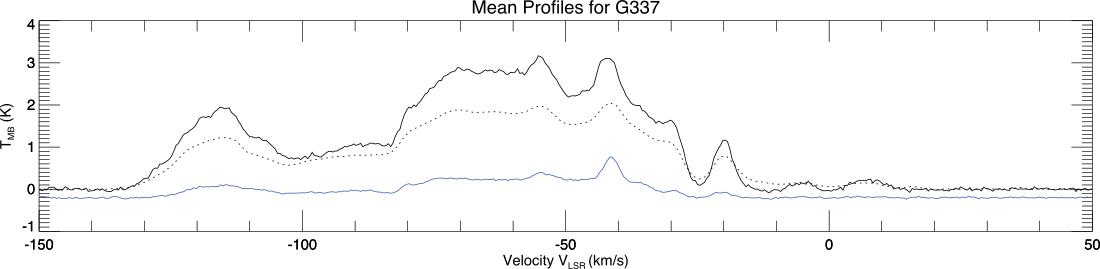}
\includegraphics[width=\textwidth]{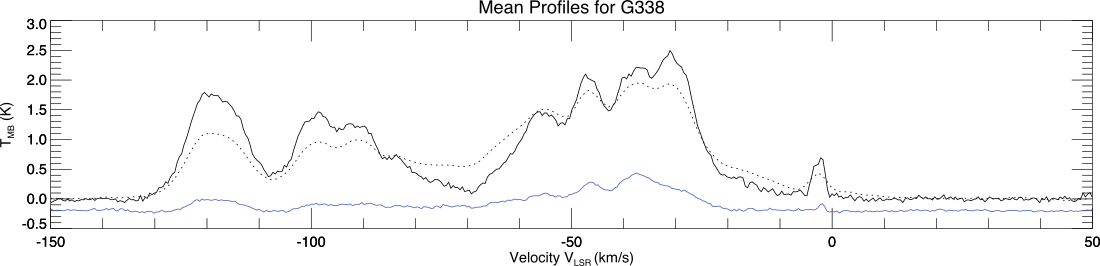}
\includegraphics[width=\textwidth]{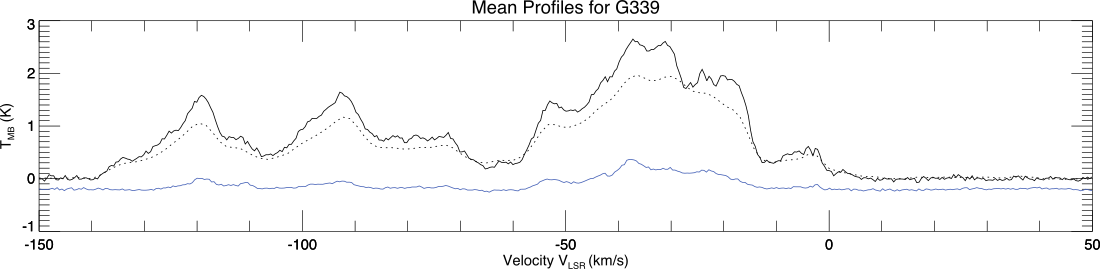}
	\caption{Average spectra for each square degree along the Galactic Plane between $l=335$--$340\degree$, labelled by their lower longitude limit, from $-150 < V_\textrm{LSR} < +50$\kms. The dark solid lines are the Mopra $^{12}$CO data, while the blue is the $^{13}$CO emission offset by $-0.2$\,K. The dashed lines are the equivalent average $^{12}$CO spectra from the Columbia CO Survey \citep{Dame:2001}, which show systematic lower line intensities by a factor of $\sim1.35$ (see Table\,\ref{tab-ratio} for degree-specific values).}\label{Spectra-335}
\end{center}
\end{figure*}

\begin{figure*}[ht]
\begin{center}
\includegraphics[width=\textwidth]{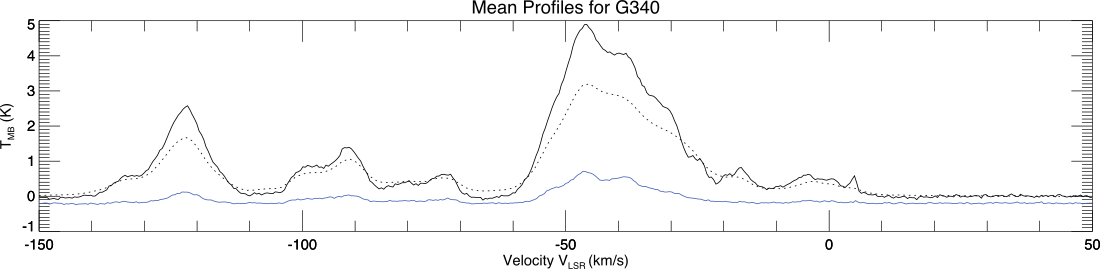}
\includegraphics[width=\textwidth]{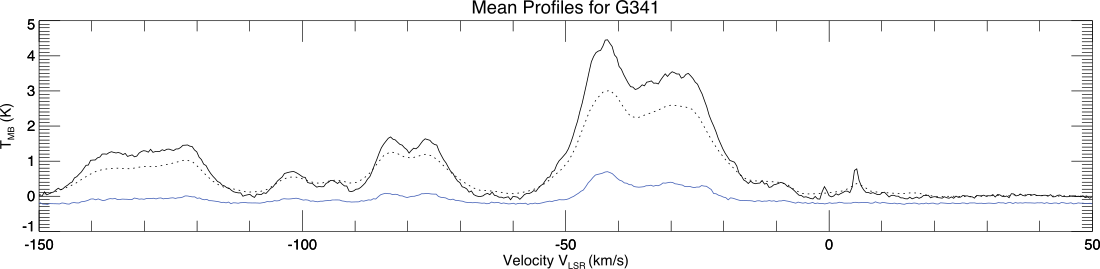}
\includegraphics[width=\textwidth]{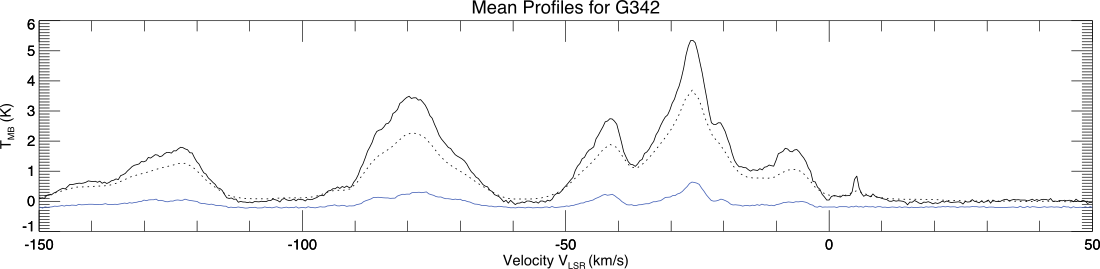}
\includegraphics[width=\textwidth]{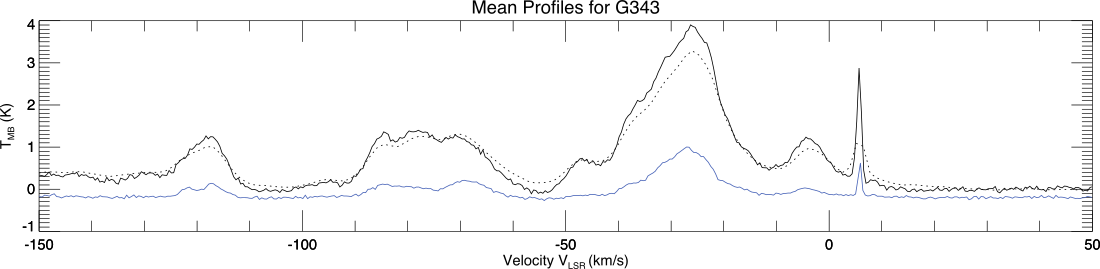}
\includegraphics[width=\textwidth]{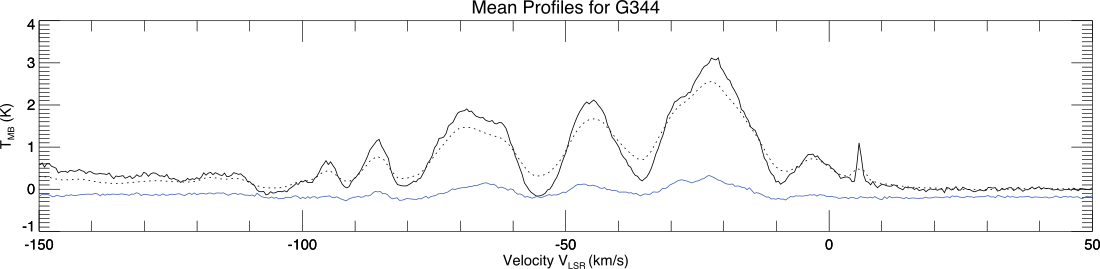}
	\caption{Average spectra for each square degree along the Galactic Plane between $l=340$--$345\degree$, labelled by their lower longitude limit, from $-150 < V_\textrm{LSR} < +50$\kms. The dark solid lines are the Mopra $^{12}$CO data, while the blue is the $^{13}$CO emission offset by $-0.2$\,K. The dashed lines are the equivalent average $^{12}$CO spectra from the Columbia CO Survey \citep{Dame:2001}, which show systematic lower line intensities by a factor of $\sim1.35$ (see Table\,\ref{tab-ratio} for degree-specific values).}\label{Spectra-340}
\end{center}
\end{figure*}

\begin{figure*}[ht]
\begin{center}
\includegraphics[width=\textwidth]{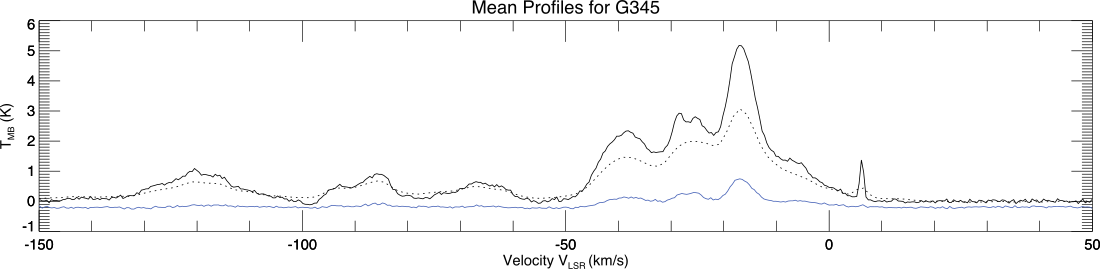}
\includegraphics[width=\textwidth]{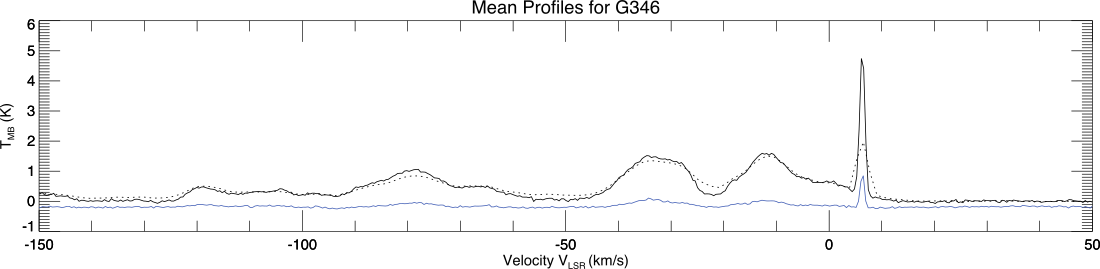}
\includegraphics[width=\textwidth]{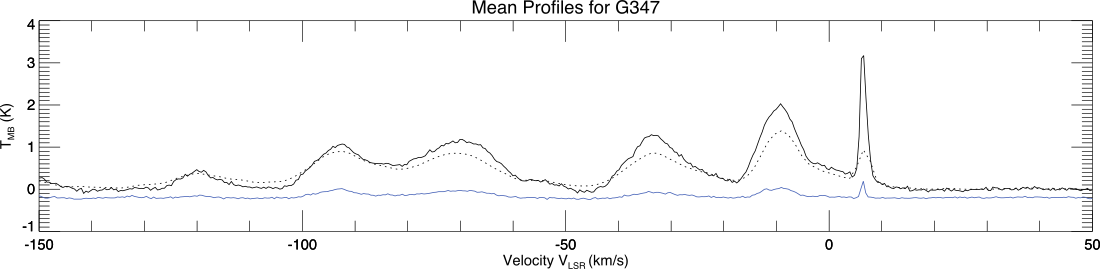}
\includegraphics[width=\textwidth]{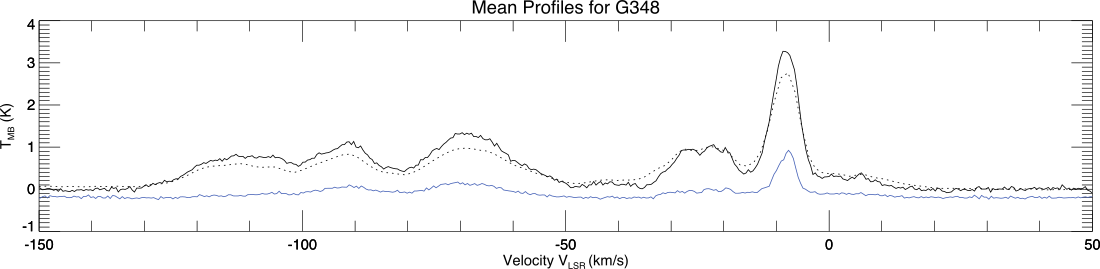}
\includegraphics[width=\textwidth]{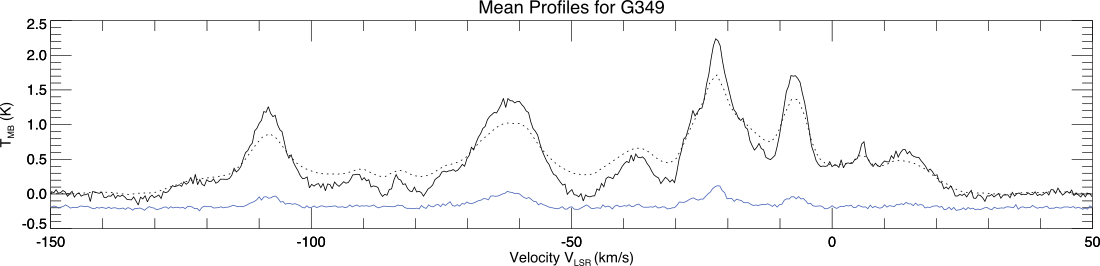}
	\caption{Average spectra for each square degree along the Galactic Plane between $l=345$--$350\degree$, labelled by their lower longitude limit, from $-150 < V_\textrm{LSR} < +50$\kms. The dark solid lines are the Mopra $^{12}$CO data, while the blue is the $^{13}$CO emission offset by $-0.2$\,K. The dashed lines are the equivalent average $^{12}$CO spectra from the Columbia CO Survey \citep{Dame:2001}, which show systematic lower line intensities by a factor of $\sim1.35$ (see Table\,\ref{tab-ratio} for degree-specific values).}\label{Spectra-345}
\end{center}
\end{figure*}

The average line profiles for the $^{12}$CO, $^{13}$CO, and $^{12}$CO from the Columbia survey \citep[][]{Dame:2001}, averaged over each square degree are displayed in Figures\,\ref{Spectra-300}--\ref{Spectra-345}. A distinct progression towards increasingly negative velocities is seen from longitudes $l=300$--$350\degree$. At $l=300\degree$, the line of sight spectral line velocities are all higher than $-50$\,km/s, consistent with the lower limit on velocity set by the Scutum-Crux arm tangent in the Galactic model adopted in this paper (see Figure\,\ref{fig:GalRot}). Positive-velocity spectral lines typically indicate emission seen from the far-side of the Galaxy (beyond the Solar Circle).

\begin{table*}
	\caption{Intensity ratios between the average spectra for each square degree of Mopra CO and the Columbia CO Survey \citep{Dame:2001}, \textbf{calculated using the integrated intensities of each spectrum}. The mean over the 50 square degrees is 1.35, and the median 1.36.}
\begin{center}
\begin{tabular}{cc|cc|cc|cc|cc}
\hline\hline
	Cube & Ratio & Cube & Ratio & Cube & Ratio & Cube & Ratio & Cube & Ratio\\
\hline%
	G300  & 1.12 & G310  & 1.44 & G320  & 1.30 & G330  & 1.39 & G340  & 1.41\\
	G301  & 1.25 & G311  & 1.48 & G321  & 1.20 & G331  & 1.47 & G341  & 1.38\\
	G302  & 1.63 & G312  & 1.45 & G322  & 1.18 & G332  & 1.43 & G342  & 1.39\\
	G303  & 1.42 & G313  & 1.43 & G323  & 1.23 & G333  & 1.45 & G343  & 1.13\\
	G304  & 1.44 & G314  & 1.35 & G324  & 1.28 & G334  & 1.45 & G344  & 1.44\\
	G305  & 1.59 & G315  & 1.23 & G325  & 1.28 & G335  & 1.23 & G345  & 1.49\\ 
	G306  & 1.19 & G316  & 1.27 & G326  & 1.27 & G336  & 1.32 & G346  & 1.14\\ 
	G307  & 1.67 & G317  & 1.34 & G327  & 1.27 & G337  & 1.54 & G347  & 1.40\\
	G308  & 1.38 & G318  & 1.26 & G328  & 1.33 & G338  & 1.47 & G348  & 1.26\\
	G309  & 1.40 & G319  & 1.01 & G329  & 1.33 & G339  & 1.41 & G349  & 1.30\\
\hline\hline
\end{tabular}
\end{center}
\label{tab-ratio}
\end{table*}
With increasing longitude, the range of spectral line velocities increases, and the negative extent of the range decreases beyond $-100$\,km/s, consistent with expectations of the inner-Galaxy (see Figure\,\ref{fig:GalRot}). The mean intensity ratio between the average spectra illustrated in Figures\,\ref{Spectra-300}--\ref{Spectra-345} from the MopraCO and the Dame survey is 1.35 (with a median value of 1.36), with the individual values for each square degree listed in Table \ref{tab-ratio} below. 

\subsection{Integrated intensity maps}\label{app:IntMaps}
\begin{figure*}[htp]
\begin{center}
\includegraphics[width=\textwidth]{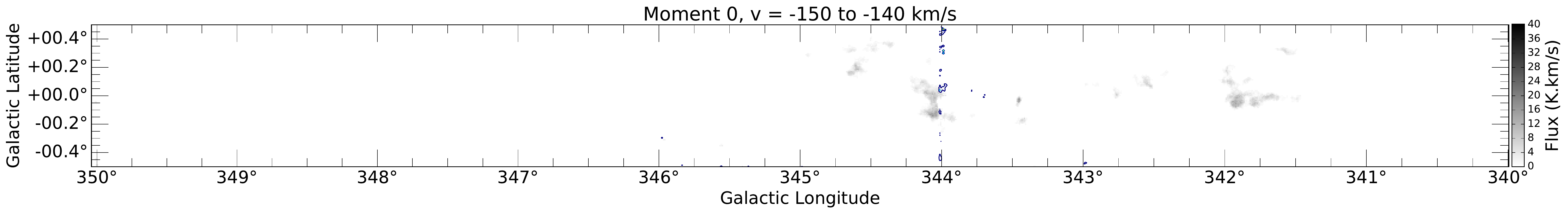}
	\caption{Moment 0 image for $l = 340$--$350^\circ$ calculated over the velocity interval $v = -150$ to $-140$ \kms\ using the main beam intensity, T$_\textrm{MB}$. The grayscale $^{12}$CO image runs from 0 to 40 K.km/s, while the $^{13}$CO contours are at 1, 6, 11, 16 K.km/s. None of the other data cubes demonstrate significant emission in this velocity range. Note that some scanning artefacts are present in this and following images, evident as 1$^{\circ}$-long line features in longitude or latitude.}\label{Mom-150}
\end{center}
\end{figure*}

\begin{figure*}[htp]
\begin{center}
\includegraphics[width=\textwidth]{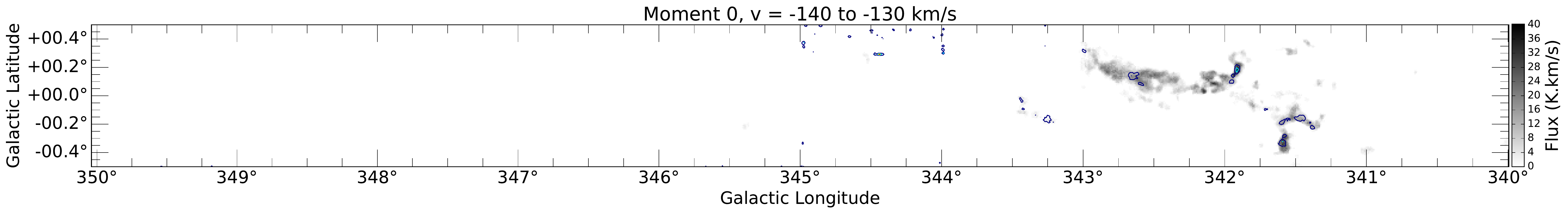}
\includegraphics[width=\textwidth]{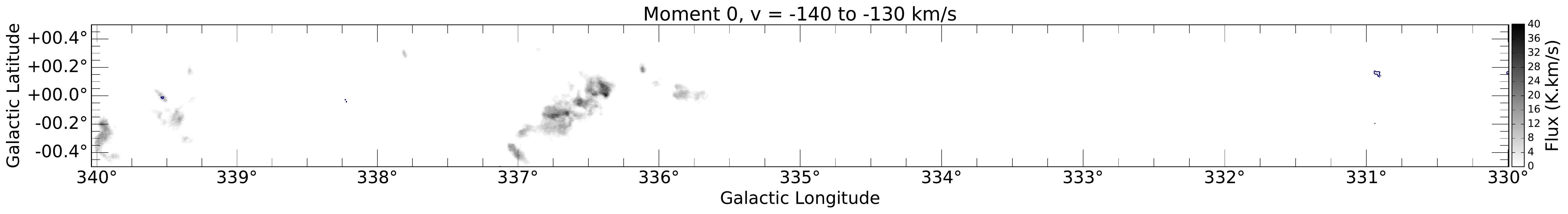}
	\caption{Moment 0 images for $l = 330$--$350^\circ$ calculated over the velocity interval $v = -140$ to $-130$ \kms. The grayscale $^{12}$CO image runs from 0 to 40 K.km/s, while the $^{13}$CO contours are at 1, 6, 11, 16 K.km/s. None of the other data cubes demonstrate significant emission in this velocity range.}\label{Mom-140}
\end{center}
\end{figure*}

\begin{figure*}[htp]
\begin{center}
\includegraphics[width=\textwidth]{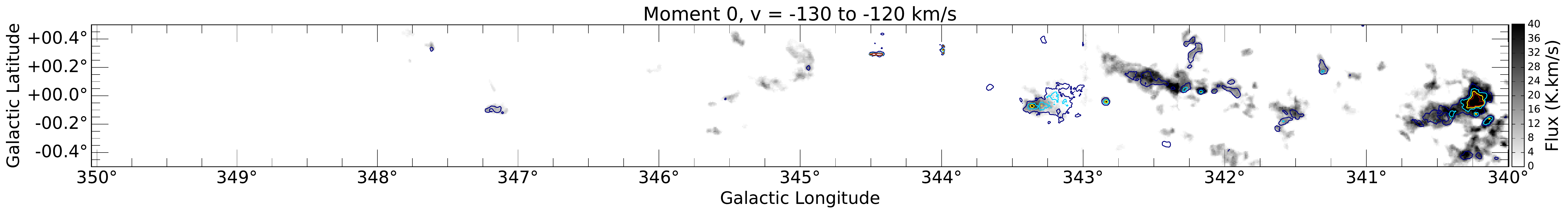}
\includegraphics[width=\textwidth]{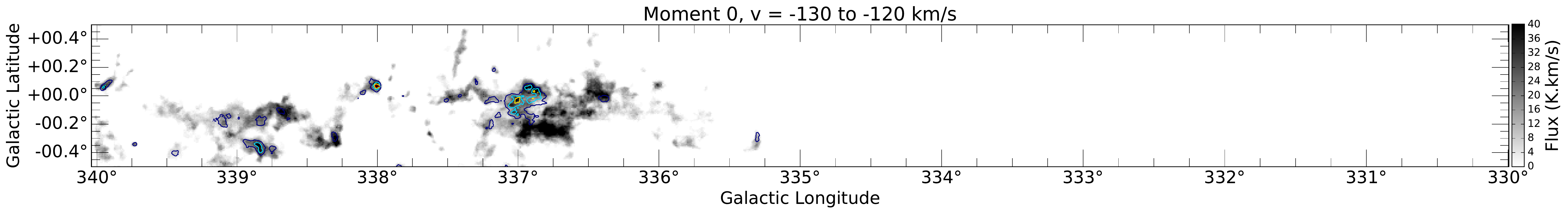}
	\caption{Moment 0 images for $l = 330$--$350^\circ$ calculated over the velocity interval $v = -130$ to $-120$ \kms. The grayscale $^{12}$CO image runs from 0 to 40 K.km/s, while the $^{13}$CO contours are at 1, 6, 11, 16 K.km/s. None of the other data cubes demonstrate significant emission in this velocity range.}\label{Mom-130}
\end{center}
\end{figure*}

\begin{figure*}[htp]
\begin{center}
\includegraphics[width=\textwidth]{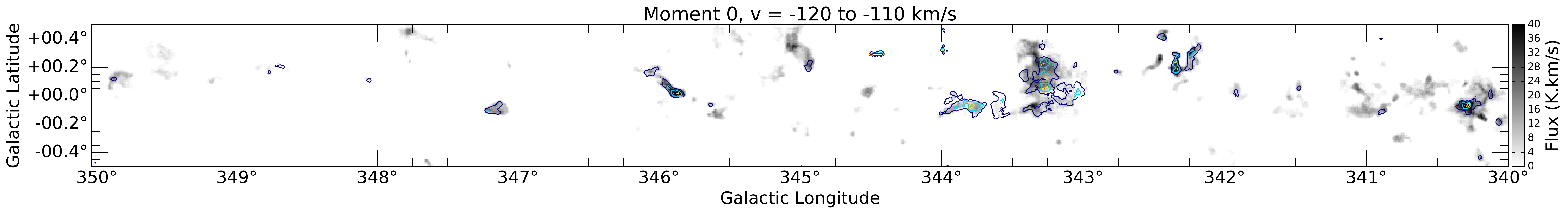}
\includegraphics[width=\textwidth]{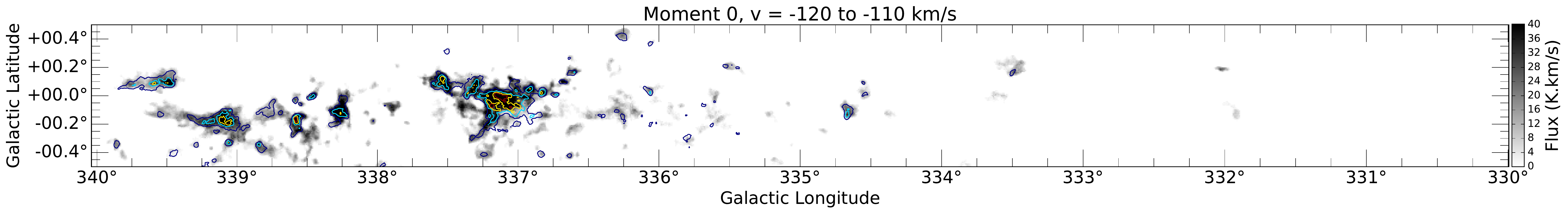}
	\caption{Moment 0 images for $l = 330$--$350^\circ$ calculated over the velocity interval $v = -120$ to $-110$ \kms. The grayscale $^{12}$CO image runs from 0 to 40 K.km/s, while the $^{13}$CO contours are at 1, 6, 11, 16 K.km/s. None of the other data cubes demonstrate significant emission in this velocity range.}\label{Mom-120}
\end{center}
\end{figure*}

\begin{figure*}[htp]
\begin{center}
\includegraphics[width=\textwidth]{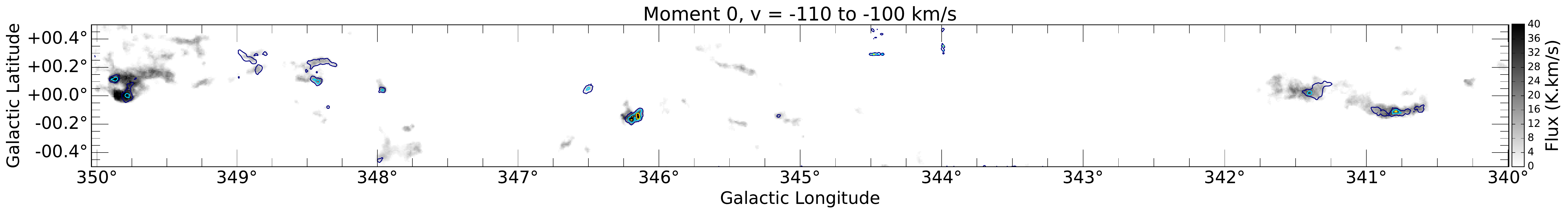}
\includegraphics[width=\textwidth]{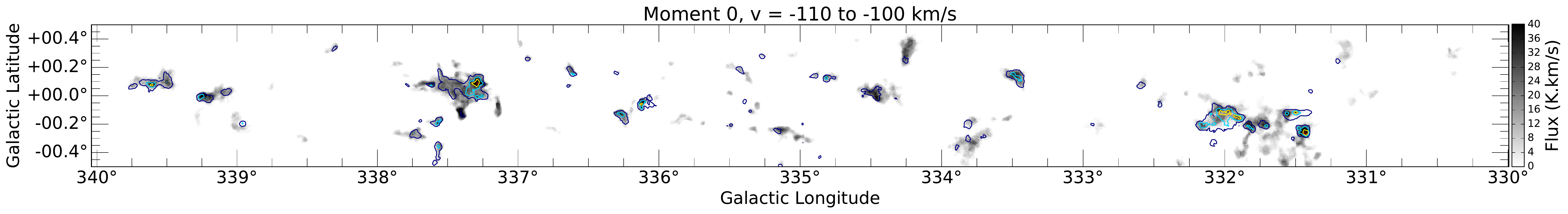}
\includegraphics[width=\textwidth]{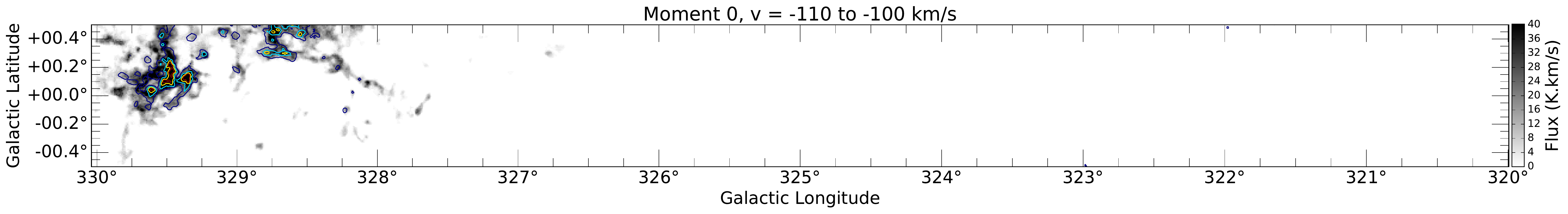}
	\caption{Moment 0 images for $l = 320$--$350^\circ$ calculated over the velocity interval $v = -110$ to $-100$ \kms. The grayscale $^{12}$CO image runs from 0 to 40 K.km/s, while the $^{13}$CO contours are at 1, 6, 11, 16 K.km/s. None of the other data cubes demonstrate significant emission in this velocity range.}\label{Mom-110}
\end{center}
\end{figure*}

\begin{figure*}[htp]
\begin{center}
\includegraphics[width=\textwidth]{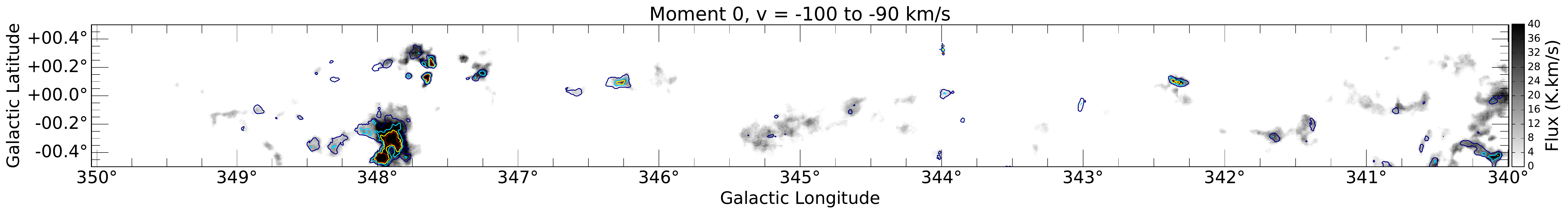}
\includegraphics[width=\textwidth]{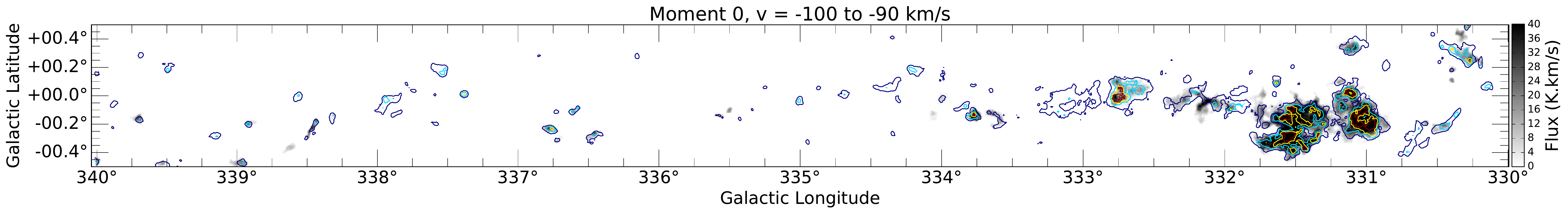}
\includegraphics[width=\textwidth]{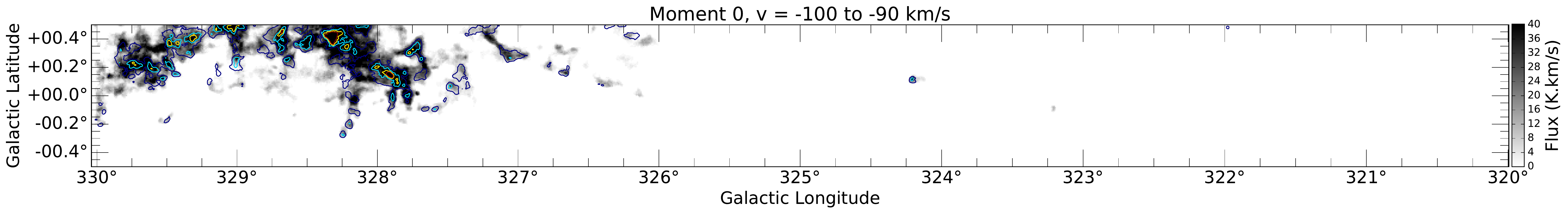}
	\caption{Moment 0 images for $l = 320$--$350^\circ$ calculated over the velocity interval $v = -100$ to $-90$ \kms. The grayscale $^{12}$CO image runs from 0 to 40 K.km/s, while the $^{13}$CO contours are at 1, 6, 11, 16 K.km/s. None of the other data cubes demonstrate significant emission in this velocity range.}\label{Mom-100}
\end{center}
\end{figure*}

\begin{figure*}[htp]
\begin{center}
\includegraphics[width=\textwidth]{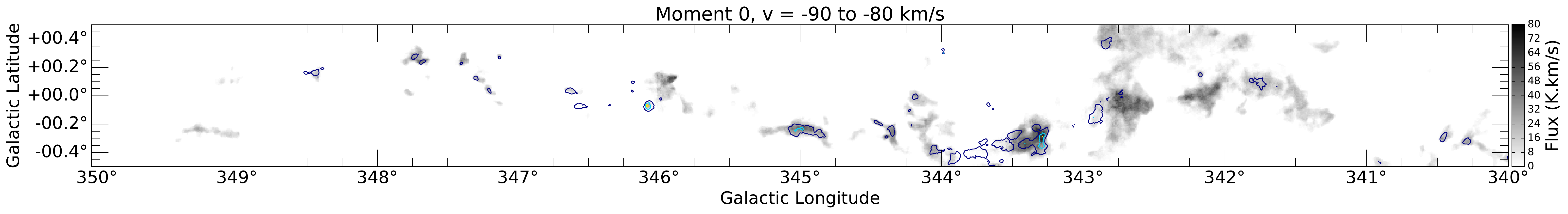}
\includegraphics[width=\textwidth]{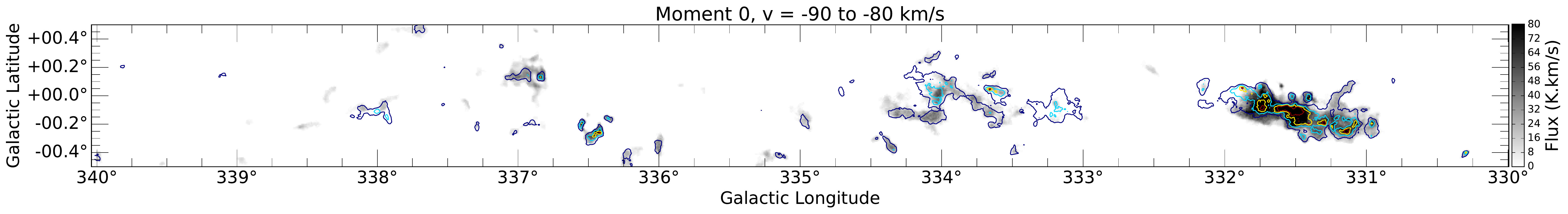}
\includegraphics[width=\textwidth]{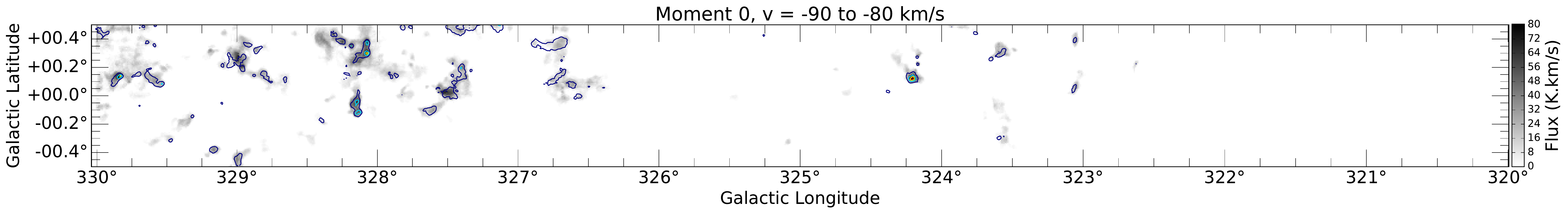}
	\caption{Moment 0 images for $l = 320$--$350^\circ$ calculated over the velocity interval $v = -90$ to $-80$ \kms. Note that the scales have changed here from previous images: the grayscale $^{12}$CO image now runs from 0 to 80 K.km/s, while the $^{13}$CO contours are at 4, 12, 20, 28 K.km/s. None of the other data cubes demonstrate significant emission in this velocity range.}\label{Mom-90}
\end{center}
\end{figure*}

\begin{figure*}[htp]
\begin{center}
\includegraphics[width=\textwidth]{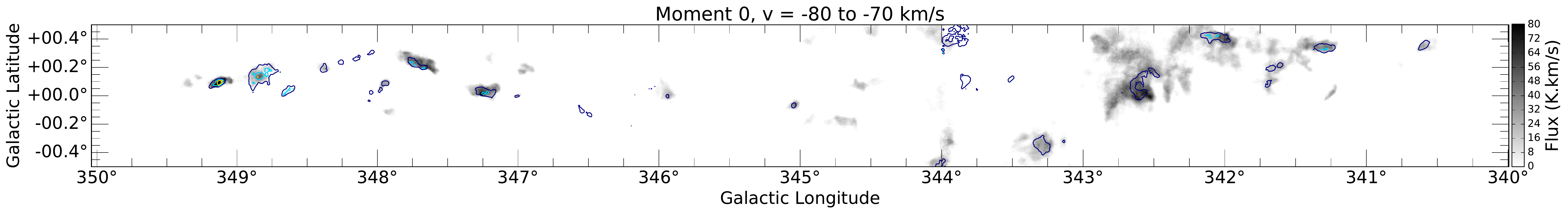}
\includegraphics[width=\textwidth]{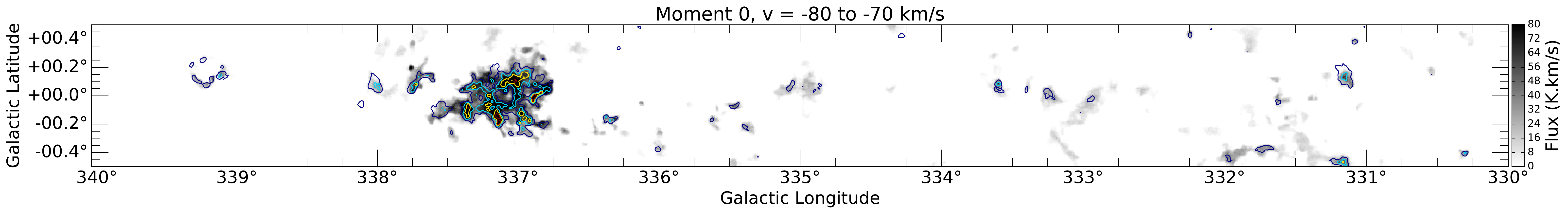}
\includegraphics[width=\textwidth]{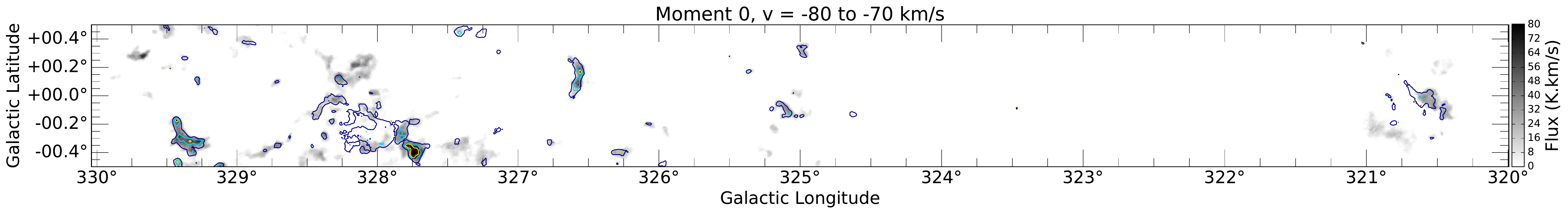}
\includegraphics[width=\textwidth]{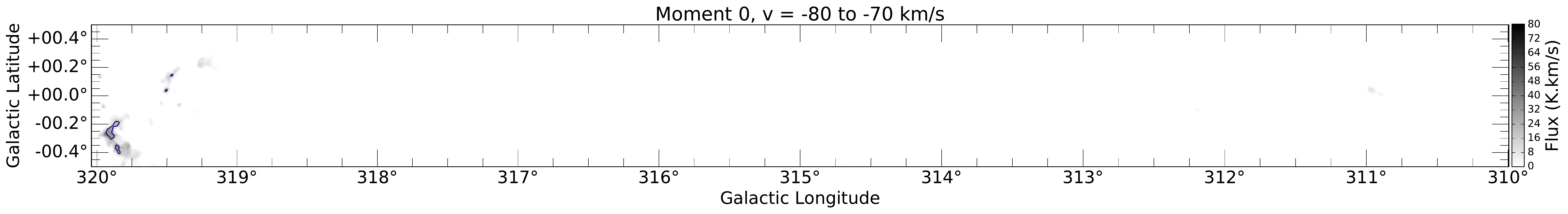}
	\caption{Moment 0 images for $l = 310$--$350^\circ$ calculated over the velocity interval $v = -80$ to $-70$ \kms. The grayscale $^{12}$CO image runs from 0 to 80 K.km/s, while the $^{13}$CO contours are at 4, 12, 20, 28 K.km/s. None of the other data cubes demonstrate significant emission in this velocity range.}\label{Mom-80}
\end{center}
\end{figure*}

\begin{figure*}[htp]
\begin{center}
\includegraphics[width=\textwidth]{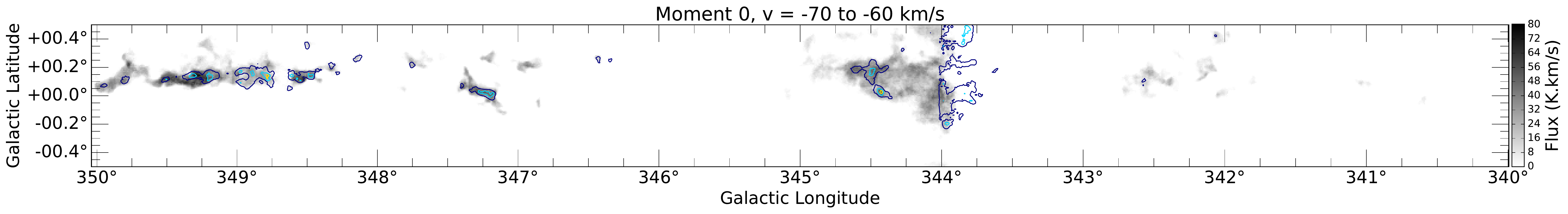}
\includegraphics[width=\textwidth]{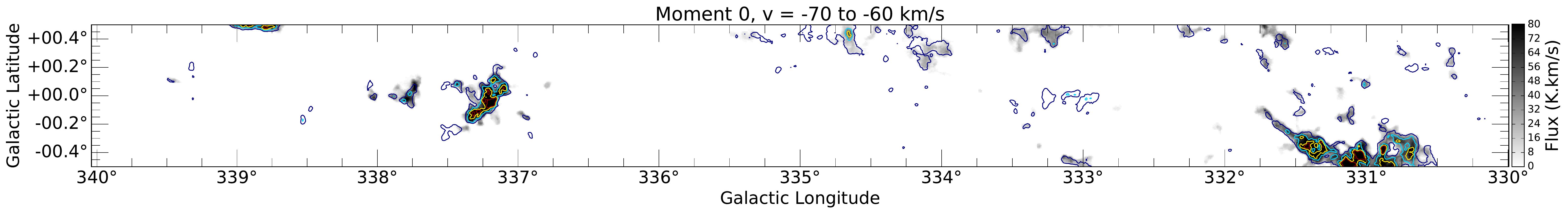}
\includegraphics[width=\textwidth]{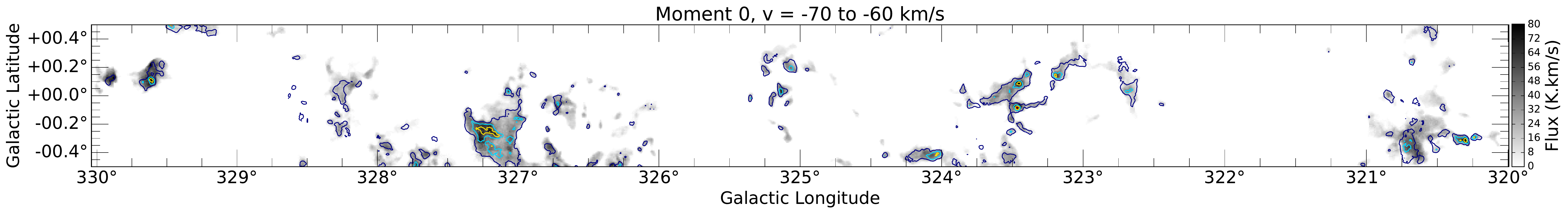}
\includegraphics[width=\textwidth]{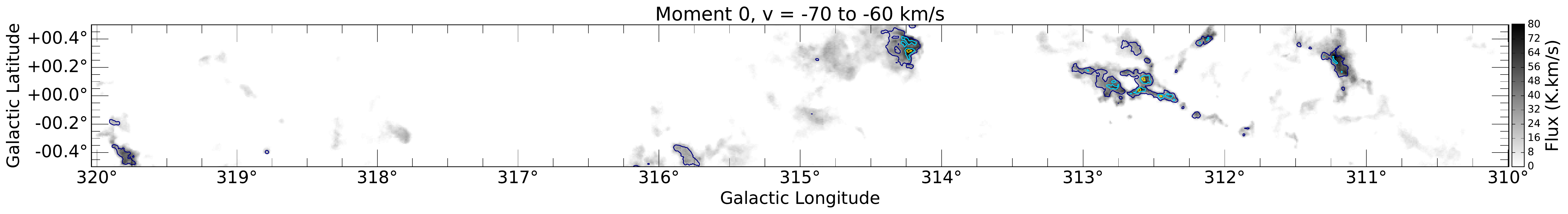}
\includegraphics[width=\textwidth]{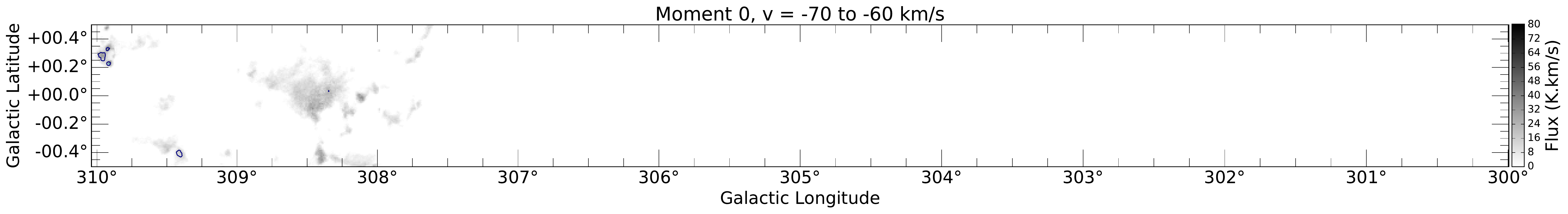}
	\caption{Moment 0 images for $l = 300$--$350^\circ$ calculated over the velocity interval $v = -70$ to $-60$ \kms. The grayscale $^{12}$CO image runs from 0 to 80 K.km/s, while the $^{13}$CO contours are at 4, 12, 20, 28 K.km/s.}\label{Mom-70}
\end{center}
\end{figure*}

\begin{figure*}[htp]
\begin{center}
\includegraphics[width=\textwidth]{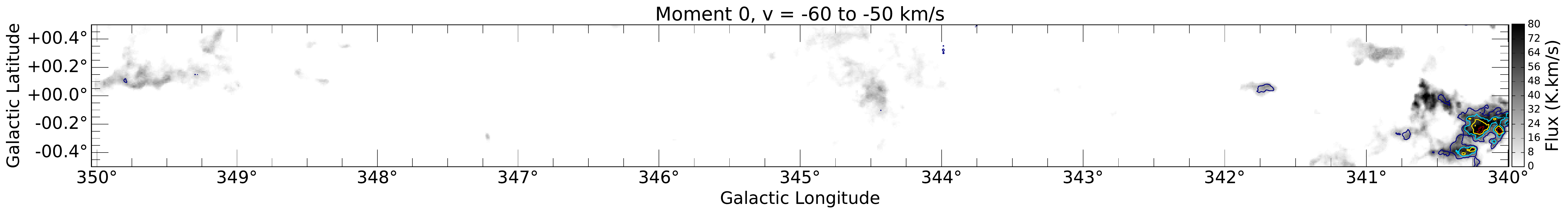}
\includegraphics[width=\textwidth]{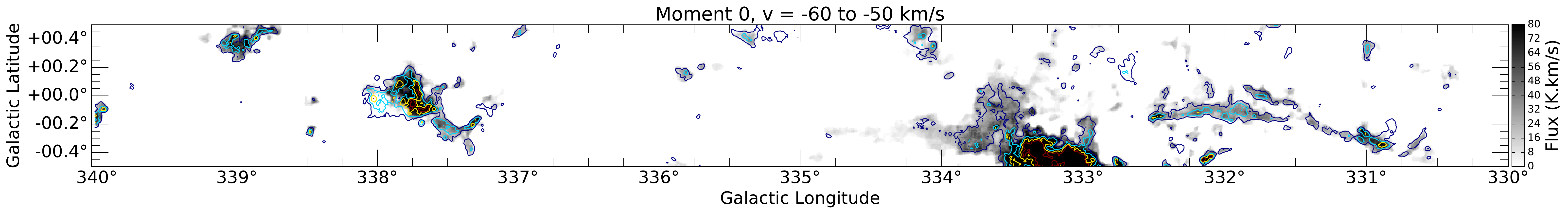}
\includegraphics[width=\textwidth]{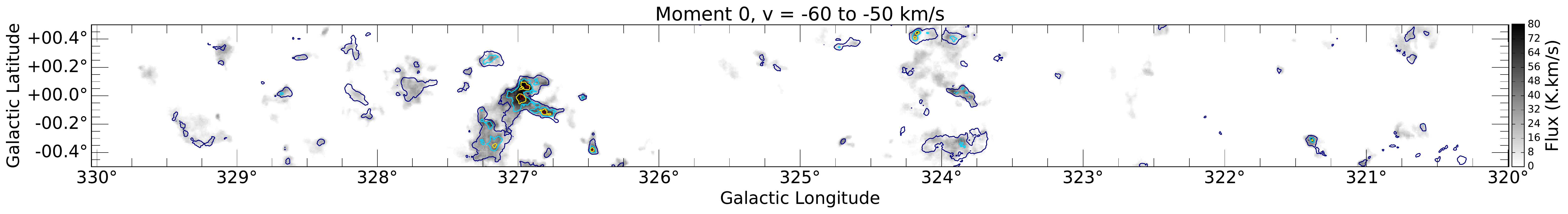}
\includegraphics[width=\textwidth]{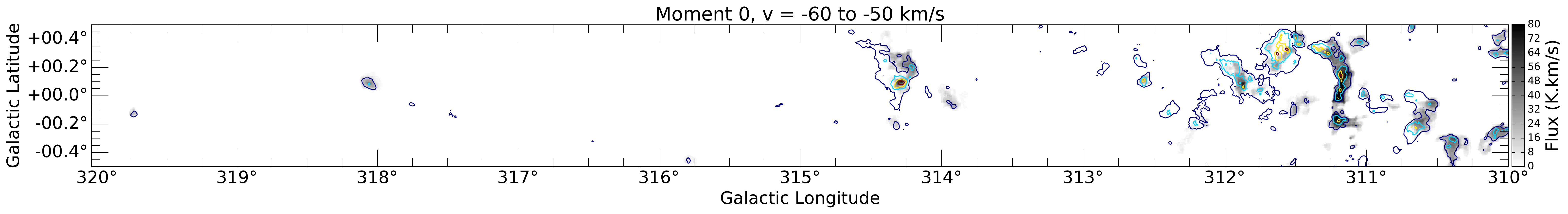}
\includegraphics[width=\textwidth]{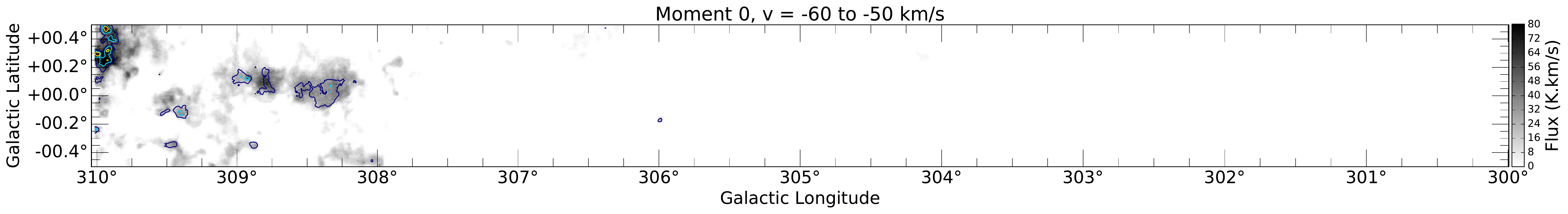}
	\caption{Moment 0 images for $l = 300$--$350^\circ$ calculated over the velocity interval $v = -60$ to $-50$ \kms. The grayscale $^{12}$CO image runs from 0 to 80 K.km/s, while the $^{13}$CO contours are at 4, 12, 20, 28 K.km/s.}\label{Mom-60}
\end{center}
\end{figure*}

\begin{figure*}[htp]
\begin{center}
\includegraphics[width=\textwidth]{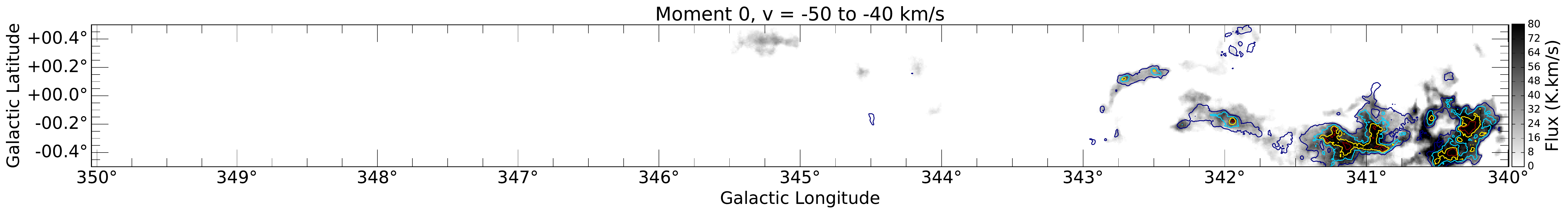}
\includegraphics[width=\textwidth]{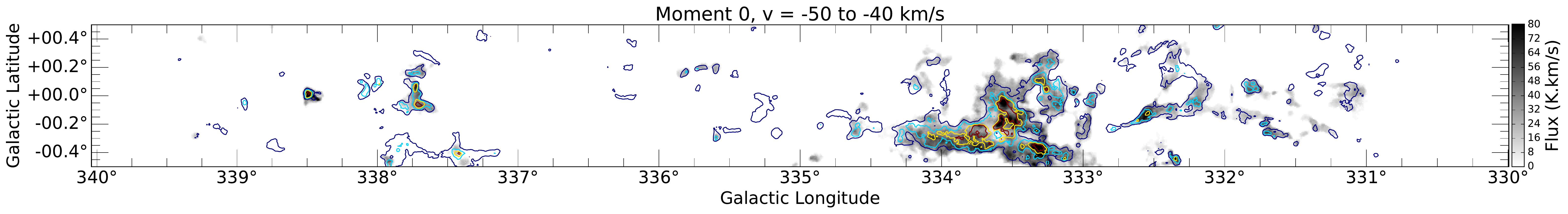}
\includegraphics[width=\textwidth]{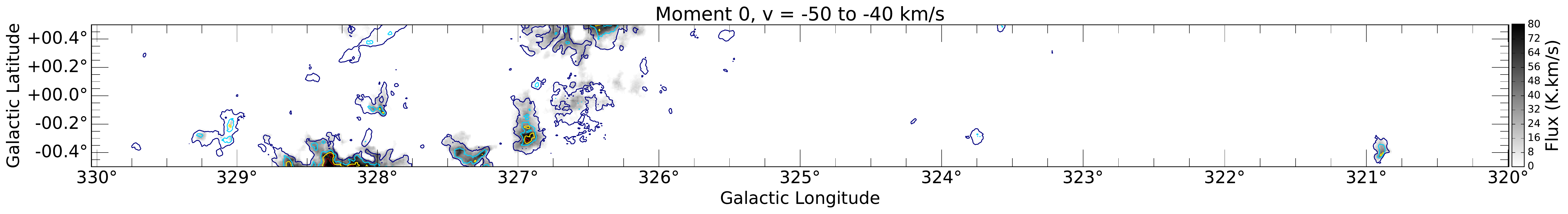}
\includegraphics[width=\textwidth]{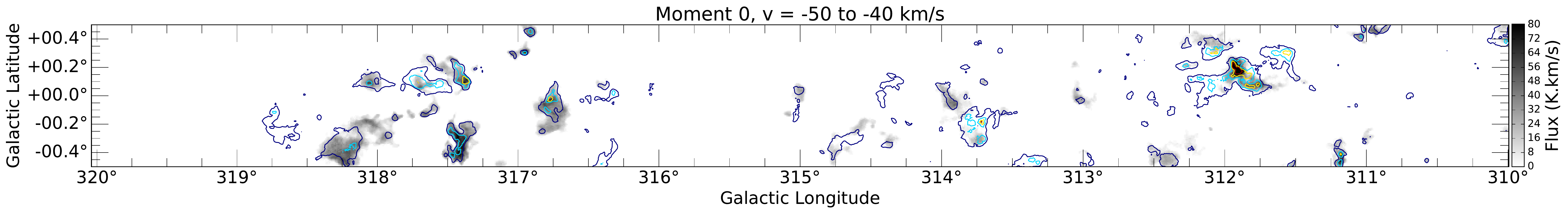}
\includegraphics[width=\textwidth]{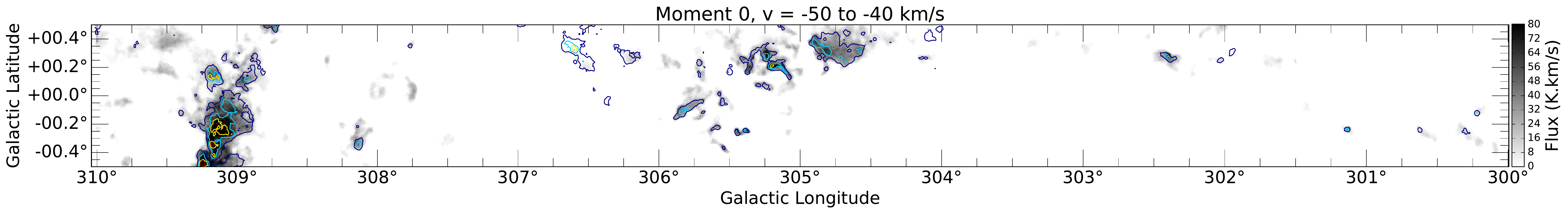}
	\caption{Moment 0 images for $l = 300$--$350^\circ$ calculated over the velocity interval $v = -50$ to $-40$ \kms. The grayscale $^{12}$CO image runs from 0 to 80 K.km/s, while the $^{13}$CO contours are at 4, 12, 20, 28 K.km/s.}\label{Mom-50}
\end{center}
\end{figure*}

\begin{figure*}[htp]
\begin{center}
\includegraphics[width=\textwidth]{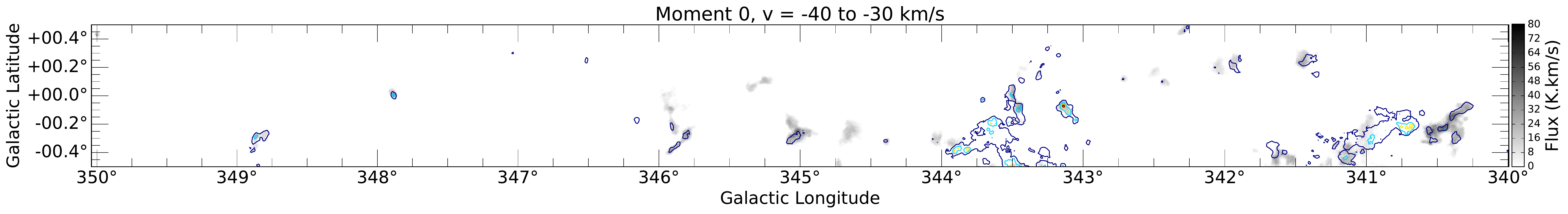}
\includegraphics[width=\textwidth]{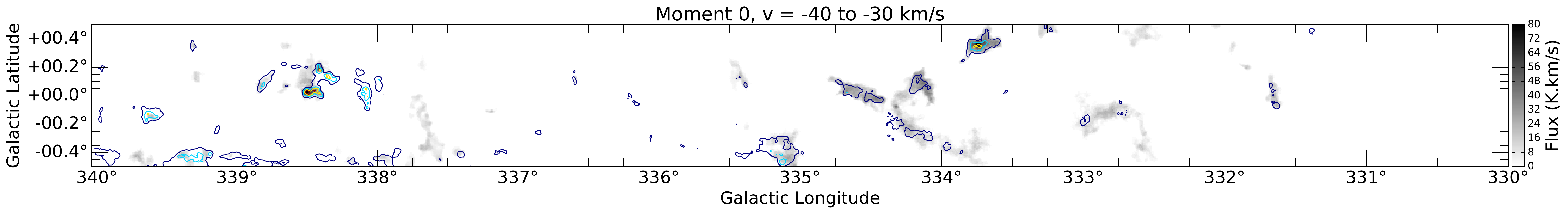}
\includegraphics[width=\textwidth]{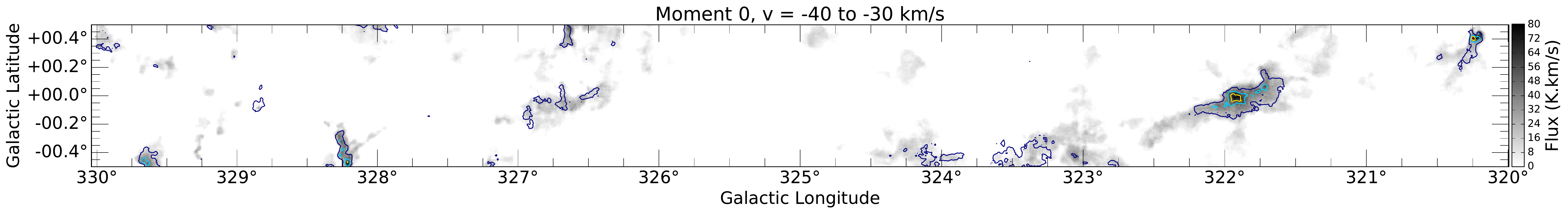}
\includegraphics[width=\textwidth]{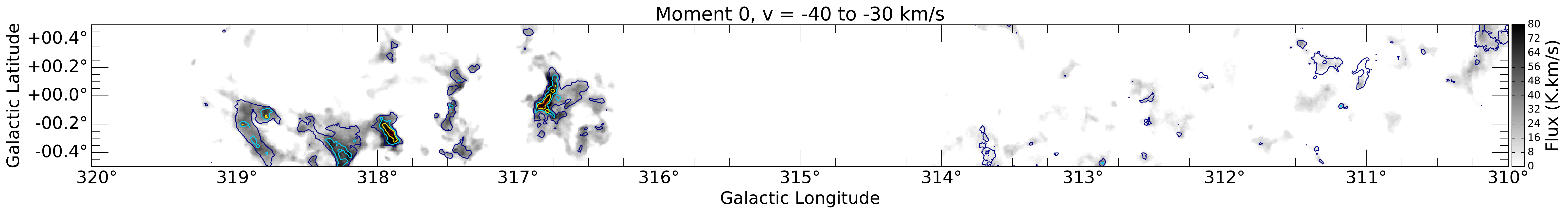}
\includegraphics[width=\textwidth]{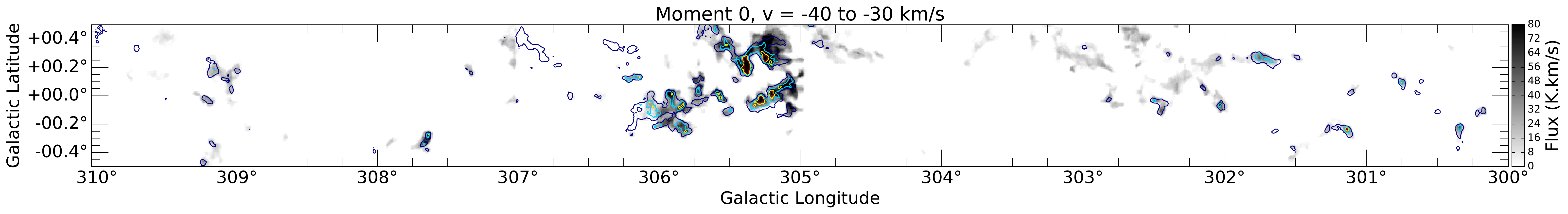}
	\caption{Moment 0 images for $l = 300$--$350^\circ$ calculated over the velocity interval $v = -40$ to $-30$ \kms. The grayscale $^{12}$CO image runs from 0 to 80 K.km/s, while the $^{13}$CO contours are at 4, 12, 20, 28 K.km/s.}\label{Mom-40}
\end{center}
\end{figure*}

\begin{figure*}[htp]
\begin{center}
\includegraphics[width=\textwidth]{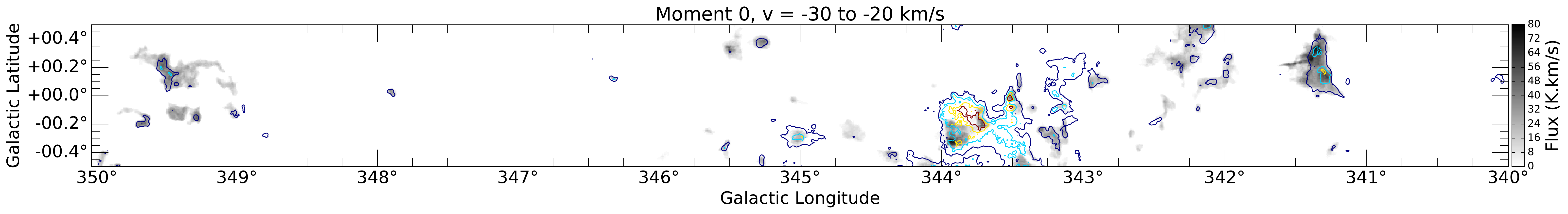}
\includegraphics[width=\textwidth]{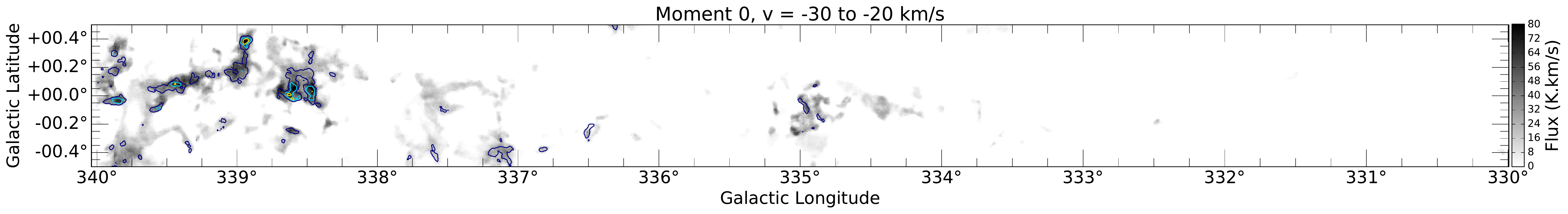}
\includegraphics[width=\textwidth]{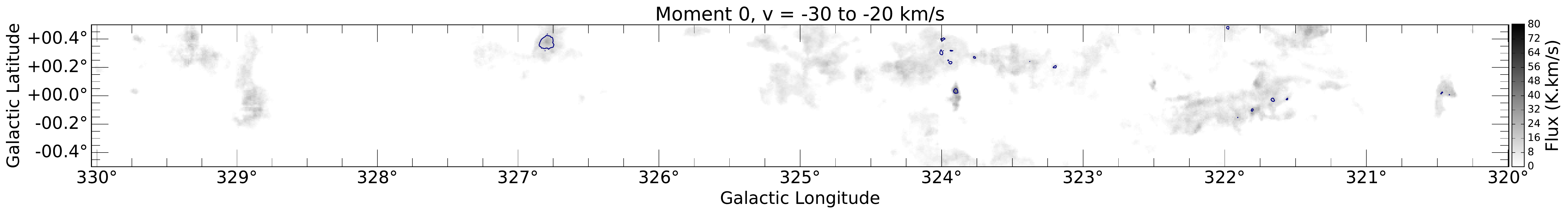}
\includegraphics[width=\textwidth]{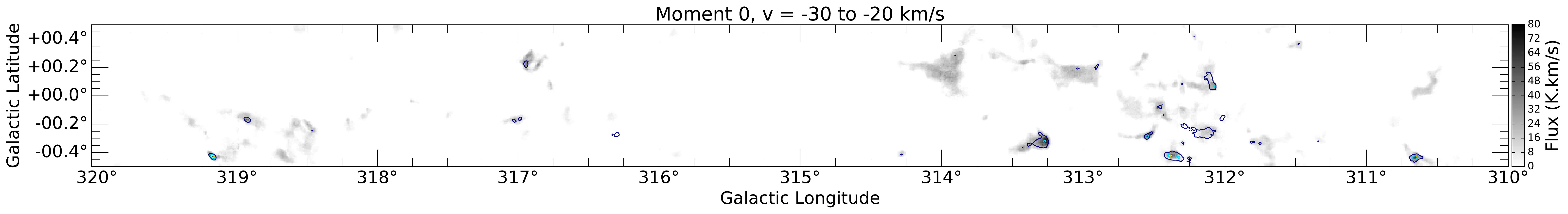}
\includegraphics[width=\textwidth]{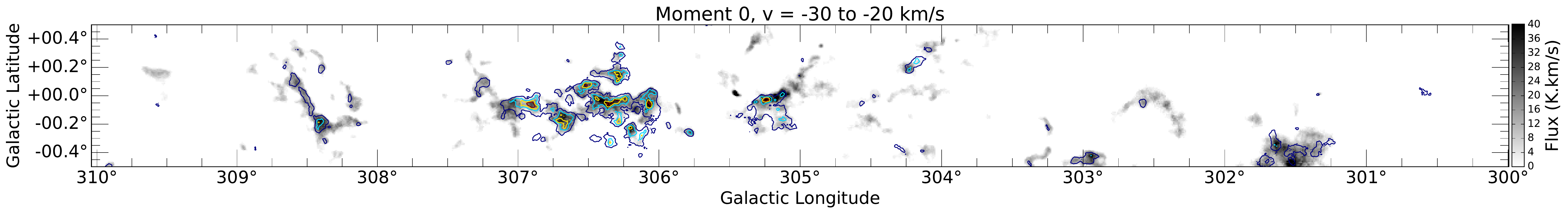}
	\caption{Moment 0 images for $l = 300$--$350^\circ$ calculated over the velocity interval $v = -30$ to $-20$ \kms. The grayscale $^{12}$CO image runs from 0 to 80 K.km/s, while the $^{13}$CO contours are at 4, 12, 20, 28 K.km/s.}\label{Mom-30}
\end{center}
\end{figure*}

\begin{figure*}[htp]
\begin{center}
\includegraphics[width=\textwidth]{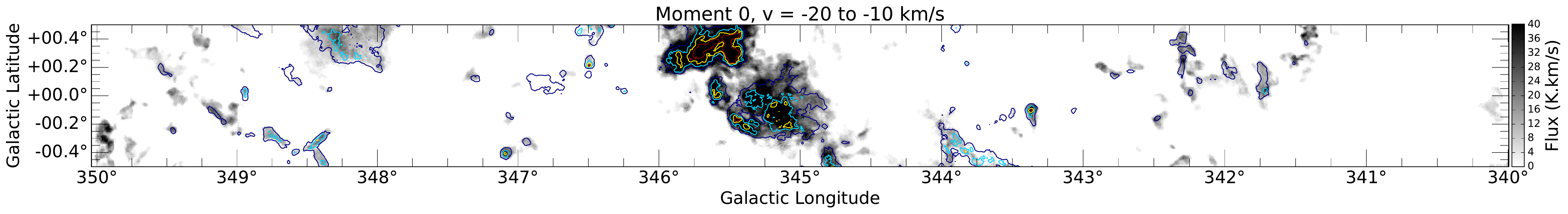}
\includegraphics[width=\textwidth]{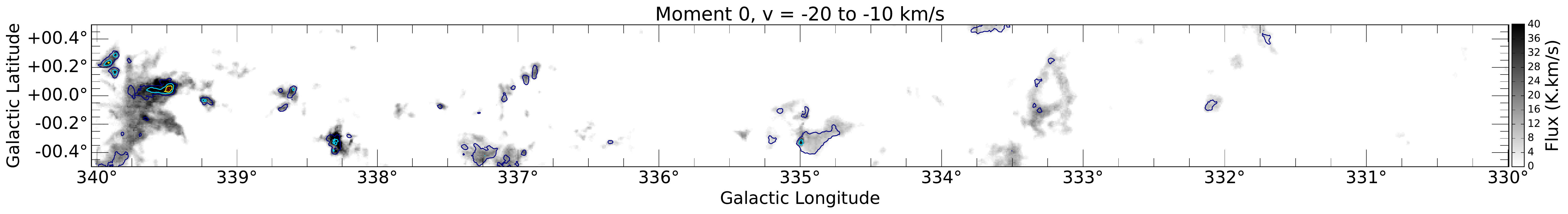}
\includegraphics[width=\textwidth]{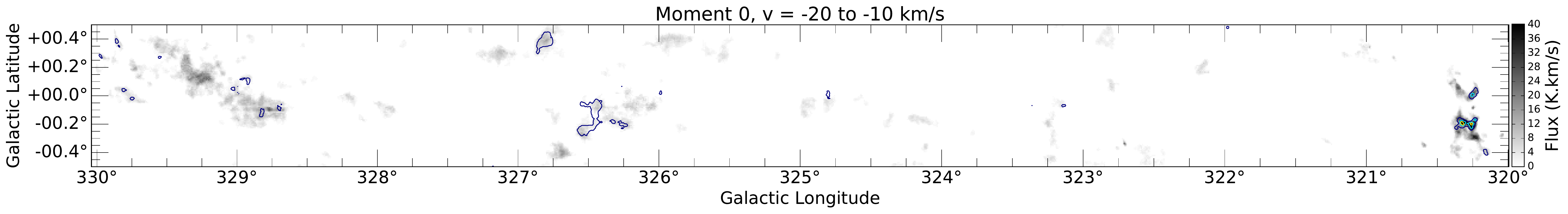}
\includegraphics[width=\textwidth]{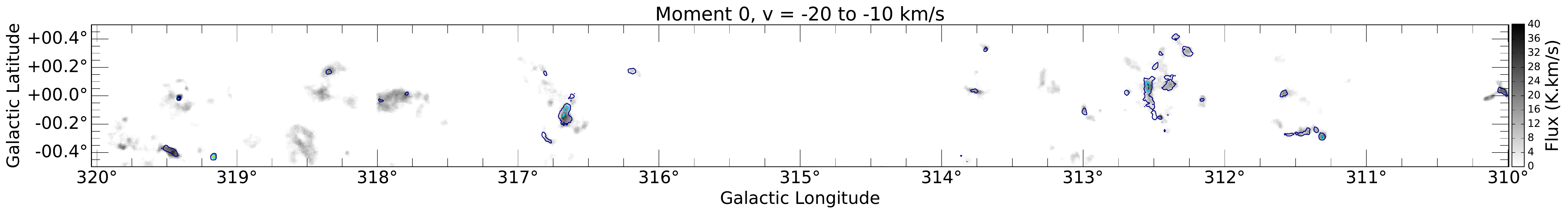}
\includegraphics[width=\textwidth]{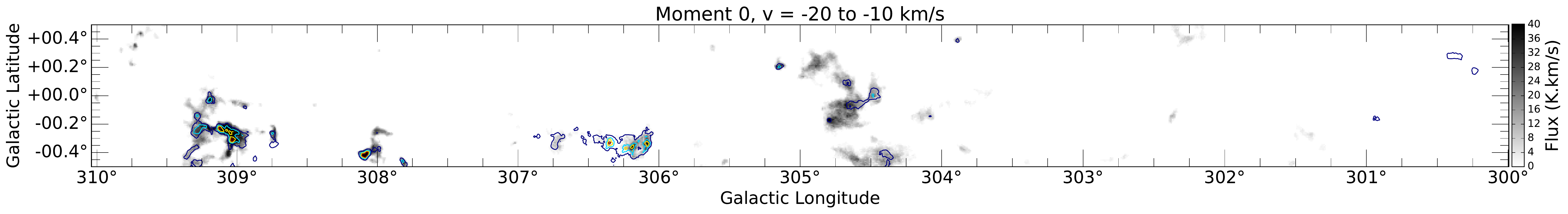}
	\caption{Moment 0 images for $l = 300$--$350^\circ$ calculated over the velocity interval $v = -20$ to $-10$ \kms. The scales have changed again, the grayscale $^{12}$CO image runs from 0 to 40 K.km/s, while the $^{13}$CO contours are at 1, 6, 11, 16 K.km/s.}\label{Mom-20}
\end{center}
\end{figure*}

\begin{figure*}[htp]
\begin{center}
\includegraphics[width=\textwidth]{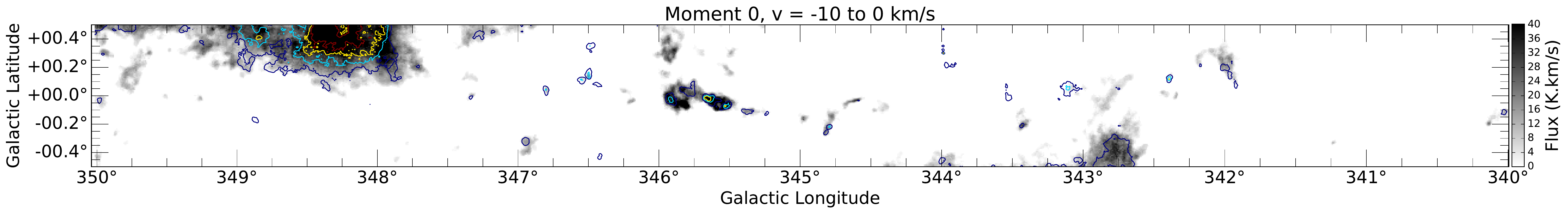}
\includegraphics[width=\textwidth]{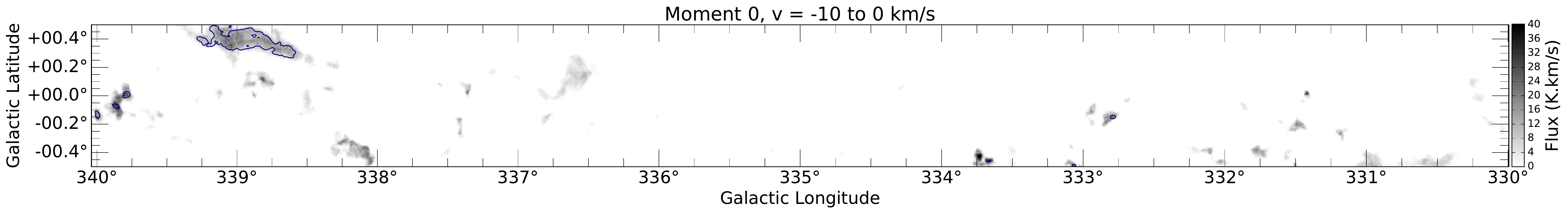}
\includegraphics[width=\textwidth]{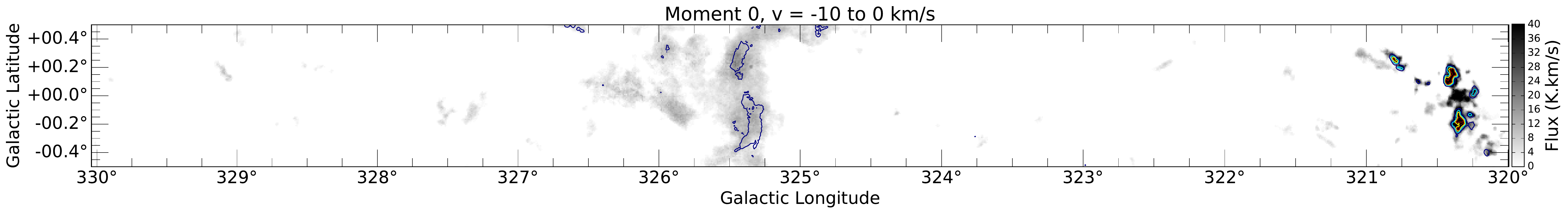}
\includegraphics[width=\textwidth]{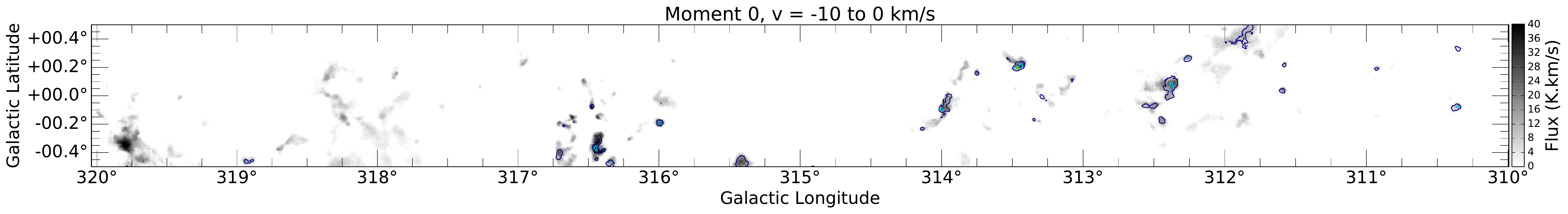}
\includegraphics[width=\textwidth]{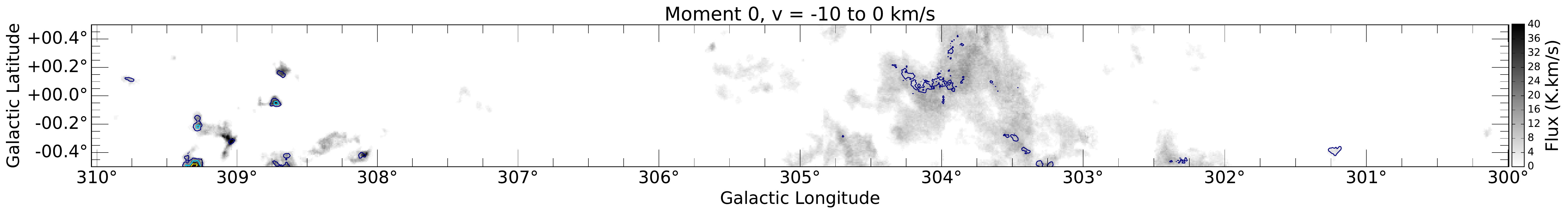}
	\caption{Moment 0 images for $l = 300$--$350^\circ$ calculated over the velocity interval $v = -10$ to $0$ \kms. The grayscale $^{12}$CO image runs from 0 to 40 K.km/s, while the $^{13}$CO contours are at 1, 6, 11, 16 K.km/s.}\label{Mom-10}
\end{center}
\end{figure*}

\begin{figure*}[htp]
\begin{center}
\includegraphics[width=\textwidth]{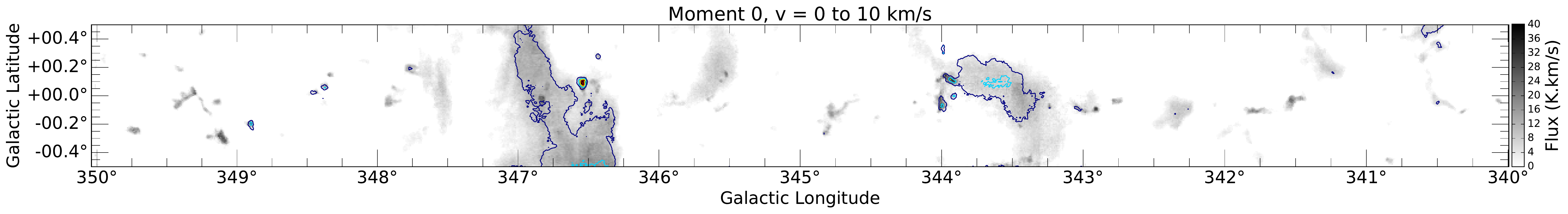}
\includegraphics[width=\textwidth]{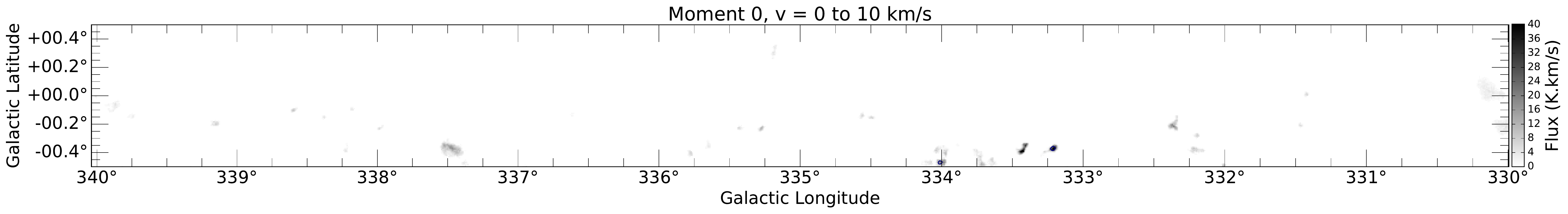}
\includegraphics[width=\textwidth]{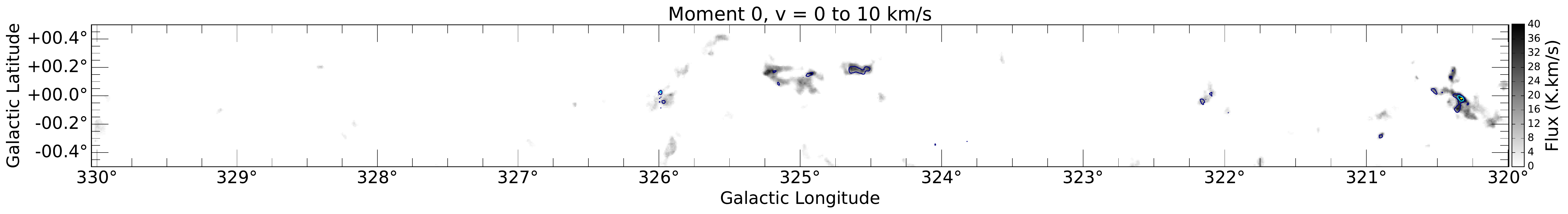}
\includegraphics[width=\textwidth]{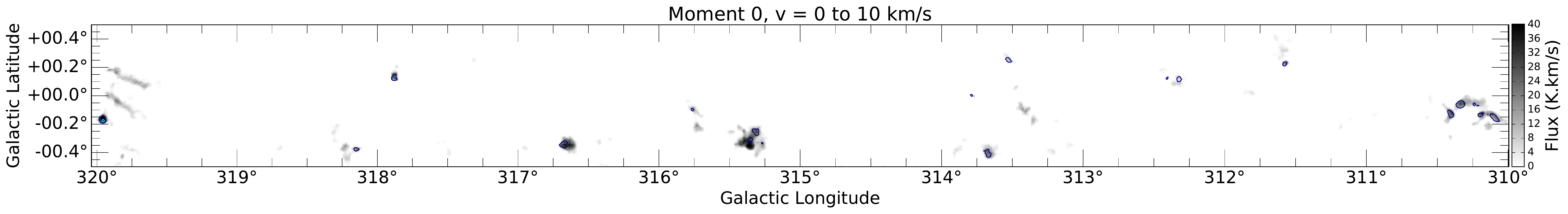}
\includegraphics[width=\textwidth]{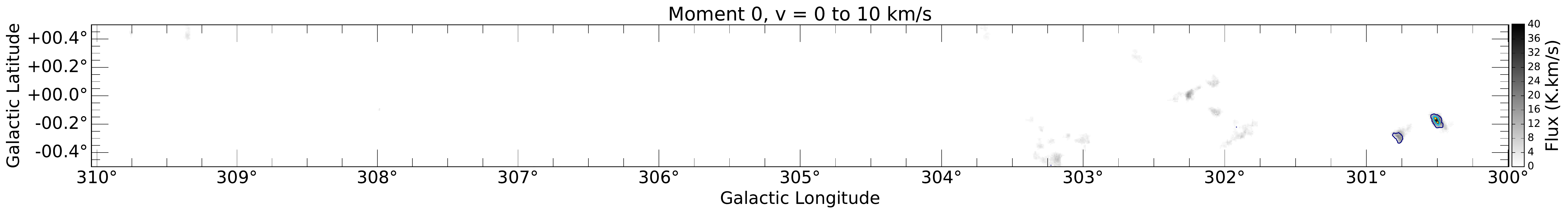}
	\caption{Moment 0 images for $l = 300$--$350^\circ$ calculated over the velocity interval $v = 0$ to $+10$ \kms. The grayscale $^{12}$CO image runs from 0 to 40 K.km/s, while the $^{13}$CO contours are at 1, 6, 11, 16 K.km/s.}\label{Mom+0}
\end{center}
\end{figure*}

\begin{figure*}[htp]
\begin{center}
\includegraphics[width=\textwidth]{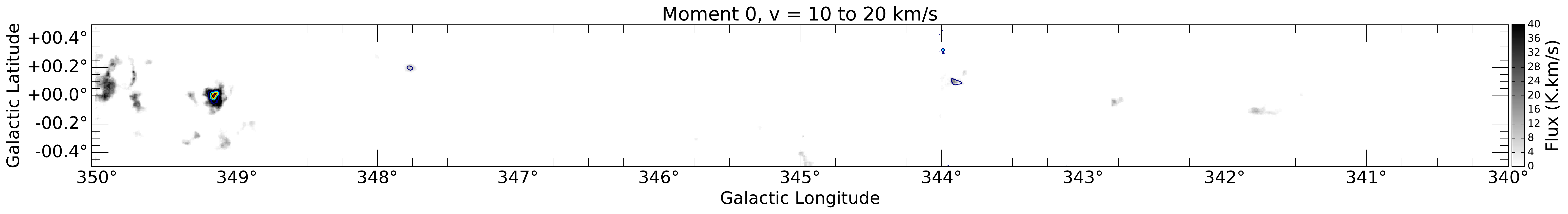}
\includegraphics[width=\textwidth]{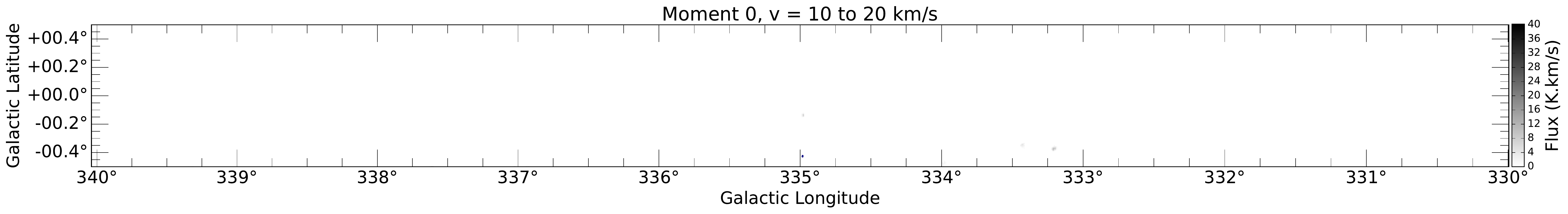}
\includegraphics[width=\textwidth]{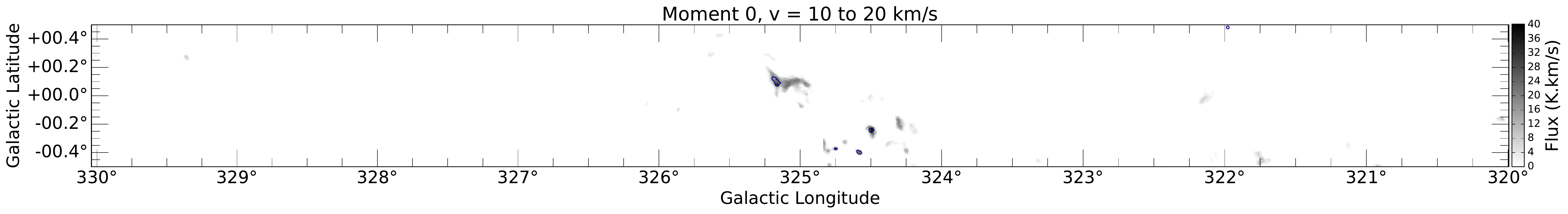}
\includegraphics[width=\textwidth]{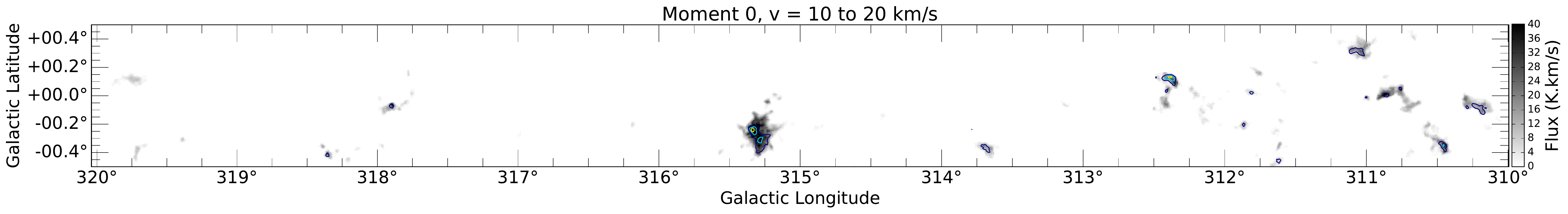}
\includegraphics[width=\textwidth]{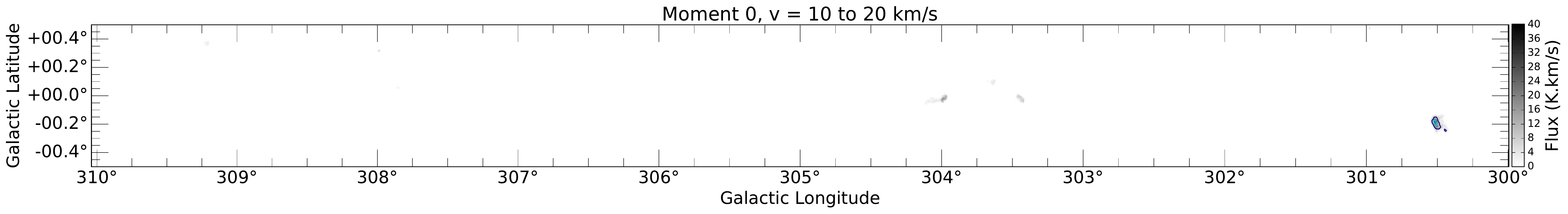}
	\caption{Moment 0 images for $l = 300$--$350^\circ$ calculated over the velocity interval $v = +10$ to $+20$\kms. The grayscale $^{12}$CO image runs from 0 to 40 K.km/s, while the $^{13}$CO contours are at 1, 6, 11, 16 K.km/s.}\label{Mom+10}
\end{center}
\end{figure*}

\begin{figure*}
\begin{center}
\includegraphics[width=\textwidth]{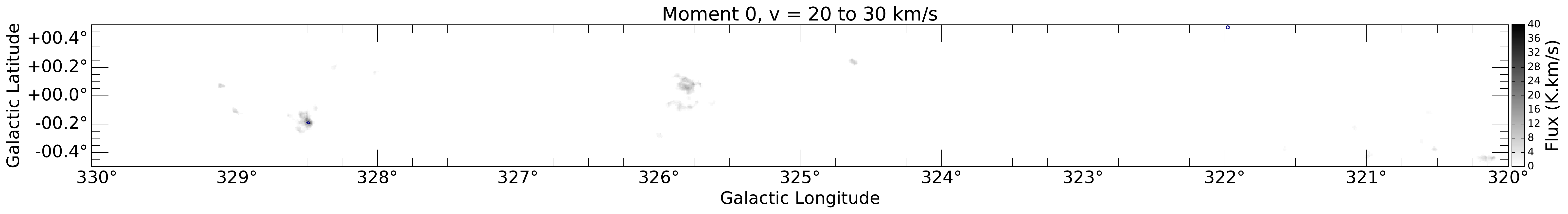}
\includegraphics[width=\textwidth]{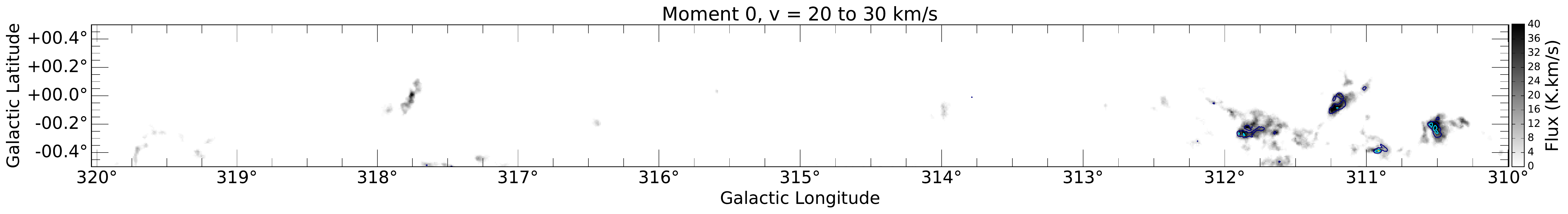}
\includegraphics[width=\textwidth]{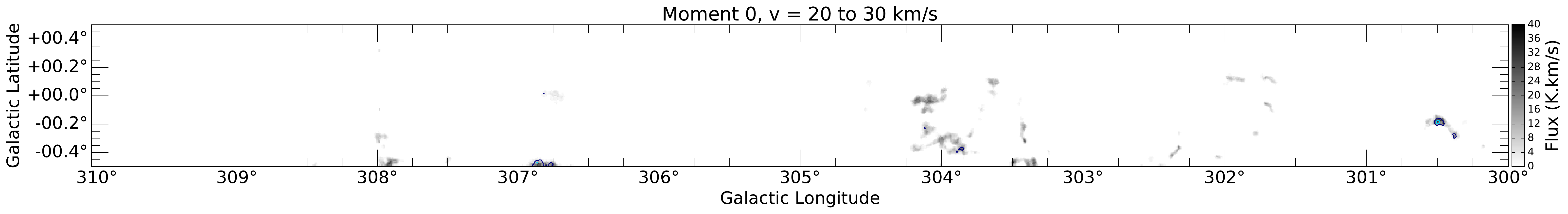}
	\caption{Moment 0 images for $l = 300$--$330^\circ$ calculated over the velocity interval $v = +20$ to $+30$\kms. The grayscale $^{12}$CO image runs from 0 to 40 K.km/s, while the $^{13}$CO contours are at 1, 6, 11, 16 K.km/s. None of the other data cubes demonstrate significant emission in this velocity range.}\label{Mom+20}
\end{center}
\end{figure*}

\begin{figure*}
\begin{center}
\includegraphics[width=\textwidth]{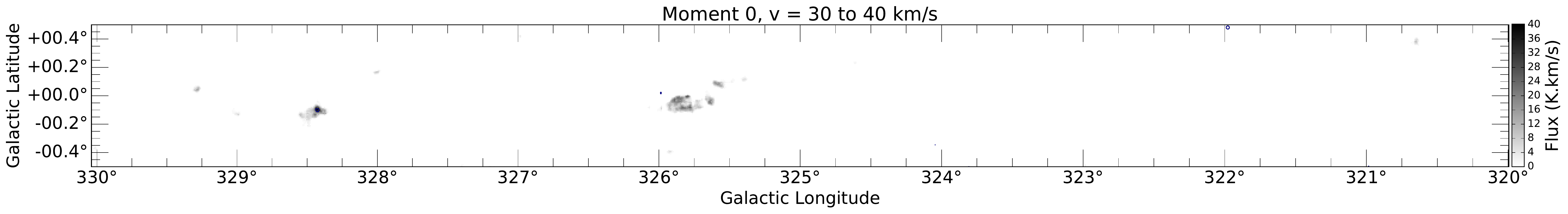}
\includegraphics[width=\textwidth]{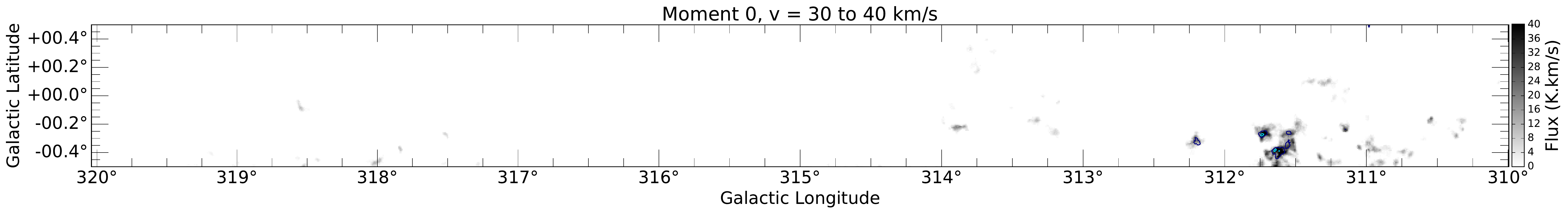}
\includegraphics[width=\textwidth]{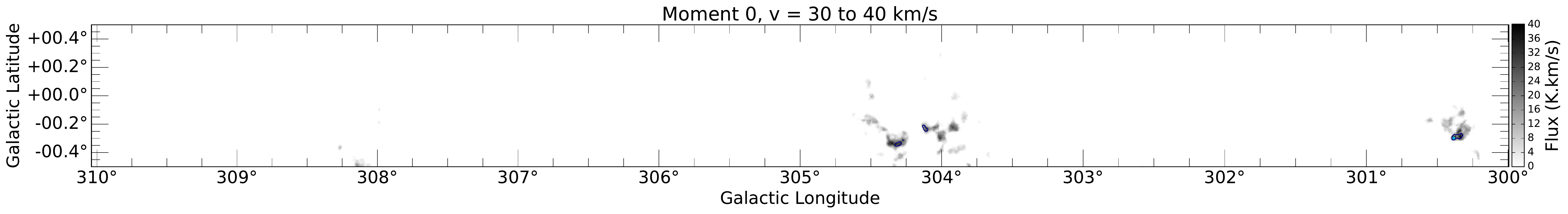}
	\caption{Moment 0 images for $l = 300$--$330^\circ$ calculated over the velocity interval $v = +30$ to $+40$\kms. The grayscale $^{12}$CO image runs from 0 to 40 K.km/s, while the $^{13}$CO contours are at 1, 6, 11, 16 K.km/s. None of the other data cubes demonstrate significant emission in this velocity range.}\label{Mom+30}
\end{center}
\end{figure*}

\begin{figure*}
\begin{center}
\includegraphics[width=\textwidth]{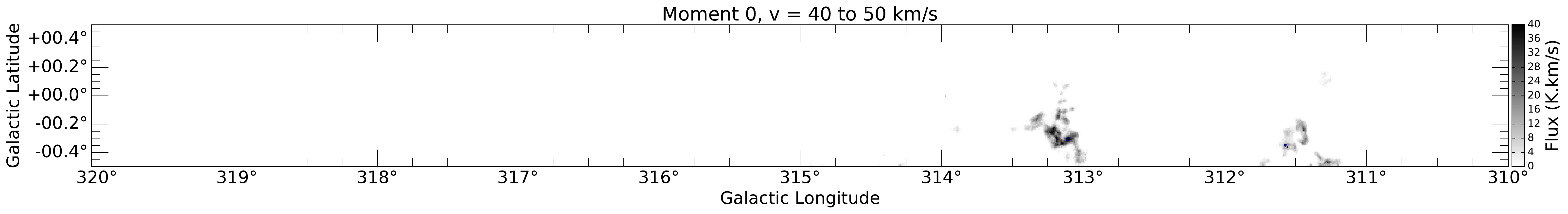}
	\caption{Moment 0 images for $l = 310$--$320^\circ$ calculated over the velocity interval $v = +40$ to $+50$\kms. The grayscale $^{12}$CO image runs from 0 to 40 K.km/s, while the $^{13}$CO contours are at 1, 6, 11, 16\kms. None of the other data cubes demonstrate significant emission in this velocity range. }\label{Mom+40}
\end{center}
\end{figure*}

Figs.~\ref{Mom-150} to \ref{Mom+40} are integrated intensity maps of T$_\textrm{MB}\Delta{V}$ from DR3. Grey scale images are taken from the $^{12}$CO data and the contours are from the $^{13}$CO data, between the velocity range of -150 and +50\,km\,s$^{-1}$  with an interval of 10\,km\,s$^{-1}$. Note that some scanning artefacts are present in these images, evident as $1\degree$-long line features in longitude or latitude, particularly near the edges of individual square degrees. 

\subsection{C$^{18}$O clump catalogue}\label{app:c18o}
Table \ref{tab-c18o} presents a sample of \cootto molecular clumps extracted from the Galactic longitude range $330 \le l \le 340$\degree data cubes as described in \S\ref{ss:c18o}. These are the first items from a wider database that is in preparation and will describe all the C$^{18}$O clumps comprised in the Mopra CO survey data ($300 \le l \le 350^{\circ}$ and beyond). 

Two methods are used to compute the cloud areas: In the first (labelled A in Table \ref{tab-c18o}), we define the clump area as the number of non-zero pixels in the integrated intensity map multiplied by the area of the single pixel in square arcseconds. The second (labelled A$_\textrm{fwhm}$) is the area of an ellipse with semi-axes defined by the FWHM angular extensions $\Delta{l}$ and $\Delta{b}$.

Note that for nine clumps we used the tangent point distance for both near and far distances, since their coordinates have no solution in the Galactic rotation curve; as a consequence their near and far masses are the same.



\clearpage
\clearpage
\clearpage
\clearpage
\clearpage

%% file: table_c18o_clumps_longtable.tex
\onecolumn
\begin{landscape}
\begin{longtable}{lrrrrrrrrrrrrrrrrr}
	\caption{C$^{18}$O clump statistics within the $330 \le l \le 340^{\circ}$ block (see Section\,\ref{ss:c18o}). For each clump we provide a Clump ID based upon its Galactic coordinates $l$, $b$ and $V_\textrm{LSR}$, while $\Delta l$, $\Delta b$ and $\Delta V$ indicate the clump FWHM extensions along the Galactic longitude, latitude, and velocity axes respectively. The clump radius, $R$, is reported in arcsec; the area A is that covered by the clump in the moment zero map; and A$_{fwhm}$ is the area of an ellipse with semi-axes $\Delta l$ and $\Delta b$. The near and far distances to the cloud centroid are listed in kpc as  D$^{N}$ and D$^{F}$ respectively, calculated using the model in \S\ref{ss:mass}. The mean and peak clump brightness temperatures are given by $\overline{T_\textrm{MB}^{18}}$ and $\widehat{T_\textrm{MB}^{18}}$ respectively; the mean and peak column densities, $\overline{N_{\textrm{H}_2}}$ and $\widehat{N_{\textrm{H}_2}}$, are calculated as described in \S\ref{ss:c18o}. Finally we report the mass of each clump, calculated at both the near (M$_{18}^{N}$) and far (M$_{18}^{F}$) distances. \label{tab-c18o}}\\
    \hline\hline
	Clump ID	 	&	\multicolumn{2}{c}{Coordinates} 			&	V$_\textrm{LSR}$	&	$\Delta l$	&	$\Delta b$	&	$\Delta V$	&	R	&	A 	& A$_\textrm{fwhm}$ &	D$^{N}$	& D$^{F}$	&	$\overline{T_\textrm{MB}^{18}}$	&	$\widehat{T_\textrm{MB}^{18}}$ & $\overline{N_{\textrm{H}_2}}$	&	$\widehat{N_{\textrm{H}_2}}$	&	M$_{18}^{N}$ &	M$_{18}^{F}$ \\
	&		$l$ deg.	&	$b$ deg.	&	 km/s	&	\multicolumn{2}{c}{arcsec}	&	km/s	&	arcsec	&	\multicolumn{2}{c}{10$^3$arcsec$^2$}		&	\multicolumn{2}{c}{kpc}		&	\multicolumn{2}{c}{K}	& \multicolumn{2}{c}{10$^{21}$\,cm$^{-2}$}	&	\multicolumn{2}{c}{10$^2\,$M$_{\odot}$}	\\		\hline\endfirsthead
		\multicolumn{11}{l}{{\bfseries Table \thetable.} Continued from previous page.}\\\hline
	Clump ID	 	&	\multicolumn{2}{c}{Coordinates} 			&	V$_\textrm{LSR}$	&	$\Delta l$	&	$\Delta b$	&	$\Delta V$	&	R	&	A 	& A$_\textrm{fwhm}$ &	D$^{N}$	& D$^{F}$	&	$\overline{T_\textrm{MB}^{18}}$	&	$\widehat{T_\textrm{MB}^{18}}$ & $\overline{N_{\textrm{H}_2}}$	&	$\widehat{N_{\textrm{H}_2}}$	&	M$_{18}^{N}$ &	M$_{18}^{F}$ \\
	&		$l$ deg.	&	$b$ deg.	&	 km/s	&	\multicolumn{2}{c}{arcsec}	&	km/s	&	arcsec	&	\multicolumn{2}{c}{10$^3$arcsec$^2$}		&	\multicolumn{2}{c}{kpc}		&	\multicolumn{2}{c}{K}	& \multicolumn{2}{c}{10$^{21}$\,cm$^{-2}$}	&	\multicolumn{2}{c}{10$^2\,$M$_{\odot}$}	\\	\hline\endhead
MCO\_330.227-0.019	&			330.227 	&	-0.019 	&	-42.7 	&	120 	&	86 	&	1.0 	&	76 	&	18	&	32 	&	3.3 	&	11.5 	&	0.9 	&	1.1 	&	5.9 	&	7.0 	&	5 	&		62 		\\
MCO\_330.394-0.052	&			330.394 	&	-0.052 	&	-41.8 	&	218 	&	129 	&	1.0 	&	107 	&	36 	&	88 	&	3.2 	&	11.6 	&	0.9 	&	1.1 	&	5.8 	&	7.0 	&	9 	&		123 		\\
MCO\_330.410+0.323	&			330.410 	&	0.323 	&	-97.0 	&	187 	&	76 	&	1.5 	&	91 	&	26 	&	44 	&	6.8 	&	8.0 	&	1.0 	&	1.5 	&	9.2 	&	18.9 	&	49 	&		67 		\\
MCO\_330.685-0.377	&			330.685 	&	-0.377 	&	-67.6 	&	374 	&	280 	&	3.3 	&	169 	&	90 	&	329 	&	4.7 	&	10.1 	&	1.1 	&	1.7 	&	12.1 	&	33.5 	&	104 	&		491 		\\
MCO\_330.685+0.181	&			330.685 	&	0.181 	&	-63.0 	&	115 	&	78 	&	1.0 	&	74 	&	17 	&	28 	&	4.4 	&	10.4 	&	1.0 	&	1.2 	&	9.3 	&	13.3 	&	14 	&		76 		\\
MCO\_330.710-0.419	&			330.710 	&	-0.419 	&	-65.7 	&	209 	&	213 	&	2.3 	&	160 	&	81 	&	140 	&	4.6 	&	10.3 	&	1.1 	&	2.0 	&	15.4 	&	38.5 	&	115 	&		577 		\\
MCO\_330.802+0.231	&			330.802 	&	0.231 	&	-45.5 	&	343 	&	229 	&	1.3 	&	170 	&	91 	&	246 	&	3.4 	&	11.4 	&	1.0 	&	1.5 	&	7.6 	&	16.1 	&	36 	&		395 		\\
MCO\_330.827-0.510	&			330.827 	&	-0.510 	&	-64.8 	&	152 	&	91 	&	1.6 	&	106 	&	35 	&	43 	&	4.5 	&	10.3 	&	1.3 	&	2.3 	&	15.8 	&	33.3 	&	50 	&		259 		\\
\hline																																							
\end{longtable}																																									
\end{landscape}

%% file: mopra_co_iii.bbl
\begin{thebibliography}{}
\makeatletter
\relax
\def\mn@urlcharsother{\let\do\@makeother \do\$\do\&\do\#\do\^\do\_\do\%\do\~}
\definecolor{darkblue}{rgb}{0,0,0.597656}
\def\mndoi{\begingroup\mn@urlcharsother \@ifnextchar [ {\mndoi@} {\mndoi@[]}}
\def\mndoi@[#1]#2{\def\@tempa{#1}\ifx\@tempa\@empty \href
  {http://dx.doi.org/#2} {\textcolor{darkblue}{doi:#2}}\else \href
  {http://dx.doi.org/#2} {\textcolor{darkblue}{#1}}\fi \endgroup}
\def\mn@eprint#1#2{\mn@eprint@#1:#2::\@nil}
\def\mn@eprint@arXiv#1{\href {http://arxiv.org/abs/#1} {{\tt arXiv:#1}}}
\def\mn@eprint@dblp#1{\href {http://dblp.uni-trier.de/rec/bibtex/#1.xml}
  {dblp:#1}}
\def\mn@eprint@#1:#2:#3:#4\@nil{\def\@tempa {#1}\def\@tempb {#2}\def\@tempc
  {#3}\ifx \@tempc \@empty \let \@tempc \@tempb \let \@tempb \@tempa \fi \ifx
  \@tempb \@empty \def\@tempb {arXiv}\fi \@ifundefined
  {mn@eprint@\@tempb}{\@tempb:\@tempc}{\expandafter \expandafter \csname
  mn@eprint@\@tempb\endcsname \expandafter{\@tempc}}}

\bibitem[\protect\citeauthoryear{{Acero} et~al.,}{{Acero}
  et~al.}{2013}]{Acero:2013}
{Acero} F.,  et~al., 2013, \mndoi [Astroparticle Physics]
  {10.1016/j.astropartphys.2012.05.024}, \href
  {http://adsabs.harvard.edu/abs/2013APh....43..276A} {43, 276}

\bibitem[\protect\citeauthoryear{{Acero} et~al.,}{{Acero}
  et~al.}{2017}]{Acero:2017}
{Acero} F.,  et~al., 2017, \mndoi [\apj] {10.3847/1538-4357/aa6d67}, \href
  {http://adsabs.harvard.edu/abs/2017ApJ...840...74A} {840, 74}

\bibitem[\protect\citeauthoryear{{Acharya} et~al.,}{{Acharya}
  et~al.}{2013}]{Acharya:2013}
{Acharya} B.~S.,  et~al., 2013, \mndoi [Astroparticle Physics]
  {10.1016/j.astropartphys.2013.01.007}, \href
  {http://adsabs.harvard.edu/abs/2013APh....43....3A} {43, 3}

\bibitem[\protect\citeauthoryear{{Aharonian} et~al.,}{{Aharonian}
  et~al.}{2006}]{Aharonian:2006}
{Aharonian} F.,  et~al., 2006, \mndoi [\aap] {10.1051/0004-6361:20065351},
  \href {http://adsabs.harvard.edu/abs/2006A%26A...457..899A} {457, 899}

\bibitem[\protect\citeauthoryear{{Ao} et~al.,}{{Ao} et~al.}{2013}]{Ao:2013}
{Ao} Y.,  et~al., 2013, \mndoi [\aap] {10.1051/0004-6361/201220096}, \href
  {http://adsabs.harvard.edu/abs/2013A%26A...550A.135A} {550, A135}

\bibitem[\protect\citeauthoryear{{Arikawa}, {Tatematsu}, {Sekimoto}  \&
  {Takahashi}}{{Arikawa} et~al.}{1999}]{Arikawa:1999}
{Arikawa} Y.,  {Tatematsu} K.,  {Sekimoto} Y.,   {Takahashi} T.,  1999, \mndoi
  [\pasj] {10.1093/pasj/51.4.L7}, \href
  {http://adsabs.harvard.edu/abs/1999PASJ...51L...7A} {51, L7}

\bibitem[\protect\citeauthoryear{{Astropy Collaboration} et~al.,}{{Astropy
  Collaboration} et~al.}{2013}]{astropy:2013}
{Astropy Collaboration} et~al., 2013, \mndoi [\aap]
  {10.1051/0004-6361/201322068}, \href
  {http://adsabs.harvard.edu/abs/2013A%26A...558A..33A} {558, A33}

\bibitem[\protect\citeauthoryear{{Atwood} et~al.,}{{Atwood}
  et~al.}{2009}]{Atwood:2009}
{Atwood} W.~B.,  et~al., 2009, \mndoi [\apj] {10.1088/0004-637X/697/2/1071},
  \href {http://adsabs.harvard.edu/abs/2009ApJ...697.1071A} {697, 1071}

\bibitem[\protect\citeauthoryear{{Benjamin} et~al.,}{{Benjamin}
  et~al.}{2003}]{Benjamin:2003}
{Benjamin} R.~A.,  et~al., 2003, \mndoi [\pasp] {10.1086/376696}, \href
  {http://adsabs.harvard.edu/abs/2003PASP..115..953B} {115, 953}

\bibitem[\protect\citeauthoryear{{Blackwell} et~al.,}{{Blackwell}
  et~al.}{2018}]{Blackwell:inprep}
{Blackwell} R.,  et~al., 2018, in prep

\bibitem[\protect\citeauthoryear{{Bolatto}, {Wolfire}  \& {Leroy}}{{Bolatto}
  et~al.}{2013}]{Bolatto:2013}
{Bolatto} A.~D.,  {Wolfire} M.,   {Leroy} A.~K.,  2013, \mndoi [\araa]
  {10.1146/annurev-astro-082812-140944}, \href
  {http://adsabs.harvard.edu/abs/2013ARA%26A..51..207B} {51, 207}

\bibitem[\protect\citeauthoryear{{Braiding} et~al.,}{{Braiding}
  et~al.}{2015}]{Braiding:2015}
{Braiding} C.,  et~al., 2015, \mndoi [\pasa] {10.1017/pasa.2015.20}, \href
  {http://adsabs.harvard.edu/abs/2015PASA...32...20B} {32, e020}

\bibitem[\protect\citeauthoryear{{Burton} et~al.,}{{Burton}
  et~al.}{2013}]{Burton:2013}
{Burton} M.~G.,  et~al., 2013, \mndoi [\pasa] {10.1017/pasa.2013.22}, \href
  {http://adsabs.harvard.edu/abs/2013PASA...30...44B} {30, e044}

\bibitem[\protect\citeauthoryear{{Burton} et~al.,}{{Burton}
  et~al.}{2014}]{Burton:2014}
{Burton} M.~G.,  et~al., 2014, \mndoi [\apj] {10.1088/0004-637X/782/2/72},
  \href {http://adsabs.harvard.edu/abs/2014ApJ...782...72B} {782, 72}

\bibitem[\protect\citeauthoryear{{Burton} et~al.,}{{Burton}
  et~al.}{2015}]{Burton:2015}
{Burton} M.~G.,  et~al., 2015, \mndoi [\apj] {10.1088/0004-637X/811/1/13},
  \href {http://adsabs.harvard.edu/abs/2015ApJ...811...13B} {811, 13}

\bibitem[\protect\citeauthoryear{{Cameron}}{{Cameron}}{2012}]{Cameron:2012}
{Cameron} R.~A.,  2012, in Ground-based and Airborne Telescopes IV. p. 844418,
  \mndoi{10.1117/12.926614}

\bibitem[\protect\citeauthoryear{{Carey} et~al.,}{{Carey}
  et~al.}{2009}]{Carey:2009}
{Carey} S.~J.,  et~al., 2009, \mndoi [\pasp] {10.1086/596581}, \href
  {http://adsabs.harvard.edu/abs/2009PASP..121...76C} {121, 76}

\bibitem[\protect\citeauthoryear{{Cherenkov Telescope Array Consortium}
  et~al.,}{{Cherenkov Telescope Array Consortium} et~al.}{2017}]{CTA:2017}
{Cherenkov Telescope Array Consortium} T.,  et~al., 2017, preprint, \href
  {http://adsabs.harvard.edu/abs/2017arXiv170907997C} {} (\mn@eprint {arXiv}
  {1709.07997})

\bibitem[\protect\citeauthoryear{{Churchwell} et~al.,}{{Churchwell}
  et~al.}{2009}]{Churchwell:2009}
{Churchwell} E.,  et~al., 2009, \mndoi [\pasp] {10.1086/597811}, \href
  {http://adsabs.harvard.edu/abs/2009PASP..121..213C} {121, 213}

\bibitem[\protect\citeauthoryear{{Dame} \& {Thaddeus}}{{Dame} \&
  {Thaddeus}}{2011}]{Dame:2011}
{Dame} T.~M.,  {Thaddeus} P.,  2011, \mndoi [\apj] {10.1088/2041-8205}, \href
  {http://adsabs.harvard.edu/abs/2001ApJ...547..792D} {734, L24}

\bibitem[\protect\citeauthoryear{{Dame}, {Hartmann}  \& {Thaddeus}}{{Dame}
  et~al.}{2001}]{Dame:2001}
{Dame} T.~M.,  {Hartmann} D.,   {Thaddeus} P.,  2001, \mndoi [\apj]
  {10.1086/318388}, \href {http://adsabs.harvard.edu/abs/2001ApJ...547..792D}
  {547, 792}

\bibitem[\protect\citeauthoryear{{Dawson} et~al.,}{{Dawson}
  et~al.}{2014}]{Dawson:2014}
{Dawson} J.~R.,  et~al., 2014, \mndoi [\mnras] {10.1093/mnras/stu032}, \href
  {http://adsabs.harvard.edu/abs/2014MNRAS.439.1596D} {439, 1596}

\bibitem[\protect\citeauthoryear{{Dempsey}, {Thomas}  \& {Currie}}{{Dempsey}
  et~al.}{2013}]{Dempsey:2013}
{Dempsey} J.~T.,  {Thomas} H.~S.,   {Currie} M.~J.,  2013, \mndoi [\apjs]
  {10.1088/0067-0049/209/1/8}, \href
  {http://adsabs.harvard.edu/abs/2013ApJS..209....8D} {209, 8}

\bibitem[\protect\citeauthoryear{{Dickey} et~al.,}{{Dickey}
  et~al.}{2013}]{Dickey:2013}
{Dickey} J.~M.,  et~al., 2013, \mndoi [\pasa] {10.1017/pasa.2012.003}, \href
  {http://adsabs.harvard.edu/abs/2013PASA...30....3D} {30, e003}

\bibitem[\protect\citeauthoryear{{Drlica-Wagner}, {G{\'o}mez-Vargas}, {Hewitt},
  {Linden}  \& {Tibaldo}}{{Drlica-Wagner} et~al.}{2014}]{Drlica:2014}
{Drlica-Wagner} A.,  {G{\'o}mez-Vargas} G.~A.,  {Hewitt} J.~W.,  {Linden} T.,
  {Tibaldo} L.,  2014, \mndoi [\apj] {10.1088/0004-637X/790/1/24}, \href
  {http://adsabs.harvard.edu/abs/2014ApJ...790...24D} {790, 24}

\bibitem[\protect\citeauthoryear{{Dubus} et~al.,}{{Dubus}
  et~al.}{2013}]{Dubus:2013}
{Dubus} G.,  et~al., 2013, \mndoi [Astroparticle Physics]
  {10.1016/j.astropartphys.2012.05.020}, \href
  {http://adsabs.harvard.edu/abs/2013APh....43..317D} {43, 317}

\bibitem[\protect\citeauthoryear{{Evoli}, {Gaggero}, {Vittino}, {Di Bernardo},
  {Di Mauro}, {Ligorini}, {Ullio}  \& {Grasso}}{{Evoli}
  et~al.}{2017}]{Evoli:2017}
{Evoli} C.,  {Gaggero} D.,  {Vittino} A.,  {Di Bernardo} G.,  {Di Mauro} M.,
  {Ligorini} A.,  {Ullio} P.,   {Grasso} D.,  2017, \mndoi [\jcap]
  {10.1088/1475-7516/2017/02/015}, \href
  {http://adsabs.harvard.edu/abs/2017JCAP...02..015E} {2, 015}

\bibitem[\protect\citeauthoryear{{Frerking}, {Langer}  \& {Wilson}}{{Frerking}
  et~al.}{1982}]{Frerking:1982}
{Frerking} M.~A.,  {Langer} W.~D.,   {Wilson} R.~W.,  1982, \mndoi [\apj]
  {10.1086/160451}, \href {http://adsabs.harvard.edu/abs/1982ApJ...262..590F}
  {262, 590}

\bibitem[\protect\citeauthoryear{{Fukuda}, {Yoshiike}, {Sano}, {Torii},
  {Yamamoto}, {Acero}  \& {Fukui}}{{Fukuda} et~al.}{2014}]{Fukuda:2014}
{Fukuda} T.,  {Yoshiike} S.,  {Sano} H.,  {Torii} K.,  {Yamamoto} H.,  {Acero}
  F.,   {Fukui} Y.,  2014, \mndoi [\apj] {10.1088/0004-637X/788/1/94}, \href
  {http://adsabs.harvard.edu/abs/2014ApJ...788...94F} {788, 94}

\bibitem[\protect\citeauthoryear{{Fukui}}{{Fukui}}{2008}]{Fukui:2008}
{Fukui} Y.,  2008, in {Aharonian} F.~A.,  {Hofmann} W.,   {Rieger} F.,  eds,
  American Institute of Physics Conference Series Vol. 1085, American Institute
  of Physics Conference Series. pp 104--111 (\mn@eprint {arXiv} {0810.5416}),
  \mndoi{10.1063/1.3076625}

\bibitem[\protect\citeauthoryear{{Fukui} et~al.,}{{Fukui}
  et~al.}{2017}]{Fukui:2017}
{Fukui} Y.,  et~al., 2017, preprint, \href
  {http://adsabs.harvard.edu/abs/2017arXiv170807911F} {} (\mn@eprint {arXiv}
  {1708.07911})

\bibitem[\protect\citeauthoryear{{Gabici}, {Aharonian}  \& {Blasi}}{{Gabici}
  et~al.}{2007}]{Gabici:2007}
{Gabici} S.,  {Aharonian} F.~A.,   {Blasi} P.,  2007, \mndoi [\apss]
  {10.1007/s10509-007-9427-6}, \href
  {http://adsabs.harvard.edu/abs/2007Ap%26SS.309..365G} {309, 365}

\bibitem[\protect\citeauthoryear{{Gooch}}{{Gooch}}{1996}]{Gooch:1996}
{Gooch} R.,  1996, in {Jacoby} G.~H.,  {Barnes} J.,  eds,  Astronomical Society
  of the Pacific Conference Series Vol. 101, Astronomical Data Analysis
  Software and Systems V. p.~80

\bibitem[\protect\citeauthoryear{{Heinz} et~al.,}{{Heinz}
  et~al.}{2015}]{Heinz:2015}
{Heinz} S.,  et~al., 2015, \mndoi [\apj] {10.1088/0004-637X/806/2/265}, \href
  {http://adsabs.harvard.edu/abs/2015ApJ...806..265H} {806, 265}

\bibitem[\protect\citeauthoryear{{Jones} et~al.,}{{Jones}
  et~al.}{2012}]{Jones:2012}
{Jones} P.~A.,  et~al., 2012, \mndoi [\mnras]
  {10.1111/j.1365-2966.2011.19941.x}, \href
  {http://adsabs.harvard.edu/abs/2012MNRAS.419.2961J} {419, 2961}

\bibitem[\protect\citeauthoryear{{Jones}, {Burton}, {Cunningham}, {Tothill}  \&
  {Walsh}}{{Jones} et~al.}{2013}]{Jones:2013}
{Jones} P.~A.,  {Burton} M.~G.,  {Cunningham} M.~R.,  {Tothill} N.~F.~H.,
  {Walsh} A.~J.,  2013, \mndoi [\mnras] {10.1093/mnras/stt717}, \href
  {http://adsabs.harvard.edu/abs/2013MNRAS.433..221J} {433, 221}

\bibitem[\protect\citeauthoryear{{Kavanagh}, {Sasaki}, {Bozzetto},
  {Filipovi{\'c}}, {Points}, {Maggi}  \& {Haberl}}{{Kavanagh}
  et~al.}{2015}]{Kavanagh:2015}
{Kavanagh} P.~J.,  {Sasaki} M.,  {Bozzetto} L.~M.,  {Filipovi{\'c}} M.~D.,
  {Points} S.~D.,  {Maggi} P.,   {Haberl} F.,  2015, \mndoi [\aap]
  {10.1051/0004-6361/201424354}, \href
  {http://adsabs.harvard.edu/abs/2015A%26A...573A..73K} {573, A73}

\bibitem[\protect\citeauthoryear{{Kissmann}}{{Kissmann}}{2014}]{Kissmann:2014}
{Kissmann} R.,  2014, \mndoi [Astroparticle Physics]
  {10.1016/j.astropartphys.2014.02.002}, \href
  {http://adsabs.harvard.edu/abs/2014APh....55...37K} {55, 37}

\bibitem[\protect\citeauthoryear{{Kruijssen}, {Dale}  \&
  {Longmore}}{{Kruijssen} et~al.}{2015}]{Kruijssen:2015}
{Kruijssen} J.~M.~D.,  {Dale} J.~E.,   {Longmore} S.~N.,  2015, \mndoi [\mnras]
  {10.1093/mnras/stu2526}, \href
  {http://adsabs.harvard.edu/abs/2015MNRAS.447.1059K} {447, 1059}

\bibitem[\protect\citeauthoryear{{Ladd}, {Purcell}, {Wong}  \&
  {Robertson}}{{Ladd} et~al.}{2005}]{Ladd:2005}
{Ladd} N.,  {Purcell} C.,  {Wong} T.,   {Robertson} S.,  2005, \mndoi [\pasa]
  {10.1071/AS04068}, \href {http://adsabs.harvard.edu/abs/2005PASA...22...62L}
  {22, 62}

\bibitem[\protect\citeauthoryear{{Landsman}}{{Landsman}}{1993}]{IDLastro:1993}
{Landsman} W.~B.,  1993, in {Hanisch} R.~J.,  {Brissenden} R.~J.~V.,   {Barnes}
  J.,  eds,  Astronomical Society of the Pacific Conference Series Vol. 52,
  Astronomical Data Analysis Software and Systems II. p.~246

\bibitem[\protect\citeauthoryear{{Lau} et~al.,}{{Lau}
  et~al.}{2017a}]{Lau:2017PASA}
{Lau} J.~C.,  et~al., 2017a, \mndoi [\pasa] {10.1017/pasa.2017.59}, \href
  {http://adsabs.harvard.edu/abs/2017PASA...34...64L} {34, e064}

\bibitem[\protect\citeauthoryear{{Lau} et~al.,}{{Lau} et~al.}{2017b}]{Lau:2017}
{Lau} J.~C.,  et~al., 2017b, \mndoi [\mnras] {10.1093/mnras/stw2692}, \href
  {http://adsabs.harvard.edu/abs/2017MNRAS.464.3757L} {464, 3757}

\bibitem[\protect\citeauthoryear{{Mathews} et~al.,}{{Mathews}
  et~al.}{2013}]{Mathews:2013}
{Mathews} G.~S.,  et~al., 2013, \mndoi [\aap] {10.1051/0004-6361/201321600},
  \href {http://adsabs.harvard.edu/abs/2013A%26A...557A.132M} {557, A132}

\bibitem[\protect\citeauthoryear{{Maxted} et~al.,}{{Maxted}
  et~al.}{2012}]{Maxted:2012}
{Maxted} N.,  et~al., 2012, \mndoi [\mnras] {10.1111/j.1365-2966.2012.20766.x},
  \href {http://adsabs.harvard.edu/abs/2012MNRAS.422.2230M} {422, 2230}

\bibitem[\protect\citeauthoryear{{Maxted} et~al.,}{{Maxted}
  et~al.}{2013}]{Maxted:2013rxj}
{Maxted} N.,  et~al., 2013, \mndoi [\pasa] {10.1017/pasa.2013.35}, \href
  {http://adsabs.harvard.edu/abs/2013PASA...30...55M} {30, e055}

\bibitem[\protect\citeauthoryear{{Maxted} et~al.,}{{Maxted}
  et~al.}{2018}]{Maxted:2018}
{Maxted} N.,  et~al., 2018, \mndoi [\mnras] {10.1093/mnras/stx2727}, \href
  {http://adsabs.harvard.edu/abs/2018MNRAS.474..662M} {474, 662}

\bibitem[\protect\citeauthoryear{{McClure-Griffiths}, {Dickey}, {Gaensler},
  {Green}, {Haverkorn}  \& {Strasser}}{{McClure-Griffiths}
  et~al.}{2005}]{McClure:2005}
{McClure-Griffiths} N.~M.,  {Dickey} J.~M.,  {Gaensler} B.~M.,  {Green} A.~J.,
  {Haverkorn} M.,   {Strasser} S.,  2005, \mndoi [\apjs] {10.1086/430114},
  \href {http://adsabs.harvard.edu/abs/2005ApJS..158..178M} {158, 178}

\bibitem[\protect\citeauthoryear{{Minamidani} et~al.,}{{Minamidani}
  et~al.}{2015}]{Minamidani:2015}
{Minamidani} T.,  et~al., 2015, in EAS Publications Series. pp 193--194,
  \mndoi{10.1051/eas/1575036}

\bibitem[\protect\citeauthoryear{{Molinari} et~al.,}{{Molinari}
  et~al.}{2010}]{Molinari:2010}
{Molinari} S.,  et~al., 2010, \mndoi [\pasp] {10.1086/651314}, \href
  {http://adsabs.harvard.edu/abs/2010PASP..122..314M} {122, 314}

\bibitem[\protect\citeauthoryear{{Moriguchi}, {Tamura}, {Tawara}, {Sasago},
  {Yamaoka}, {Onishi}  \& {Fukui}}{{Moriguchi} et~al.}{2005}]{Moriguchi:2005}
{Moriguchi} Y.,  {Tamura} K.,  {Tawara} Y.,  {Sasago} H.,  {Yamaoka} K.,
  {Onishi} T.,   {Fukui} Y.,  2005, \mndoi [\apj] {10.1086/432653}, \href
  {http://adsabs.harvard.edu/abs/2005ApJ...631..947M} {631, 947}

\bibitem[\protect\citeauthoryear{{Norris} et~al.,}{{Norris}
  et~al.}{2011}]{Norris:2011}
{Norris} R.~P.,  et~al., 2011, \mndoi [\pasa] {10.1071/AS11021}, \href
  {http://adsabs.harvard.edu/abs/2011PASA...28..215N} {28, 215}

\bibitem[\protect\citeauthoryear{{Pierce-Price} et~al.,}{{Pierce-Price}
  et~al.}{2000}]{Pierce:2000}
{Pierce-Price} D.,  et~al., 2000, \mndoi [\apjl] {10.1086/317884}, \href
  {http://adsabs.harvard.edu/abs/2000ApJ...545L.121P} {545, L121}

\bibitem[\protect\citeauthoryear{{Pineda}, {Goldsmith}, {Chapman}, {Snell},
  {Li}, {Cambr{\'e}sy}  \& {Brunt}}{{Pineda} et~al.}{2010}]{Pineda:2010}
{Pineda} J.~L.,  {Goldsmith} P.~F.,  {Chapman} N.,  {Snell} R.~L.,  {Li} D.,
  {Cambr{\'e}sy} L.,   {Brunt} C.,  2010, \mndoi [\apj]
  {10.1088/0004-637X/721/1/686}, \href
  {http://adsabs.harvard.edu/abs/2010ApJ...721..686P} {721, 686}

\bibitem[\protect\citeauthoryear{{Planck Collaboration} et~al.,}{{Planck
  Collaboration} et~al.}{2015}]{Ade:2015}
{Planck Collaboration} et~al., 2015, \mndoi [\aap]
  {10.1051/0004-6361/201424955}, \href
  {http://adsabs.harvard.edu/abs/2015A%26A...582A..31P} {582, A31}

\bibitem[\protect\citeauthoryear{{Rebolledo} et~al.,}{{Rebolledo}
  et~al.}{2016}]{Rebolledo:2016}
{Rebolledo} D.,  et~al., 2016, \mndoi [\mnras] {10.1093/mnras/stv2776}, \href
  {http://adsabs.harvard.edu/abs/2016MNRAS.456.2406R} {456, 2406}

\bibitem[\protect\citeauthoryear{{Risacher} et~al.,}{{Risacher}
  et~al.}{2016}]{Risacher:2016}
{Risacher} C.,  et~al., 2016, \mndoi [\aap] {10.1051/0004-6361/201629045},
  \href {http://adsabs.harvard.edu/abs/2016A%26A...595A..34R} {595, A34}

\bibitem[\protect\citeauthoryear{{Rowell}}{{Rowell}}{2017}]{Rowell:2017sofia}
{Rowell} G. e.~a.,  2017, SOFIA Proposal P06-0122

\bibitem[\protect\citeauthoryear{{Sano}}{{Sano}}{2017}]{Sano:2017alma}
{Sano} H. e.~a.,  2017, ALMA Proposal XXXXXX

\bibitem[\protect\citeauthoryear{{Sano} et~al.,}{{Sano}
  et~al.}{2013}]{Sano:2013}
{Sano} H.,  et~al., 2013, \mndoi [\apj] {10.1088/0004-637X/778/1/59}, \href
  {http://adsabs.harvard.edu/abs/2013ApJ...778...59S} {778, 59}

\bibitem[\protect\citeauthoryear{{Sano} et~al.,}{{Sano}
  et~al.}{2017a}]{Sano:2017arXiv}
{Sano} H.,  et~al., 2017a, preprint, \href
  {http://adsabs.harvard.edu/abs/2017arXiv170605763S} {} (\mn@eprint {arXiv}
  {1706.05763})

\bibitem[\protect\citeauthoryear{{Sano} et~al.,}{{Sano}
  et~al.}{2017b}]{Sano:2017ApJ}
{Sano} H.,  et~al., 2017b, \mndoi [\apj] {10.3847/1538-4357/aa73e0}, \href
  {http://adsabs.harvard.edu/abs/2017ApJ...843...61S} {843, 61}

\bibitem[\protect\citeauthoryear{{Schuller} et~al.,}{{Schuller}
  et~al.}{2017}]{Schuller:2017}
{Schuller} F.,  et~al., 2017, \mndoi [\aap] {10.1051/0004-6361/201628933},
  \href {http://adsabs.harvard.edu/abs/2017A%26A...601A.124S} {601, A124}

\bibitem[\protect\citeauthoryear{{Strong} \& {Moskalenko}}{{Strong} \&
  {Moskalenko}}{1998}]{Strong:1998}
{Strong} A.~W.,  {Moskalenko} I.~V.,  1998, \mndoi [\apj] {10.1086/306470},
  \href {http://adsabs.harvard.edu/abs/1998ApJ...509..212S} {509, 212}

\bibitem[\protect\citeauthoryear{{Strong}, {Moskalenko}  \& {Reimer}}{{Strong}
  et~al.}{2004}]{Strong:2004}
{Strong} A.~W.,  {Moskalenko} I.~V.,   {Reimer} O.,  2004, \mndoi [\apj]
  {10.1086/423193}, \href {http://adsabs.harvard.edu/abs/2004ApJ...613..962S}
  {613, 962}

\bibitem[\protect\citeauthoryear{{Umemoto} et~al.,}{{Umemoto}
  et~al.}{2017}]{Umemoto:2017}
{Umemoto} T.,  et~al., 2017, \mndoi [\pasj] {10.1093/pasj/psx061}, \href
  {http://adsabs.harvard.edu/abs/2017PASJ...69...78U} {69, 78}

\bibitem[\protect\citeauthoryear{{Urquhart} et~al.,}{{Urquhart}
  et~al.}{2018}]{Urquhart:2018}
{Urquhart} J.~S.,  et~al., 2018, \mndoi [\mnras] {10.1093/mnras/stx2258}, \href
  {http://adsabs.harvard.edu/abs/2018MNRAS.473.1059U} {473, 1059}

\bibitem[\protect\citeauthoryear{{Vall{\'e}e}}{{Vall{\'e}e}}{2014}]{Vallee:2014}
{Vall{\'e}e} J.~P.,  2014, \mndoi [\aj] {10.1088/0004-6256/148/1/5}, \href
  {http://adsabs.harvard.edu/abs/2014AJ....148....5V} {148, 5}

\bibitem[\protect\citeauthoryear{{Vall{\'e}e}}{{Vall{\'e}e}}{2016}]{Vallee:2016}
{Vall{\'e}e} J.~P.,  2016, \mndoi [\aj] {10.3847/0004-6256/151/3/55}, \href
  {http://adsabs.harvard.edu/abs/2016AJ....151...55V} {151, 55}

\bibitem[\protect\citeauthoryear{{Voisin}, {Rowell}, {Burton}, {Walsh}, {Fukui}
   \& {Aharonian}}{{Voisin} et~al.}{2016}]{Voisin:2016}
{Voisin} F.,  {Rowell} G.,  {Burton} M.~G.,  {Walsh} A.,  {Fukui} Y.,
  {Aharonian} F.,  2016, \mndoi [\mnras] {10.1093/mnras/stw473}, \href
  {http://adsabs.harvard.edu/abs/2016MNRAS.458.2813V} {458, 2813}

\bibitem[\protect\citeauthoryear{{Walker} et~al.,}{{Walker}
  et~al.}{2004}]{Walker:2004}
{Walker} C.~K.,  et~al., 2004, in {Oschmann} Jr. J.~M.,  ed.,  \procspie Vol.
  5489, Ground-based Telescopes. pp 470--480, \mndoi{10.1117/12.551424}

\bibitem[\protect\citeauthoryear{{Williams}, {de Geus}  \& {Blitz}}{{Williams}
  et~al.}{1994}]{Williams:1994}
{Williams} J.~P.,  {de Geus} E.~J.,   {Blitz} L.,  1994, \mndoi [\apj]
  {10.1086/174279}, \href {http://adsabs.harvard.edu/abs/1994ApJ...428..693W}
  {428, 693}

\bibitem[\protect\citeauthoryear{{Wilson}, {Jefferts}  \& {Penzias}}{{Wilson}
  et~al.}{1970}]{Wilson:1970}
{Wilson} R.~W.,  {Jefferts} K.~B.,   {Penzias} A.~A.,  1970, \mndoi [\apjl]
  {10.1086/180567}, \href {http://adsabs.harvard.edu/abs/1970ApJ...161L..43W}
  {161, L43}

\bibitem[\protect\citeauthoryear{{Wilson}, {Rohlfs}  \&
  {H{\"u}ttemeister}}{{Wilson} et~al.}{2009}]{Wilson:2009}
{Wilson} T.~L.,  {Rohlfs} K.,   {H{\"u}ttemeister} S.,  2009, {Tools of Radio
  Astronomy}.
Springer-Verlag, \mndoi{10.1007/978-3-540-85122-6}

\bibitem[\protect\citeauthoryear{{Yan} et~al.,}{{Yan} et~al.}{2017}]{Yan:2017}
{Yan} Q.-Z.,  et~al., 2017, \mndoi [\mnras] {10.1093/mnras/stx1724}, \href
  {http://adsabs.harvard.edu/abs/2017MNRAS.471.2523Y} {471, 2523}

\makeatother
\end{thebibliography}
